\documentclass[journal,onecolumn]{IEEEtran}
\usepackage[T1]{fontenc}
\usepackage{cite}
\usepackage{amsmath,amssymb,amsfonts,amsthm}
\usepackage{algorithmic}
\usepackage{graphicx}
\usepackage{mathtools}
\usepackage{balance}
\usepackage{hyperref}
\hypersetup{hidelinks}
\usepackage{multirow}
\usepackage{enumerate}
\usepackage{bbm}
\usepackage{bm}
\usepackage{url}

\usepackage{booktabs}

\usepackage{textcomp}
 
\def\BibTeX{{\rm B\kern-.05em{\sc i\kern-.025em b}\kern-.08em
    T\kern-.1667em\lower.7ex\hbox{E}\kern-.125emX}}

\newtheorem{theorem}{Theorem}
\newtheorem{lemma}[theorem]{Lemma}
\newtheorem{corollary}[theorem]{Corollary}
\newtheorem{proposition}[theorem]{Proposition}
\newtheorem{definition}{Definition}
\newtheorem{remark}{Remark}
\newtheorem{assumption}{Assumption}
\newtheorem{example}{Example}

\newcommand{\Fq}{\mathbb{F}_q}
\newcommand{\N}{\mathbf{N}}
\newcommand{\A}{\mathbf{A}}
\newcommand{\B}{\mathbf{B}}
\newcommand{\C}{\mathbf{C}}
\newcommand{\btheta}{B^{(\theta)}}
\newcommand{\Ctheta}{C^{(\theta)}}
\newcommand{\CPIR}{C_{\mathrm{PIR}}}

\newcommand{\indep}{\perp\!\!\!\perp}
\newcommand{\Hb}{H_2}
\newcommand{\HcondBA}{H(\btheta \mid \A)}
\newcommand{\Yans}{\mathsf{Y}}
\newcommand{\Ntheta}{\N^{(\theta)}}
\newcommand{\Nell}{\N^{(\ell)}}
    
\begin{document}

\title{Functional Converses for Private Matrix-Product Retrieval under
Library-Wide Decoder-Side Correlation}
\author{Malihe~Aliasgari,~\IEEEmembership{Member,~IEEE,}
        and~Yousef~Nejatbakhsh%
\thanks{M.~Aliasgari is with the Department of Mathematical Sciences, Kean
University, Union, NJ 07083 USA (e-mail: malihe.aliasgari@kean.edu).}%
\thanks{Y.~Nejatbakhsh is with Rutgers University, New Brunswick, NJ USA
(e-mail: nejatbakhsh.y@gmail.com).}%
\thanks{Corresponding author: Malihe Aliasgari.}}

\markboth{}%
{Aliasgari \MakeLowercase{\textit{et al.}}: Functional Converses for Private Matrix-Product Retrieval}

\maketitle

\begin{abstract}
We consider private retrieval of the matrix product $\A^\top\B^{(\theta)}$ from
replicated $P_c$-colluding servers when $\B^{(\ell)}=\A+\N^{(\ell)}$, with the
innovations mutually independent and independent of the user-held
$\A\in\Fq^{m\times m}$. The servers store the full library, correlated
with $\A$, which changes both the communication problem and the security
definition: a secrecy condition including the stored library cannot hold, as
the library itself partly reveals $\A$. We therefore measure only additional
leakage and require index privacy conditioned on $\A$;
both reduce to classical notions under independence.
For $\A$ of rank $k<m$ we prove, with no structural hypothesis on the map,
$D^*\geq H(\A^\top\B^{(\theta)}\mid\A)/C_{\mathrm{PIR}}(N,P_c,L)-o(km)$,
where $C_{\mathrm{PIR}}(N,P_c,L)$ is the replicated $P_c$-colluding
private-information-retrieval capacity. The proof rests on a colluding converse for servers storing arbitrary
refinements of the demanded messages, whose answers need not be measurable
functions of the functional-message library, and a privacy-transfer
characterization.
Publishing a canonical row-space representative makes the bound achievable for
arbitrary independent and identically distributed innovations under a
public-setup measure. Under a strengthened observational model---exact correctness, common
conditional library support, and full-transcript security---concealment forces
$D\geq H(G_m\N^{(\theta)})/C_{\mathrm{PIR}}$, where $G_m$ spans the linear
\emph{join} of all row spaces in the support of the operand distribution; this
extends to
vanishing error under a simultaneous-decodability assumption.
In the full-rank case product recovery reduces to innovation recovery:
$D^*=H(\N^{(\theta)})/C_{\mathrm{PIR}}+o(m^2)$.
\end{abstract}


\section{Introduction}
\label{sec:intro}

Secure distributed matrix multiplication protects the user's operand from the
workers that compute on it~\cite{chang2018capacity,aliasgari2020private}.
Private information retrieval protects a requested file
index~\cite{sun2017capacity,sun2018colluding}. Fully private matrix-computation
models combine variants of these
objectives~\cite{ulukus2022private,zhu2023information,tauz2023fully,jia2021capacity,kakar2019capacity}.
Coded-computation techniques underlie these
lines~\cite{yu2018polynomial,yu2019lagrange,li2020coded,yu2022entangled,malak2024distributed}.
\emph{These protocols assume the user's data and the library are statistically
independent}---yet in iterative optimization, federated pipelines, and data
updates, the desired matrix is correlated with data the user already
holds~\cite{mcmahan2017communication,vithana2024pruw}, and we are not aware of a
characterization of how such decoder-side
correlation affects the fundamental communication cost.
We address this gap.

\subsection{This Work and Main Results}
\label{subsec:thiswork}

We study a model in which \emph{every} library matrix is correlated with the
user's data through a common latent state. Every library matrix has the form
$\B^{(\ell)}=\A+\Nell$, where the innovations are mutually independent,
independent of~$\A$, and identically distributed. This models candidate updates
around a common state. Unlike target-only side-information PIR, the correlation
structure does not depend on the requested index. In the full-rank base case we
prove the tight characterization
$D^*=H(\btheta\mid\A)/\CPIR+o(m^2)=H(\Ntheta)/\CPIR+o(m^2)$: the PIR capacity
factor determines the privacy overhead, while the conditional entropy
determines the gain from correlation. The reduction to colluding PIR requires care because every file is correlated
with~$\A$. The side-information converse of~\cite{kadhe2017private} cannot be
applied to the raw library. We therefore reduce to the $\A$-independent
innovation library, with achievability by \emph{linear} Slepian--Wolf
sketching. Two further obstacles are handled along the way: privacy under
conditioning, and the provably false pointwise transfer
(Remark~\ref{rem:singleton}).

The paper's \emph{main} results lie in the rank-deficient regime
$\mathrm{rank}(\A)=k<m$ (Section~\ref{subsec:extension}), where invertibility
fails and recovering $\Ctheta$ no longer determines $\btheta$: a
functional-rate converse valid for every rank-$k$ operand, via a universal
privacy-transfer characterization and a refined-storage colluding converse, and
a \emph{published/concealed contrast} in which the communication requirement
depends on whether the functional structure is public or concealed.

One concrete interpretation of the model is a \emph{private update-selection
loop}. The user holds a current private state matrix~$\A$, while the replicated
library contains candidate matrices generated as perturbations of that same
state, $\B^{(\ell)}=\A+\Nell$. The user wishes to evaluate or apply one
candidate through $\A^\top\B^{(\theta)}$ without revealing the selected
index~$\theta$, and without leaking additional information about~$\A$ beyond
what the correlated library already stored at the servers discloses. Because
every candidate shares the same baseline state, the correlation is library-wide
rather than target-only.

\noindent\textit{Scope.} The full-rank result applies to the
replicated-storage, memoryless, library-wide additive model; coded storage is
open.

\subsection{Main Contributions}
\label{subsec:contributions}

\begin{enumerate}
  \item \textbf{A functional converse for arbitrary rank-$k$ operands (main result).}
        Our main result concerns operands $\A$ of rank $k<m$. For every such
        operand the download satisfies
        $D^*\geq H(\A^\top\btheta\mid\A)/\CPIR-o(km)$, with no structural
        hypothesis on the map (Theorem~\ref{prop:rankdef_lb}). The proof has two
        ingredients. Proposition~\ref{thm:transfer} transfers privacy from the
        raw library to the induced functional library.
        Theorem~\ref{thm:refined_converse} then supplies a colluding-PIR
        converse when the servers store refinements of the requested functional
        messages---a fixed-function, private-\emph{index} counterpart of private
        computation~\cite{sun2019capacity,obead2022private}.
  \item \textbf{Published versus concealed structure.}
        Under the public-setup metric $D_{\mathrm{pub}}$, publishing a
        canonical row-space representative yields a \emph{capacity theorem} at
        the functional rate (Theorem~\ref{thm:rankk_public}). Under stronger
        full-transcript, common-support, and exact-correctness assumptions,
        Theorem~\ref{thm:rankk_impossible} establishes a \emph{conditional
        join-subspace converse} for concealed functional structure. These are
        not symmetric capacity branches: one is a capacity theorem under
        published structure, the other a strengthened-observation converse,
        under different hypotheses and \emph{different download
        accounting}---$D_{\mathrm{pub}}$ conditioned on the announced
        representative versus the unconditional $D$
        (Remark~\ref{rem:dichotomy_consistency}).
  \item \textbf{Full-rank base case.}
        In the full-rank regime the optimal download is characterized to
        first order, $D^*=\HcondBA/\CPIR+o(m^2)$ (Theorems~\ref{thm:converse}
        and~\ref{thm:achievability}). The problem reduces to innovation
        retrieval: any admissible protocol induces an innovation-PIR instance
        (Theorem~\ref{thm:equivalence}), with achievability by linear source
        coding composed with colluding PIR.
  \item \textbf{Security model.}
        Under genuine library-wide correlation the standard library-inclusive
        security condition is infeasible (Proposition~\ref{prop:infeasible}). We work
        instead under conditioned data security
        (Definitions~\ref{def:security} and~\ref{def:security_public}) and
        side-information-robust index privacy (Definition~\ref{def:index}),
        both of which reduce to the classical notions when the operand and
        library are independent.
  \item \textbf{Finite-length accounting.}
        We give a finite-length design accounting for the composition of a
        linear syndrome code with an explicit colluding-PIR scheme at exact
        subpacketization (Proposition~\ref{thm:finiteblock},
        Appendix~\ref{app:finiteblock}), quantifying the padding and dispersion
        costs and their $O(1/m)$ decay.
\end{enumerate}

\smallskip
\noindent\textbf{Structure of the full-rank proof.} The full-rank converse
proceeds by conditioning on $\A$, reducing to the innovation library via
Lemma~\ref{lem:innovation}, and averaging the resulting per-realization bounds
(Section~\ref{subsec:converse}, Appendix~\ref{app:converse}).

\subsection{Relationship to Prior Work}
\label{subsec:priorwork}

PIR with side information~\cite{kadhe2017private} and PIR with private side
information~\cite{chen2018private} assume an independent message library and give
the user a subset of library messages as a decoder-side resource, and they
assume the retrieval structure from the outset; here the retrieval structure is
a \emph{consequence} of robust index privacy, the additive innovation model,
and full-rank invertibility, not a
starting assumption. Prior SDMM and private matrix multiplication
works~\cite{zhu2023information,tauz2023fully,jia2021capacity,kim2023fully,
tajeddine2024modular,cartor2024secure} neither consider correlated inputs nor
address the infeasibility of the library-inclusive security condition under
correlation. Recent Reed--Solomon-based PSDMM
schemes~\cite{wang2025reedsolomon} improve storage and download efficiency for a
confidentially selected matrix from an MDS-coded public library; they retain the
conventional independence between the user's private matrix and the library, and
therefore do not address library-wide decoder-side correlation or the relative
security definitions considered here.

Malak~\cite{malak2024distributed} studies functional compression for
$\A^\top\B$ when two correlated sources are held at separate encoders and the
receiver minimizes their sum rate, achieving rates below the Slepian--Wolf
point through structured codes. That model has neither a replicated candidate
library nor a private retrieval index. In our setting, $\A$ is decoder side
information, the servers store a replicated collection $\{\B^{(\ell)}\}$, and
the communication problem includes colluding-server index privacy and
conditioned data security. Thus~\cite{malak2024distributed} addresses
distributed functional source coding of a single product, whereas our results
characterize the additional PIR cost of private retrieval from a correlated
library and the effect of published versus concealed decoder-private functional
structure.

The closest structural relatives are private federated submodel learning (FSL)
and private
read--update--write~\cite{ulukus2022private,jia2022xsecure,vithana2024pruw}. In
those settings the server-side stored objects are independent of the reader's
side information during the read phase, because correlations are hidden through
secure ($X$-secure or noise-protected) writes. Here the correlation is \emph{cleartext and library-wide}: the stored library
itself is correlated with~$\A$. This makes the library-inclusive security
condition infeasible (Proposition~\ref{prop:infeasible}), and forces both the
conditioned security definition and the innovation reduction. Neither problem
contains the other.

\emph{Private function retrieval, private computation, and side
information.} A closely adjacent line studies retrieval of a \emph{linear
function} of the library while hiding the function itself: private
computation~\cite{sun2019capacity} and private linear/function
computation~\cite{obead2022private} over replicated or coded databases, where
the user's coefficient vector is the secret and the capacity is governed by the
number or rank of the independent underlying datasets; the nonlinear case over
replicated databases is treated in~\cite{obead2019nonlinear}. PIR with ordinary or private
side information~\cite{kadhe2017private,chen2018private} considers users
possessing \emph{subsets} of database messages---a cache of ordinary messages,
or of privately-held ones. A more closely related variant is PIR with
\emph{coded} side information, where the user possesses a linear combination of
a subset of library messages~\cite{heidarzadeh2021role,kazemi2019multiserver,kazemi2019private}. In those
models the underlying library remains independent and the coded side
information is a user-side resource. Private function retrieval and private linear
computation hide the requested function or its coefficients over an otherwise
specified message library. Our setting differs in three respects: the function
$\A^\top(\cdot)$ is \emph{fixed} by private decoder-side state rather than
chosen from a candidate set; the requested \emph{file index} remains private;
and the servers store correlated \emph{refinements} of the functional messages
rather than the functional messages themselves. The refined-storage feature
alone is not absent from all neighboring private-computation models:
\cite{obead2022private} and the general formulation of~\cite{obead2019nonlinear}
permit answers to depend on underlying stored messages that need not be
determined by the candidate-function family. Our distinction is the combination
of $P_c$-collusion, a private file index under a map induced by decoder-private
state, conditionally independent per-file functional messages, and arbitrary
possibly cross-file-correlated refinements.
Theorem~\ref{thm:refined_converse} establishes a converse under this
combination. In our setting the correlation is
library-wide and arises from a common private state $\A$, rather than being a
user-side resource over an independent library, and the requested function is
itself induced by that decoder-private state.

To the best of our knowledge, no prior work combines library-wide additive
correlation, a private file index, a decoder-private matrix-product function,
and refined storage. Our main contribution is the functional converse for
arbitrary rank-$k$ operands in Theorem~\ref{prop:rankdef_lb}. We also
characterize the full-rank and published-structure regimes and establish a
concealed-structure converse under the stronger assumptions of
Theorem~\ref{thm:rankk_impossible}.

Technically, the closest converses are colluding
PIR~\cite{sun2018colluding}, private computation~\cite{sun2019capacity},
private linear/function computation~\cite{obead2022private}, and the
dependent-PIR line~\cite{chen2020asymptotic}. Existing private-computation and
dependent-PIR converses allow the requested objects to be dependent functions
of underlying stored data; in several of those formulations, however, the
private index selects the \emph{function} itself, or the candidate function
collection contains sufficient information to recover the underlying datasets.
Our setting differs on several axes. The per-file function is fixed by the
decoder-private $\A$, and the private index is the \emph{file}, not the
function. The servers cannot evaluate the $\A$-dependent function directly, and
the scheme permits $P_c$-collusion. Crucially, the servers store the raw
matrices $\B^{(\ell)}$, not only the functional messages
$\A^\top\B^{(\ell)}$. The raw matrices may carry residual information absent
from the complete functional library, and the functional messages are
independent across files only after conditioning on $\A$. The server answers
therefore need not be deterministic functions of the functional messages, so
standard message-level converses do not directly cover this setting.
Remark~\ref{rem:prior_converse_comparison} identifies, for each neighbouring
converse, the specific hypothesis that our refined-storage model violates.
Theorem~\ref{thm:refined_converse} supplies the corresponding colluding
converse, and Proposition~\ref{thm:transfer} the accompanying transfer of
privacy from the raw correlated library to the $\A$-dependent coarsening.
Together they make the rank-$k$ bound possible for every fixed-rank operand.

\smallskip
\noindent\textbf{Organization.} Section~\ref{sec:formulation} formalizes the
correlated-library model, privacy and security constraints, and rank-profile
assumptions. Section~\ref{sec:results} treats the full-rank base case and
establishes the innovation-retrieval equivalence and tight download
characterization. Section~\ref{subsec:extension} contains the main functional
results for the rank-deficient regime: the privacy-transfer characterization,
the refined-storage converse, the functional lower bound, and the
published/concealed contrast. Section~\ref{sec:examples} gives
correlation-model examples and finite-length accounting.
Section~\ref{sec:discussion} discusses interpretation, limitations, and open
directions; Section~\ref{sec:conclusion} concludes. The appendices contain the
detailed converse, achievability, finite-length, and refined-storage proofs.

\section{Problem Formulation}
\label{sec:formulation}

\subsection{Notation}

$\Fq$: finite field, $q$ prime power.
$[k]\triangleq\{1,\ldots,k\}$.
Bold uppercase: matrices.
$H(\cdot)$: Shannon entropy; all entropies and mutual informations are
measured in bits, with $\log\triangleq\log_2$, so one $\Fq$-symbol carries
$\log q$ bits.
$I(\cdot\,;\cdot)$: mutual information.
$X\indep Y$: independence; $X\indep Y\mid Z$: conditional independence.
$\Yans_i^{(\theta)}$: server answers (avoids clash with matrix $\A$);
the demand is written $\Yans_i^{(\theta)}$ in the main text and
$\Yans_i^{[\theta]}$ in Appendix~\ref{app:kadhe}, where a fixed demand
value indexes the induced PIR instance.
\emph{Asymptotic convention:} all asymptotic statements ($o(\cdot)$, $O(\cdot)$,
and limits) are taken as $m\to\infty$ with the library size $L$, the number of
servers $N$, the collusion parameter $P_c$, and the field size $q$ held
\emph{fixed}, so $\CPIR(N,P_c,L)$ is a constant. The factor $L$
appears in the Fano bookkeeping of the converses, so the error penalties are
sublinear at the stated scale for fixed $L$ but not uniformly in~$L$.
All large-library statements therefore use the iterated limit ($m\to\infty$
first, then $L\to\infty$) specified in Remark~\ref{rem:iterated_limit}.

\subsection{Data Model}

Fix prime power $q$ and integer $m\geq 1$.
The user holds $\A\in\Fq^{m\times m}$.

\begin{assumption}[Full-rank regime]
\label{ass:fullrank}
In Sections~\ref{sec:formulation}--\ref{sec:results}, the rank-profile
axiom~(M1) is specialized to $r=m$: $P_A$ is supported on
$\mathrm{GL}_m(\Fq)$, so $\A$ is invertible almost surely.
\end{assumption}

\begin{remark}[Role of full rank]
\label{rem:fullrank}
Invertibility of~$\A$ is what reduces the matrix-product problem to retrieval:
recovering $\Ctheta$ then determines $\btheta$. The assumption cannot be
dropped---if $\A=\mathbf{0}$ a.s.\ then $D^*=0$ while $\HcondBA>0$, and for
$\mathrm{rank}(\A)=k<m$ only $\Ctheta$, not $\btheta$, need be recovered, so
functional compression may beat $\HcondBA/\CPIR$. The rank-deficient regime is
the subject of Section~\ref{subsec:extension}. Random matrices over $\Fq$ are
full-rank with probability $\prod_{i=1}^{m}(1-q^{i-m-1})\to c_q>0$, and
conditioning on full rank yields an admissible~$P_A$.
\end{remark}

There exists a library $\{\B^{(1)},\ldots,\B^{(L)}\}$,
$\B^{(\ell)}\in\Fq^{m\times m}$, replicated at $N$ servers.
The user wishes to compute
\begin{equation}
  \Ctheta = \A^\top\btheta, \qquad \theta\in[L].
  \label{eq:goal}
\end{equation}
We treat the invertible case first (Section~\ref{sec:results}) because it
identifies the innovation-retrieval baseline against which the functional
regime is measured. The main functional regime of the paper is the
rank-deficient case $r=k<m$, developed in Section~\ref{subsec:extension}.
\subsubsection{Library-Wide Additive Correlation Model}
Every library matrix is a noisy version of the common user-side state~$\A$:
\begin{equation}
  \B^{(\ell)} = \A + \Nell, \qquad \ell\in[L],
  \label{eq:additive}
\end{equation}
subject to the following assumptions.
\begin{enumerate}[(M1)]
  \item \emph{Rank profile.} $\A\sim P_A$ supported on the $m\times m$
        matrices over~$\Fq$ of fixed rank~$r$; we impose \emph{no} i.i.d.-entry
        assumption on~$\A$. In Sections~\ref{sec:formulation}--\ref{sec:results}
        we take $r=m$ (the full-rank regime, Assumption~\ref{ass:fullrank}); in
        Section~\ref{subsec:extension} we take $r=k<m$. (An i.i.d.-entry law
        on~$\A$ would be incompatible with a fixed rank profile: for i.i.d.\
        uniform entries the full-rank probability is
        $\prod_{i=1}^{m}(1-q^{-i})<1$.)
  \item \emph{Independent i.i.d.\ innovations.} $\N^{(1)},\ldots,\N^{(L)}$ are
        mutually independent and independent of~$\A$, with i.i.d.\ entries across
        matrix positions and across files, $N^{(\ell)}_{i,j}\overset{\mathrm{i.i.d.}}{\sim}P_N$.
        This is the \emph{library-wide} correlation: every file is a noisy copy
        of the same~$\A$. The i.i.d.\ structure is used by the achievability
        (to invoke an asymptotically optimal lossless source code); the converse
        uses only that the innovations are independent of~$\A$ and identically
        distributed across files (Remark~\ref{rem:assumption_split}).
  \item \emph{Identical marginals.} All $\Nell$ share the law $P_N$, hence all
        $\B^{(\ell)}$ share an identical marginal law; with (M2) the library is
        exchangeable across files, so no file is distinguished by its
        distribution. This is a statement about the library alone: the
        independence $I(\theta;\{\B^{(\ell)}\})=0$ does \emph{not} follow from
        identical marginals and is imposed separately by~(M4).
  \item \emph{Independent, symmetric retrieval demand.} $\theta$ is independent of
        $(\A,\B^{(1)},\ldots,\B^{(L)})$ and drawn by the user after the library is
        fixed and stored; we take $\theta$ uniform on $[L]$ (the download
        criterion is insensitive to this choice under index privacy;
        Remark~\ref{rem:download_criterion}). Queries depend on $(\theta,\A,\mathcal{R})$ and answers on the
        query and stored library (Section~\ref{subsec:system}).
  \item \emph{Local user randomness.} The private randomness $\mathcal{R}$ is an
        independently generated seed:
        $\mathcal{R}\indep(\theta,\A,\{\B^{(\ell)}\})$. This is a
        protocol-design assumption---the seed is drawn from a source unrelated
        to the demand, the operand, and the library---and we state it as a
        model axiom because every converse
        below uses it when quantifying over admissible protocols
        (e.g., hypothesis~(i) of Theorem~\ref{thm:refined_converse}).
\end{enumerate}

\begin{remark}[Which assumptions the converse versus the achievability needs]
\label{rem:assumption_split}
The converse (Theorem~\ref{thm:converse}) uses (M1) with $r=m$, the cross-file
and operand independence conditions of (M2)--(M3), and the demand and
randomness axioms (M4)--(M5); it does \emph{not} require the innovations to be memoryless i.i.d.\
\emph{within} a file---only that $\{\Nell\}$ are independent across files and
independent of~$\A$, so that the induced innovation library is an
independent-message library. The achievability (Theorem~\ref{thm:achievability})
additionally uses the i.i.d.\ entry structure of~(M2), which guarantees an
asymptotically optimal lossless linear source code at rate $H(\Ntheta)/m^2+o(1)$ bits per matrix entry.
\end{remark}

\begin{remark}[Model scope]
\label{rem:quantization}
In the motivating applications (Section~\ref{sec:intro}) candidate
matrices are noisy versions of a common state, so the
correlation is naturally \emph{library-wide} with $\theta$-independent
statistics; a target-only correlation model would make the correlation pattern
itself depend on the hidden index. The finite-field model is the standard
quantized representation for information-theoretic PIR and coded computation:
real-valued updates are quantized to $\Fq$-symbols, and the characterization
governs the communication cost after quantization; continuous-alphabet
analogues are outside our scope.
\end{remark}

\begin{lemma}[Innovation Equivalence]
\label{lem:innovation}
Under Assumption~\ref{ass:fullrank}, $\A$ is invertible, and
\begin{equation*}
  \Ctheta = \A^\top\btheta = \A^\top\A + \A^\top\Ntheta,
\end{equation*}
where $\A^\top\A$ is a function of $\A$ alone. Hence, given~$\A$, the map
$\Ntheta\mapsto\Ctheta$ is a bijection, so recovering $\Ctheta$ with vanishing
error is equivalent to recovering the innovation $\Ntheta=\btheta-\A$ with
vanishing error, and
\begin{equation}
  \HcondBA = H(\A+\Ntheta\mid\A) = H(\Ntheta\mid\A) = H(\Ntheta).
  \label{eq:innovation_entropy}
\end{equation}
The effective retrieval target is the innovation~$\Ntheta$, and the effective
message library $\{\Nell\}$ is independent of~$\A$ with identical marginals.
\end{lemma}
\begin{proof}
$\A^\top\A$ and the additive shift are deterministic given~$\A$; composing with
the invertible left-multiplication by~$\A^\top$ yields a bijection
$\Ntheta\mapsto\Ctheta$ for each fixed full-rank~$\A$. Equation
\eqref{eq:innovation_entropy} follows since conditioning on~$\A$ removes the
additive~$\A$ and $\Ntheta\indep\A$ by~(M2).
\end{proof}

\subsection{System Model}
\label{subsec:system}

The library is replicated across $N$ servers, and privacy is required against
any colluding set of at most $P_c$ servers, where
\begin{equation}
  0\leq P_c < N.
  \label{eq:pc_range}
\end{equation}
The constraint $P_c<N$ is necessary for meaningful private retrieval: at $P_c=N$
all servers collude, the $P_c$-colluding PIR capacity~\eqref{eq:pir_cap}
vanishes, and the download formula $\HcondBA/\CPIR$ diverges (no private scheme
exists). We therefore assume $P_c<N$ throughout. We focus on replicated storage
because it isolates the effect of library-wide correlation from storage-coding
effects; it is also the canonical baseline of the PIR-capacity
literature~\cite{sun2017capacity,sun2018colluding}. Coded storage is left open.

\textbf{Queries:} $Q_i^{(\theta)}=f_i(\theta,\A,\mathcal{R})$,
$i\in[N]$, $\mathcal{R}$ private user randomness.
In general we permit queries to depend on the user side information~$\A$, as is
standard in PIR with side information~\cite{kadhe2017private}; the converse
(Section~\ref{subsec:converse}) handles this generality. The achievability
scheme (Section~\ref{subsec:achievability}) uses queries that depend only on
$(\theta,\mathcal{R})$ and a public map, never on~$\A$, which is what makes the
side-information-robust privacy of Definition~\ref{def:index} hold and the
conditioned data security of Definition~\ref{def:security} automatic.

\textbf{Answers:}
$\Yans_i^{(\theta)}=g_i(Q_i^{(\theta)},\{\B^{(\ell)}\}_{\ell=1}^L)$.

\textbf{Correctness:}
\begin{equation}
  \Pr\!\left[\hat{\C}^{(\theta)}\neq\A^\top\btheta\right]
  \xrightarrow{m\to\infty}0.
  \label{eq:correctness}
\end{equation}
Under Assumption~\ref{ass:fullrank}, $\A$ is invertible almost surely,
so $\btheta=(\A^\top)^{-1}\Ctheta$ is uniquely determined by $(\Ctheta,\A)$.
Therefore recovering $\Ctheta=\A^\top\btheta$ with vanishing error is
equivalent to recovering $\btheta$ with vanishing error.
This equivalence holds only when $\A$ is full rank; it fails in the
rank-deficient regime, which is the regime where functional compression
becomes relevant (Section~\ref{subsec:extension}).

\textbf{Download cost:}
$D\triangleq\sum_{i=1}^{N}H(\Yans_i^{(\theta)})$.

\begin{remark}[Download criterion: realized demand versus worst case]
\label{rem:download_criterion}
Two criteria are common: the realized-demand download $D$ above ($\theta$
random inside the marginal entropies) and the worst case
$D_{\mathrm{wc}}\triangleq\max_{t\in[L]}\sum_i H(\Yans_i^{(t)})$. Under index
privacy (Definition~\ref{def:index} with $P_c\geq 1$) they coincide exactly:
given $\A=\mathbf{a}$ each single answer's law is invariant to the demand, hence
so is its unconditional law, and $H(\Yans_i^{(\theta)})=H(\Yans_i^{(t)})$ for
every~$t$. Independently of privacy, every converse in this paper uses only the
direction $H(\Yans_i^{(\theta)})\geq H(\Yans_i^{(\theta)}\mid\theta,\cdot)$
(e.g., in Proposition~\ref{prop:app_converse}), so the bounds hold for the
realized-demand criterion and a fortiori for the worst case; and every
achievability scheme constructed here is symmetric in~$\theta$, achieving both
criteria simultaneously.
\end{remark}

\subsection{Privacy and Security Constraints}

\begin{definition}[Side-Information-Robust Index Privacy]
\label{def:index}
For every $\mathcal{T}\subseteq[N]$ with $|\mathcal{T}|\leq P_c$:
\begin{equation}
  I\!\left(\theta;\bigl\{Q_i^{(\theta)},\Yans_i^{(\theta)}\bigr\}_{i\in\mathcal{T}},
  \{\B^{(\ell)}\}\;\Big|\;\A\right)=0,
  \label{eq:index_privacy}
\end{equation}
equivalently, for $P_A$-almost every realization $\mathbf{a}$,
$I(\theta;\{Q_i^{(\theta)},\Yans_i^{(\theta)}\}_{i\in\mathcal{T}},\{\B^{(\ell)}\}
\mid\A=\mathbf{a})=0$.
\end{definition}

\begin{remark}[Why privacy must be conditioned on $\A$]
\label{rem:index_equiv}
The converse (Section~\ref{subsec:converse}) builds, for each full-rank
realization~$\mathbf{a}$, a per-realization PIR instance by fixing $\A=\mathbf{a}$.
Ordinary unconditional index privacy
$I(\theta;\{Q_i,\Yans_i\}_{i\in\mathcal{T}},\{\B^{(\ell)}\})=0$ need
\emph{not} survive this conditioning. A simple XOR-type obstruction shows the
gap: if $V=\theta\oplus\A$ with $\theta,\A$ independent and uniform, then
$I(\theta;V)=0$ but $I(\theta;V\mid\A)=1$~bit---fixing~$\A$ can create
$\theta$-dependence in the server view. Definition~\ref{def:index} therefore
imposes the conditional form directly, which is what the per-realization
converse requires. This obstruction is the same construction used in
Proposition~\ref{thm:transfer}(b) to delimit the
boundary of universal privacy transfer.

The conditional form implies the unconditional one by averaging, and the two
coincide precisely for schemes whose query distribution is privacy-preserving
for each fixed side-information realization---in particular the
$\A$-independent PIR schemes used in our achievability
(Proposition~\ref{prop:sir_satisfiable}), for which the strengthened definition
costs nothing.
\end{remark}

\begin{proposition}[Standard PIR with $\A$-independent or public-parameter queries is side-information-robust]
\label{prop:sir_satisfiable}
Let $\pi$ be a $P_c$-colluding PIR scheme satisfying the independence assumptions
\begin{equation}
  \mathcal{R}\indep(\A,\{\B^{(\ell)}\},\theta),
  \qquad \theta\indep(\A,\{\B^{(\ell)}\})
  \label{eq:sir_indep}
\end{equation}
(automatic under (M4)--(M5)),
whose queries depend only on $(\theta,\mathcal{R})$ and a public map---not
on~$\A$, or on~$\A$ only through a \emph{declared public parameter}
$\mathsf{P}=f(\A)$ fixed before the protocol runs---and whose colluding-set
query distribution is independent of~$\theta$
for every fixed library realization (the standard per-realization privacy
property, required for every fixed value of the public parameter in the second
case). Then $\pi$ satisfies side-information-robust index privacy
(Definition~\ref{def:index}).
\end{proposition}
\begin{proof}
Fix $|\mathcal{T}|\leq P_c$ and $\A=\mathbf{a}$. If the queries depend on a
declared public parameter $\mathsf{P}=f(\A)$, conditioning on $\A=\mathbf{a}$
fixes $\mathsf{P}=f(\mathbf{a})$ as a constant, so in either case the queries
are, given $\A=\mathbf{a}$, functions of $(\theta,\mathcal{R})$ and constants.
Since answers are deterministic
in $(Q_{\mathcal{T}},\{\B^{(\ell)}\})$, it suffices that
$I(\theta;Q_{\mathcal{T}}^{(\theta)},\{\B^{(\ell)}\}\mid\A{=}\mathbf{a})=0$,
which splits by the chain rule into
$I(\theta;\{\B^{(\ell)}\}\mid\A{=}\mathbf{a})$---zero by (M4)---and
$I(\theta;Q_{\mathcal{T}}^{(\theta)}\mid\{\B^{(\ell)}\},\A{=}\mathbf{a})$---zero
because the queries are functions of $(\theta,\mathcal{R})$ with
$\mathcal{R}\indep(\A,\{\B^{(\ell)}\})$, so their conditional law is the law
given $\theta$ alone, which per-realization privacy (instantiated at the fixed
public parameter) makes $\theta$-independent.
\end{proof}

\noindent Thus side-information-robust privacy adds no design burden beyond
the already standard requirement that queries not depend on the user's side
information.

Prior works on SDMM~\cite{aliasgari2020private,chang2018capacity,tauz2023fully}
define data security as
\begin{equation}
  I\!\left(\A;\bigl\{Q_i^{(\theta)},\Yans_i^{(\theta)}\bigr\}_{i\in\mathcal{T}}\right)=0
  \quad\text{for every }|\mathcal{T}|\leq P_c.
  \label{eq:classical_security}
\end{equation}
Under independence of the complete library, $\A\indep\B^{(1:L)}$, the
transcript-only form~\eqref{eq:classical_security}, the library-inclusive form
$I(\A;\{Q_i,\Yans_i\}_{i\in\mathcal{T}},\{\B^{(\ell)}\})=0$, and the
library-conditioned form of Definition~\ref{def:security} coincide, and the
equivalence uses the protocol structure, not the chain rule alone. The chain
rule
$$I(\A;\{Q_i,\Yans_i\},\{\B^{(\ell)}\})$$
\begin{equation}
  =\underbrace{I(\A;\{\B^{(\ell)}\})}_{=\,0}
  +I(\A;\{Q_i,\Yans_i\}\mid\{\B^{(\ell)}\})
  \label{eq:chain_forms}
\end{equation}
identifies the inclusive and conditioned forms; the substantive direction is
that the transcript-only form implies them. Suppose
$I(\A;Q_{\mathcal{T}},\Yans_{\mathcal{T}})=0$, where
$Q_{\mathcal{T}}=f_{\mathcal{T}}(\A,\theta,\mathcal{R})$ and
$\Yans_{\mathcal{T}}=g_{\mathcal{T}}(Q_{\mathcal{T}},\B^{(1:L)})$ with
$\B^{(1:L)}\indep(\A,\theta,\mathcal{R})$. Monotonicity gives
$I(\A;Q_{\mathcal{T}})=0$; since $(\A,Q_{\mathcal{T}})$ is a function of
$(\A,\theta,\mathcal{R})$, the pair $(\A,Q_{\mathcal{T}})$ is independent of
$\B^{(1:L)}$, and combined with $\A\indep Q_{\mathcal{T}}$ the joint law
factorizes,
$P(\mathbf{a},q_{\mathcal{T}},b^{(1:L)})
=P(\mathbf{a})P(q_{\mathcal{T}})P(b^{(1:L)})$, i.e.\
$\A\indep(Q_{\mathcal{T}},\B^{(1:L)})$. Deterministic answer generation then
makes $(Q_{\mathcal{T}},\Yans_{\mathcal{T}},\B^{(1:L)})$ a function of
$(Q_{\mathcal{T}},\B^{(1:L)})$, so
$\A\indep(Q_{\mathcal{T}},\Yans_{\mathcal{T}},\B^{(1:L)})$ and in particular
$I(\A;Q_{\mathcal{T}},\Yans_{\mathcal{T}}\mid\B^{(1:L)})=0$. The converse
direction is~\eqref{eq:chain_forms} read right to left. (Without the protocol
structure the implication fails: $V=\A\oplus\B$ with independent uniform bits
has $I(\A;V)=0$ yet $I(\A;V\mid\B)=1$.)
Under correlation, the library-inclusive unconditional form becomes infeasible,
as we now state formally; the transcript-only form, while not formally ruled out,
ignores the adversary's actual side information---the correlated library the
servers hold---so conditioning on the library is necessary
(Remark~\ref{rem:transcript_only}).

\begin{proposition}[Infeasibility of Library-Inclusive Data Security under Correlation]
\label{prop:infeasible}
Consider the model~\eqref{eq:additive} with genuine library correlation:
$I(\A;\btheta)>0$, or equivalently---since
$H(\btheta\mid\A)=H(\Ntheta)$ by (M2)---$H(\Ntheta)<H(\btheta)$. (This requires
in particular that $\A$ be non-degenerate \emph{and} that $\Ntheta$ be
non-uniform; neither alone suffices. E.g., if $P_A$ is a point mass on some
$\mathbf{a}\in\mathrm{GL}_m(\Fq)$, which (M1) permits, then $\btheta$ is a
deterministic shift of $\Ntheta$ and $H(\btheta)=H(\Ntheta)$ however non-uniform
the innovation law is, the hypothesis fails, and the library-inclusive condition
is trivially feasible.) Then the library-inclusive
unconditional data-security condition
$I(\A;\{Q_i,\Yans_i\}_{i\in\mathcal{T}},\{\B^{(\ell)}\})=0$ is infeasible: it is
violated by the joint distribution alone, independently of any protocol.
\end{proposition}
\begin{proof}
By hypothesis and (M2),
\begin{align*}
  I(\A;\{\B^{(\ell)}\})&\;\geq\;I(\A;\B^{(\theta)})
  \;=\;H(\B^{(\theta)})-H(\B^{(\theta)}\mid\A)\\
  &\;=\;H(\B^{(\theta)})-H(\Ntheta)\;>\;0,
\end{align*}
and the chain rule gives
$I(\A;\{Q_i,\Yans_i\}_{i\in\mathcal{T}},\{\B^{(\ell)}\})\geq I(\A;\{\B^{(\ell)}\})>0$,
which already violates the constraint regardless of the protocol.
\end{proof}

\begin{remark}[Why the security definition must condition on the library]
\label{rem:security}
\label{rem:transcript_only}
Since the servers already store $\{\B^{(\ell)}\}$, any meaningful data-security
definition must condition on the stored library; otherwise it falls into one of
two failure modes. The library-inclusive unconditional form is
\emph{impossible}, because the library itself is correlated with~$\A$
(Proposition~\ref{prop:infeasible}). The transcript-only unconditional form
$I(\A;\{Q_i,\Yans_i\}_{i\in\mathcal{T}})=0$ (without the library) is \emph{not}
ruled out by Proposition~\ref{prop:infeasible}---a linear sketch
$T_\ell=h(\B^{(\ell)})=h(\A)+h(\Nell)$ can be marginally independent of~$\A$ while
still being useful to the user, who subtracts the locally known $h(\A)$---but it
\emph{ignores the adversary's actual side information}: the colluding servers hold
the correlated library $\{\B^{(\ell)}\}$, so a transcript that leaks nothing in
isolation may still leak about~$\A$ jointly with the library the adversary
possesses. We condition the security definition on the stored library
(Definition~\ref{def:security}) because the library is part of the adversary's
observation; the resulting requirement isolates the \emph{additional} leakage
caused by the protocol beyond what the library already discloses, and it is
feasible under correlation where the library-inclusive form is not. Other
formal security metrics---computational, distinguishability-based, or
leakage-budgeted variants---could be posed; this is the one the remainder of
the paper uses.
\end{remark}

\begin{definition}[Conditioned Data Security --- Relative No-Additional-Leakage Form]
\label{def:security}
For every $\mathcal{T}\subseteq[N]$ with $|\mathcal{T}|\leq P_c$:
\begin{equation}
  I\!\left(\A;\bigl\{Q_i^{(\theta)},\Yans_i^{(\theta)}\bigr\}_{i\in\mathcal{T}}
  \;\Big|\;\{\B^{(\ell)}\}\right)=0.
  \label{eq:data_security}
\end{equation}
\end{definition}

\begin{remark}[Threat model, guarantee scope, and exclusions]
\label{rem:threat_model}\label{rem:security_scope}
The guarantees are for honest-but-curious, replicated-storage adversaries: any
$\leq P_c$ colluding servers follow the protocol. The security guarantee is
\emph{relative}: in a correlated-library model the stored library itself
reveals information about~$\A$ ($I(\A;\{\B^{(\ell)}\})$ is nondecreasing in $L$;
in standard parametrized degraded-noise families it also typically increases as
the noise level decreases), and this leakage is unavoidable for any correct scheme.
Definition~\ref{def:security} guarantees that the transcript leaks \emph{no
additional} information beyond it---a strong attainable no-additional-leakage
guarantee, reducing to the classical unconditional one under independence. Explicitly
outside the model: Byzantine servers, coded or encrypted storage, upload-side
constraints, and demand--content dependence (excluded by~(M4)).
\end{remark}

A protocol may also involve \emph{public parameters}: design objects declared at
setup and known to every server, such as the linear sketch map $h$ of
Theorem~\ref{thm:achievability}, the parity-check matrix $H_n$ of
Proposition~\ref{thm:finiteblock} (Appendix~\ref{app:finiteblock}), or---in the rank-deficient setting of
Section~\ref{subsec:extension}---a published functional-structure
representative. Such parameters are
part of the adversary's baseline knowledge exactly as the stored library is, and
the relative-leakage principle behind Definition~\ref{def:security} extends to
them.

\begin{definition}[Conditioned Data Security with Public Parameters]
\label{def:security_public}
Let $\mathsf{P}$ denote the tuple of public protocol parameters declared at setup.
The scheme satisfies \emph{conditioned data security relative to $\mathsf{P}$} if
for every $\mathcal{T}\subseteq[N]$ with $|\mathcal{T}|\leq P_c$:
\begin{equation}
  I\!\left(\A;\bigl\{Q_i^{(\theta)},\Yans_i^{(\theta)}\bigr\}_{i\in\mathcal{T}}
  \;\Big|\;\{\B^{(\ell)}\},\,\mathsf{P}\right)=0.
  \label{eq:data_security_public}
\end{equation}
\end{definition}

\begin{remark}[Public-design form; relation to Definition~\ref{def:security}]
\label{rem:security_public}
Definition~\ref{def:security_public} adopts a public-design convention: whatever
the protocol legitimately publishes is counted as part of the adversary's
baseline, and the guarantee isolates the transcript's leakage \emph{beyond} the
library and the declared-public parameters. The convention is our own
formulation for this model; it fixes the adversary's baseline observation
explicitly, in the spirit of information-theoretic secrecy
formulations~\cite{wyner1975wire,goldwasser1984probabilistic}.
Two regimes must be distinguished.
\emph{(i)~$\A$-independent public parameters.} When $\mathsf{P}$ is a fixed
(deterministic) design object, \eqref{eq:data_security_public} is literally
identical to~\eqref{eq:data_security}; when $\mathsf{P}$ is generated from a
public seed independent of $(\A,\{\B^{(\ell)}\})$, the chain rule gives
$I(\A;V_{\mathcal{T}},\mathsf{P}\mid\{\B^{(\ell)}\})
=I(\A;\mathsf{P}\mid\{\B^{(\ell)}\})+I(\A;V_{\mathcal{T}}\mid\{\B^{(\ell)}\},\mathsf{P})
=I(\A;V_{\mathcal{T}}\mid\{\B^{(\ell)}\},\mathsf{P})
\geq I(\A;V_{\mathcal{T}}\mid\{\B^{(\ell)}\})$, so
Definition~\ref{def:security_public} is at least as strong as
Definition~\ref{def:security}. Every full-rank or concealed-structure scheme
considered here uses queries depending only on $(\theta,\mathcal{R})$ and
$\A$-independent public parameters ($h_{m,\mathsf{P}}$ for a fixed selector,
$H_n$, padding conventions), and its transcript is a function of
$(\theta,\mathcal{R})$ and the library, so both definitions hold there
simultaneously; all full-rank results are unaffected by the distinction. The
published-structure scheme of Theorem~\ref{thm:rankk_public} is governed
separately by Definition~\ref{def:security_public} with the $\A$-dependent
public parameter $R_\A$, as described next.
\emph{(ii)~$\A$-dependent public parameters.} If the protocol publishes an
object that is a function of $\A$---as when a canonical row-space
representative $R_\A$ is announced in the published-structure regime of
Section~\ref{subsec:extension}---Definition~\ref{def:security} becomes
infeasible by the joint distribution (Proposition~\ref{prop:infeasible} one
level up), and Definition~\ref{def:security_public} is the correct relative
form; this is what reconciles the two branches of the rank-deficient contrast
(Remark~\ref{rem:dichotomy_consistency}).
\end{remark}

To see that the guarantee is relative rather than absolute, consider the scalar
binary instance $\A\sim\mathrm{Bern}(\tfrac12)$, $\B^{(\ell)}=\A\oplus\N^{(\ell)}$,
$\N^{(\ell)}\overset{\mathrm{i.i.d.}}{\sim}\mathrm{Bern}(p)$. The residual
uncertainty $H(\A\mid\B^{(1:L)})$ (in bits, with $H(\A)=1$) is the baseline the
library already leaks, before any protocol runs:
\begin{center}
\renewcommand{\arraystretch}{1.15}
\begin{tabular}{cccccc}
\toprule
$p$ & $L{=}1$ & $L{=}2$ & $L{=}4$ & $L{=}8$ & $L{=}16$ \\
\midrule
$0.05$ & $0.286$ & $0.120$ & $0.018$ & $0.001$ & $0.000$ \\
$0.15$ & $0.610$ & $0.401$ & $0.178$ & $0.038$ & $0.002$ \\
$0.3$ & $0.881$ & $0.781$ & $0.619$ & $0.398$ & $0.173$ \\
$0.45$ & $0.993$ & $0.986$ & $0.972$ & $0.944$ & $0.892$ \\
\bottomrule
\end{tabular}
\end{center}
For $0<p<\tfrac12$ and finite $L$ the baseline is strictly positive, so the
definition is not vacuous; as $L$ grows the library reveals $\A$ and the
baseline decreases toward zero. Conditioned data security guarantees zero
leakage \emph{beyond} this baseline; it does not restore uncertainty already
lost through storage.

\section{The Full-Rank Base Case}
\label{sec:results}

Throughout, all entropy quantities ($H(\cdot)$, $I(\cdot;\cdot)$,
and $D$) are measured in bits, with $\log\triangleq\log_2$, consistent with the
notation in Section~\ref{sec:formulation}; one $\Fq$-symbol carries $\log q$
bits, which is why file-scale quantities appear as multiples of $\log q$
(e.g., $\log|\mathcal{W}|=m^2\log q$ for $\mathcal{W}=\Fq^{m\times m}$).
When $q=2$, bits and $\Fq$-symbols coincide.

\subsection{Tools: The Standard Colluding PIR Converse and the Averaging Transfer}
\label{subsec:tools}

For self-containedness we restate the $P_c$-colluding PIR converse in the general
form that permits decoder side information; in our application the effective
messages are \emph{independent} of the side information, so it specializes to the
standard colluding PIR converse (Remark~\ref{rem:standard_converse}). Standard
replicated PIR formulations~\cite{sun2017capacity,sun2018colluding} assume
mutually independent \emph{uniform} messages. We restate the $P_c$-colluding
converse~\cite{sun2018colluding} in a finite-error form that permits an
independent decoder parameter $S$, and Appendix~\ref{app:kadhe} proves that
uniformity is unnecessary---the self-contained proof remains valid for
independent, identically distributed, non-uniform messages.

\begin{lemma}[Colluding PIR with side information, restated]
\label{lem:kadhe}
Let $\pi$ be a $P_c$-colluding PIR scheme retrieving one of $L$ mutually
independent, identically distributed files $W_1,\ldots,W_L$ from $N$ replicated
servers, with decoder side information $S$, in which
$(\theta,\mathcal{R},S)\indep(W_1,\ldots,W_L)$, the demand has full conditional
support, $\Pr[\theta=t\mid S]>0$ almost surely for every $t\in[L]$ (e.g.,
$\theta$ uniform on $[L]$ and independent of $S$, as (M4) supplies in every
application of this paper), the queries are generated as
$Q_i=f_i(\theta,\mathcal{R},S)$, and the answers are deterministic,
$\Yans_i=\varphi_i(Q_i,\{W_\ell\},S)$. Suppose the scheme satisfies:
\begin{enumerate}[(i)]
  \item \emph{Correctness (demand-by-demand, transcript-conditioned):}
        $H(W_t\mid \Yans_{1:N}^{(t)},Q_{1:N}^{(t)},S)\leq\varepsilon$ for every
        $t\in[L]$, for some $\varepsilon\geq 0$ \emph{not} assumed to be
        asymptotically negligible. The queries appear
        in the conditioning because the decoder observes them---they are
        functions of $(\theta,\mathcal{R},S)$ available at the user---and it is
        this form, not the query-free bound
        $H(W_t\mid S,\Yans_{1:N}^{(t)})\leq\varepsilon$, that applications can
        verify by Fano's inequality;
  \item \emph{$P_c$-privacy:} $I(\theta;\{Q_i,\Yans_i\}_{i\in\mathcal{T}}\mid\{W_\ell\},S)=0$
        for all $|\mathcal{T}|\leq P_c$;
  \item \emph{Message independence of side information:} $W_\ell\indep S$ for every $\ell$.
\end{enumerate}
Under the full-support demand hypothesis, condition~(ii) is equivalent to the
distribution-free \emph{demand-invariance} property that the conditional law of
$\{Q_i,\Yans_i\}_{i\in\mathcal{T}}$ given $(\{W_\ell\},S)$ is the same under
every demand $t\in[L]$; the proof uses only this invariance, not the particular demand distribution,
so the conclusion below holds for every fixed $\theta\in[L]$ and for any demand
distribution having full conditional support.
Then the total download satisfies the \emph{finite-error} bound
\begin{equation}
  D_S \;=\; \sum_i H(\Yans_i\mid S)
  \;\geq\; \frac{H(W_\theta\mid S)-\varepsilon}{\CPIR(N,P_c,L)},
  \label{eq:kadhe_finite}
\end{equation}
where $\CPIR(N,P_c,L)$ is the replicated $P_c$-colluding capacity~\eqref{eq:pir_cap}~\cite{sun2018colluding}. The finite-error parameter-conditioned extension used here is proved in Appendix~\ref{app:kadhe}.
In particular, if $\varepsilon=o(\log|\mathcal{W}|)$ then
$D_S\geq H(W_\theta\mid S)/\CPIR(N,P_c,L)-o(\log|\mathcal{W}|)$.
The finite-error form~\eqref{eq:kadhe_finite} is the one applied at fixed
$\A=\mathbf{a}$ in Appendix~\ref{app:converse}, where pointwise convergence of $\varepsilon_{\mathbf{a}}$ is not required and
the Fano penalty is controlled only after averaging over~$\A$.
A self-contained proof, using only message independence and entropies (never
uniformity of the message law), is given in Appendix~\ref{app:kadhe}.
\emph{Scale note for the matrix setting:} when $W_\theta=\Ntheta\in\Fq^{m\times m}$,
the file alphabet has size $|\mathcal{W}|=q^{m^2}$, so
$\log|\mathcal{W}|=m^2\log q$ and $o(\log|\mathcal{W}|)=o(m^2)$,
consistent with the $o(m^2)$ notation used throughout.
\end{lemma}

\begin{remark}[Our converse is the \emph{standard} colluding PIR converse]
\label{rem:standard_converse}
Lemma~\ref{lem:kadhe} is applied to the innovation library $\{\Nell\}$, never to
the raw library, whose files are all $\A$-correlated and violate the
independence hypothesis. Because the effective messages satisfy
$\Nell\indep\A$, the side information does not reduce the file entropy:
$H(\Ntheta\mid\A)=H(\Ntheta)$. Consequently the bound
we apply is not a side-information converse in the entropy-reducing sense---it is
the \emph{standard} $P_c$-colluding PIR converse~\cite{sun2018colluding} applied to
the induced innovation library. The decoder side information $\A=\mathbf{a}$ enters
\emph{only} as a fixed parameter of the induced query and answer maps of
Theorem~\ref{thm:equivalence} (Section~\ref{subsec:capacity}), not as a
message-correlated resource. We
state Lemma~\ref{lem:kadhe} in the general side-information form only to make the
per-realization conditioning explicit; in our regime it specializes to the
standard converse, and the paper does not rely on side-information PIR in any
regime where the side information would reduce the innovation entropy. Two
consequences follow: under independence $\A\indep\btheta$ the bound degenerates
to the uncorrelated $\Theta(m^2)$ baseline of~\cite{sun2018colluding}, and the
capacity formula~\eqref{eq:pir_cap} depends only on $(N,P_c,L)$, not on the
file distribution, so it applies identically in both regimes.
\end{remark}

\begin{lemma}[Averaging transfer]
\label{lem:averaging}
For any protocol and any random variable~$\A$, with
$D_{\mathbf{a}}:=\sum_{i=1}^{N}H(\Yans_i^{(\theta)}\mid\A{=}\mathbf{a})$,
\begin{equation}
  D=\sum_{i=1}^{N}H\bigl(\Yans_i^{(\theta)}\bigr)
  \;\geq\;\sum_{i=1}^{N}H\bigl(\Yans_i^{(\theta)}\mid\A\bigr)
  \;=\;\mathbb{E}_\A[D_\A].
  \label{eq:averaging}
\end{equation}
No hypothesis on the model---in particular none on the rank profile
of~$\A$---is used. The pointwise transfer $D\geq D_{\mathbf{a}}$ for a fixed
$\mathbf{a}$ is false in general (Remark~\ref{rem:singleton}).
\end{lemma}
\begin{proof}
Conditioning cannot increase entropy, term by term.
\end{proof}

\begin{remark}[Averaging versus pointwise transfer]
\label{rem:singleton}
The converse conditions on a fixed realization $\A=\mathbf{a}$ and applies
Lemma~\ref{lem:kadhe} to each induced instance. A natural but \emph{invalid}
shortcut is to transfer the bound pointwise via
$H(\Yans_i^{(\theta)})\geq H(\Yans_i^{(\theta)}\mid\A=\mathbf{a})$
for a fixed $\mathbf{a}$: conditioning on a
specific value can \emph{increase} entropy. For example, with
$X\sim\mathrm{Bern}(0.5)$, $Y\mid X{=}0$ deterministic,
$Y\mid X{=}1\sim\mathrm{Bern}(0.5)$, one has $H(Y)\approx 0.811<H(Y\mid X{=}1)=1$.
Only the averaged form~\eqref{eq:averaging} is valid: applying
Lemma~\ref{lem:kadhe} to each induced instance and then taking expectations
yields $D\geq\HcondBA/\CPIR$. The i.i.d.\ hypothesis is needed only for the
achievability (existence of a Slepian--Wolf code at rate $\HcondBA+\varepsilon$
\cite{slepian1973noiseless,csiszar1982linear}); the converse needs only
Lemma~\ref{lem:averaging} and,
for the library-wide model, entropy flatness $H(\Ntheta\mid\A=\mathbf{a})=H(\Ntheta)$
is automatic since $\Ntheta\indep\A$.
\end{remark}

\subsection{Converse}
\label{subsec:converse}

The converse conditions on each realization $\A=\mathbf{a}$, applies the
colluding converse (Lemma~\ref{lem:kadhe}) to the induced instance over the
$\mathbf{a}$-independent innovation library, and averages
(Lemma~\ref{lem:averaging}); Section~\ref{subsec:capacity} packages the
underlying reduction as an operational-equivalence theorem.

\begin{theorem}[Converse under Library-Wide Correlation]
\label{thm:converse}
Consider the library-wide additive correlation model~\eqref{eq:additive}
under the source assumptions (M1)--(M3) and the demand and randomness
axioms (M4)--(M5), with replicated storage of
$\{\B^{(1)},\ldots,\B^{(L)}\}$ across $N$ servers and $P_c$-collusion.
Assume $H(\Ntheta)=\Theta(m^2)$, with $\N^{(1)},\ldots,\N^{(L)}$ mutually
independent, identically distributed, and independent of~$\A$; entrywise
memorylessness is \emph{not} required for the converse
(Remark~\ref{rem:beyond_memoryless}).
Under Assumption~\ref{ass:fullrank}, any scheme satisfying
correctness~\eqref{eq:correctness}, side-information-robust index privacy
(Definition~\ref{def:index}), and conditioned data security
(Definition~\ref{def:security})---the last part of the threat model but not used
by the converse (Remark~\ref{rem:converse_security_unused})---must satisfy
\begin{equation}
  D\;\geq\;\frac{\HcondBA}{\CPIR(N,P_c,L)}-o(m^2)
  \;=\;\frac{H(\Ntheta)}{\CPIR(N,P_c,L)}-o(m^2),
  \label{eq:converse}
\end{equation}
where
\begin{equation}
  \CPIR(N,P_c,L)=\!\left(\sum_{k=0}^{L-1}\!\left(\frac{P_c}{N}\right)^{\!k}
  \right)^{\!-1}
  =\frac{1-(P_c/N)}{1-(P_c/N)^L}
  \label{eq:pir_cap}
\end{equation}
is the $P_c$-colluding replicated-PIR capacity~\cite{sun2018colluding}, reducing to the noncolluding capacity~\cite{sun2017capacity} when $P_c=1$.
In the large-library regime the bound holds as an \emph{iterated} limit,
\begin{equation}
  \liminf_{L\to\infty}\;\liminf_{m\to\infty}\;
  \frac{D}{\HcondBA}\;\geq\;\frac{N}{N-P_c},
  \label{eq:converse_largeL}
\end{equation}
see Remark~\ref{rem:iterated_limit}.
\end{theorem}

\begin{remark}[Conditioned data security is not used by the converse]
\label{rem:converse_security_unused}
Conditioned data security (Definition~\ref{def:security}) is retained because it
is part of the problem's threat model and is satisfied by the matching
construction (Theorem~\ref{thm:achievability}); it is \emph{not} needed for the
converse itself. Inspection of the proof (Appendix~\ref{app:converse}) shows
that it uses only correctness~\eqref{eq:correctness},
side-information-robust index privacy (Definition~\ref{def:index}), the
library-wide innovation independence of~(M2)--(M3), the demand and randomness
axioms~(M4)--(M5), and full rank (Assumption~\ref{ass:fullrank}). The converse
therefore remains valid for the larger class of protocols satisfying
correctness and side-information-robust index privacy under the stated source
model. Accordingly, the innovation-retrieval reduction of
Theorem~\ref{thm:equivalence} is a consequence of full-rank
invertibility, the additive innovation model, and robust index
privacy, not of the data-security constraint. The same holds for
Theorem~\ref{prop:rankdef_lb} in the rank-deficient regime. By contrast,
Theorem~\ref{thm:rankk_impossible} genuinely uses its security
hypothesis~\eqref{eq:full_transcript_security}, in the common-channel step.
\end{remark}

\begin{proof}[Proof sketch]
Full details in Appendix~\ref{app:converse}. \emph{Step~0:} by
Lemma~\ref{lem:innovation} and full rank, recovering $\Ctheta$ is equivalent to
recovering $\Ntheta$ from the innovation library $\{\N^{(\ell)}\}$, to which the
standard $P_c$-colluding converse applies (Remark~\ref{rem:standard_converse}).
\emph{Step~1:} for each fixed $\A=\mathbf{a}$, the induced instance is a
replicated $P_c$-colluding PIR problem retrieving $\N^{(\theta)}$, giving
$D_{\mathbf{a}}\geq H(\N^{(\theta)})/\CPIR-\delta_{\mathbf{a}}$ with Fano penalty
$\delta_{\mathbf{a}}$. \emph{Steps~2--3:} global vanishing-error correctness
gives $\mathbb{E}_\A[\delta_\A]=o(m^2)$ (the pointwise transfer being invalid,
Remark~\ref{rem:singleton}). \emph{Step~4:} the averaging transfer
(Lemma~\ref{lem:averaging}) yields~\eqref{eq:converse}; the large-library limit
follows from $\CPIR\to1-P_c/N$.
\end{proof}

\subsection{Achievability}
\label{subsec:achievability}

\begin{theorem}[Achievability under Library-Wide Correlation]
\label{thm:achievability}
Under the conditions of Theorem~\ref{thm:converse} and
Assumption~\ref{ass:fullrank}, and assuming additionally that the entries of
$\Ntheta$ are i.i.d.\ (so that a linear source code of rate
$H(\Ntheta)+o(m^2)$ bits---realized as
$\lceil\cdot/\log q\rceil$ $\Fq$-symbols, Appendix~\ref{app:achievability}---with
vanishing error exists), there exists a scheme
satisfying all constraints with
\begin{equation}
  D\;\leq\;\frac{\HcondBA}{\CPIR(N,P_c,L)}+o(m^2).
  \label{eq:achievability}
\end{equation}
In the large-library regime,
$\limsup_{L\to\infty}\limsup_{m\to\infty}D/\HcondBA\leq N/(N-P_c)$
(Remark~\ref{rem:iterated_limit}).
\end{theorem}

\begin{proof}[Proof sketch]
Full details in Appendix~\ref{app:achievability}. The scheme composes a fixed
public linear sketch $h$ (a Slepian--Wolf binning of rate
$H(\Ntheta)+\epsilon m^2$, computed server-side on each file, independent
of~$\A$) with a capacity-achieving $P_c$-colluding PIR retrieval of the target
sketch; the user decodes $\N^{(\theta)}$ against its side information $\A$ and
forms $\Ctheta=\A^\top\A+\A^\top\N^{(\theta)}$ locally. Index privacy and
conditioned data security follow because every query depends only on
$(\theta,\mathcal{R},h)$ and every answer is a deterministic function of the
public sketches (independent of~$\A$); download is
$H(\Ntheta)/\CPIR+o(m^2)$, matching Theorem~\ref{thm:converse}.
\end{proof}

\noindent The achievability uses standard components. The full-rank
contribution lies in the converse and in Theorem~\ref{thm:equivalence}, which
shows that every admissible protocol induces an innovation-PIR instance, so the
retrieve-then-compute architecture is a consequence of robust index privacy,
the additive innovation model, and full-rank invertibility
(Section~\ref{subsec:nontrivial}).

\subsection{First-Order Operational Equivalence and Capacity}
\label{subsec:capacity}

With the converse and achievability in hand, we package the reduction as a
single \emph{operational-equivalence} theorem: in the full-rank regime every
admissible matrix-product protocol \emph{is}, after conditioning on
$\A=\mathbf{a}$, a colluding PIR protocol over the innovation library, with
matched error, conditional privacy, and an averaged download transfer.

\begin{theorem}[Full-rank first-order operational equivalence to innovation PIR]
\label{thm:equivalence}
Consider the model~\eqref{eq:additive} under (M1)--(M5) with $r=m$
(Assumption~\ref{ass:fullrank}). Let $\{\Pi_m\}$ be any sequence of
protocols satisfying correctness~\eqref{eq:correctness}, side-information-robust
index privacy (Definition~\ref{def:index}), and conditioned data security
(Definition~\ref{def:security}). For each full-rank realization~$\mathbf{a}$, define
the induced protocol $\Pi_{m,\mathbf{a}}$ over the innovation library by the queries
$Q_i^{(\theta,\mathbf{a})}=f_i(\theta,\mathbf{a},\mathcal{R})$ and the induced answer
functions
$g_{i,\mathbf{a}}(q,\mathbf{n}^{(1)},\ldots,\mathbf{n}^{(L)})
\triangleq g_i(q,\mathbf{a}+\mathbf{n}^{(1)},\ldots,\mathbf{a}+\mathbf{n}^{(L)})$.
Then, for $P_A$-almost every~$\mathbf{a}$:
\begin{enumerate}[(i)]
  \item \emph{(Error transfer.)}
    $\Pr_{\Pi_{m,\mathbf{a}}}\{\widehat{\N}^{(\theta)}\neq\Ntheta\}
     =\Pr_{\Pi_m}\{\widehat{\C}^{(\theta)}\neq\A^\top\B^{(\theta)}\mid\A=\mathbf{a}\}$,
    because full rank makes recovering $\Ctheta$ equivalent to recovering
    $\Ntheta$ (Lemma~\ref{lem:innovation}).
  \item \emph{(Conditional privacy.)} For every $\mathcal{T}\subseteq[N]$,
    $|\mathcal{T}|\leq P_c$,
    \begin{equation}
      I\!\left(\theta;\{Q_i^{(\theta,\mathbf{a})},\Yans_i^{(\theta,\mathbf{a})}\}_{i\in\mathcal{T}}
      \;\big|\;\N^{(1)},\ldots,\N^{(L)}\right)=0,
      \label{eq:equiv_privacy}
    \end{equation}
    i.e.\ $\Pi_{m,\mathbf{a}}$ is a finite-error induced $P_c$-colluding PIR
    instance, over an innovation library whose messages are independent
    of~$\mathbf{a}$, to which the finite-error
    converse~\eqref{eq:kadhe_finite} applies.
  \item \emph{(Averaged download transfer.)}
    $D(\Pi_m)\geq\mathbb{E}_\A[D(\Pi_{m,\A})]$; the pointwise transfer is
    invalid (Remark~\ref{rem:singleton}).
\end{enumerate}
Writing
\begin{equation}
  \varepsilon_{m,\mathbf{a}}\;\triangleq\;\max_{t\in[L]}
  H\!\left(\N^{(t)}\,\middle|\,\Yans^{(t)}_{1:N},Q^{(t)}_{1:N},
  \A{=}\mathbf{a}\right),
  \label{eq:eps_ma}
\end{equation}
the finite-error converse~\eqref{eq:kadhe_finite} gives
\begin{equation}
  D(\Pi_{m,\mathbf{a}})\;\geq\;
  \frac{H(\Ntheta)-\varepsilon_{m,\mathbf{a}}}{\CPIR(N,P_c,L)}
  \label{eq:equiv_pointwise}
\end{equation}
for $P_A$-a.e.~$\mathbf{a}$, with no pointwise convergence of
$\varepsilon_{m,\mathbf{a}}$ assumed. Global
correctness~\eqref{eq:correctness} and Fano give
$\mathbb{E}_\A[\varepsilon_{m,\A}]=o(m^2)$ (Appendix~\ref{app:converse},
Step~3), so averaging~\eqref{eq:equiv_pointwise} via~(iii) yields
\begin{equation}
  D(\Pi_m)\;\geq\;\frac{H(\Ntheta)}{\CPIR(N,P_c,L)}-o(m^2).
  \label{eq:equiv_converse}
\end{equation}
Conversely, any $P_c$-colluding PIR protocol over a linearly source-coded
innovation library, composed with a public linear source code~$h$, yields an
admissible matrix-product protocol with
$D(\Pi_m)\leq H(\Ntheta)/\CPIR+o(m^2)$
(Theorem~\ref{thm:achievability}). Hence, writing $D^*_{\mathrm{MP}}$ for the optimal download over admissible
matrix-product protocols and $D^*_{\mathrm{PIR}}(\Ntheta)$ for that over
$P_c$-colluding PIR protocols retrieving from the innovation library,
\begin{equation}
  D^*_{\mathrm{MP}}=\frac{\HcondBA}{\CPIR(N,P_c,L)}+o(m^2),
  \quad
  D^*_{\mathrm{PIR}}(\Ntheta)=\frac{\HcondBA}{\CPIR(N,P_c,L)}+o(m^2),
  \label{eq:equiv_capacity}
\end{equation}
and therefore
\begin{equation}
  D^*_{\mathrm{MP}}-D^*_{\mathrm{PIR}}(\Ntheta)=o(m^2).
  \label{eq:equiv_gap}
\end{equation}
The equivalence is first-order: we do not claim exact equality at finite~$m$,
nor that every optimal innovation-PIR scheme can be implemented from the raw
$\B$-library without knowledge of~$\A$.
\end{theorem}

\begin{proof}
Parts (i)--(iii) are established by results above and are collected here. (i) is Lemma~\ref{lem:innovation} (full-rank invertibility gives a bijection
between $\Ctheta$ and $\Ntheta$ given $\A=\mathbf{a}$), applied to the induced
answer functions defined in the statement. (ii) follows directly from
Definition~\ref{def:index}, with no appeal to
Proposition~\ref{prop:sir_satisfiable}---whose hypothesis of $\A$-independent
queries would not cover the full admissible class here, since the induced
queries $Q_i^{(\theta,\mathbf{a})}=f_i(\theta,\mathbf{a},\mathcal{R})$ may
depend on~$\mathbf{a}$. The joint form of Definition~\ref{def:index} gives
$I(\theta;\{Q_i^{(\theta)},\Yans_i^{(\theta)}\}_{i\in\mathcal{T}},
\{\B^{(\ell)}\}\mid\A{=}\mathbf{a})=0$, hence by the chain rule
$I(\theta;\{Q_i^{(\theta)},\Yans_i^{(\theta)}\}_{i\in\mathcal{T}}
\mid\{\B^{(\ell)}\},\A{=}\mathbf{a})=0$; given $\A=\mathbf{a}$, the shifts
$\B^{(\ell)}=\mathbf{a}+\N^{(\ell)}$ and $\N^{(\ell)}=\B^{(\ell)}-\mathbf{a}$
are mutually inverse, so conditioning on the raw library and on the innovation
library generate the same $\sigma$-algebra, yielding~\eqref{eq:equiv_privacy}.
The induced messages $\N^{(\ell)}$ are independent of~$\mathbf{a}$ and mutually
independent by~(M2). (iii)
is the averaging transfer (Lemma~\ref{lem:averaging}); the pointwise transfer
fails (Remark~\ref{rem:singleton}). Given (i)--(iii), the induced
instance is a finite-error $P_c$-colluding PIR instance over an
$\mathbf{a}$-independent $L$-message library, so the finite-error
form~\eqref{eq:kadhe_finite} of Lemma~\ref{lem:kadhe}
gives~\eqref{eq:equiv_pointwise} for $P_A$-a.e.~$\mathbf{a}$, with the
per-realization penalty $\varepsilon_{m,\mathbf{a}}$
of~\eqref{eq:eps_ma} retained rather than assumed to vanish. Averaging
via~(iii) and using $\mathbb{E}_\A[\varepsilon_{m,\A}]=o(m^2)$
yields~\eqref{eq:equiv_converse}. The converse direction is
Theorem~\ref{thm:converse} and the achievability direction is
Theorem~\ref{thm:achievability}, giving~\eqref{eq:equiv_capacity}.
\end{proof}

\begin{corollary}[No first-order gain from the matrix-product algebra, full rank]
\label{cor:no_gain}
Under full rank, any admissible protocol must carry enough information to recover
the innovation $\Ntheta$; hence the only first-order gain over the
independence baseline comes from the entropy reduction $\HcondBA=H(\Ntheta)$, not
from any structural property of the matrix product $\A^\top\B^{(\theta)}$. In
particular no admissible protocol beats $H(\Ntheta)/\CPIR$ at first order by
exploiting the algebra of $\A^\top(\cdot)$.
\end{corollary}

\begin{corollary}[Tight Download-Cost Characterization]
\label{cor:capacity}
Under the conditions of Theorems~\ref{thm:converse}
and~\ref{thm:achievability}:
\begin{equation}
  D^*=\frac{\HcondBA}{\CPIR(N,P_c,L)}+o(m^2).
  \label{eq:capacity}
\end{equation}
In the large-library regime (Remark~\ref{rem:iterated_limit}, so
$\CPIR\to 1-P_c/N$),
\begin{equation}
  \lim_{L\to\infty}\;\lim_{m\to\infty}\;\frac{D^*}{\HcondBA}
  \;=\;\frac{N}{N-P_c}.
  \label{eq:capacity_largeL}
\end{equation}
\emph{Note on hypothesis asymmetry:} the converse bound
$D \geq \HcondBA/\CPIR$ holds for any source satisfying the model
conditions via the averaging argument (Remark~\ref{rem:singleton}).
The achievability bound $D \leq \HcondBA/\CPIR$ additionally requires
that $H(\Ntheta\mid\A=\mathbf{a})$ be constant in~$\mathbf{a}$ (entropy
flatness), so that the Slepian--Wolf code can operate at a single rate. In
the library-wide model this is automatic since $\Ntheta\indep\A$; for more
general sources where flatness fails, only the converse half of
Corollary~\ref{cor:capacity} applies and the achievable rate may exceed
$\HcondBA/\CPIR$.
\end{corollary}

\begin{remark}[The large-library limit is iterated]
\label{rem:iterated_limit}
The two limits $m\to\infty$ and $L\to\infty$ are not interchangeable here, and
the large-library statements are to be read in the order $m$ first, $L$ second.
The reason is the $L$-factor in the Fano bookkeeping: the per-realization
penalty of Appendix~\ref{app:converse} obeys
$\varepsilon_{\mathbf{a}}\leq 1+L\,P_e(\mathbf{a})\,m^2\log q$, so
$\mathbb{E}_\A[\varepsilon_\A]=o(m^2)$ holds at each fixed $L$ but the implied
constant grows with~$L$; the same $L$-factor appears in
Theorem~\ref{prop:rankdef_lb} and in
Corollary~\ref{cor:rankk_vanishing}. A joint $(m,L)\to\infty$ statement would
require an error analysis uniform in $L$---for instance a Fano penalty
$o(m^2)$ uniformly over $L\leq L(m)$ for some admissible growth $L(m)$---which
we do not supply. Accordingly~\eqref{eq:converse_largeL},
\eqref{eq:capacity_largeL}, and the achievability counterpart are iterated
limits, and no joint-limit theorem is claimed.
\end{remark}

\begin{remark}[Beyond entrywise memoryless innovations: scope of the converses]
\label{rem:beyond_memoryless}
Entrywise independence is not required by the converses. The self-contained
colluding converse (Lemma~\ref{lem:kadhe} and its refined-storage extension,
Theorem~\ref{thm:refined_converse}) treats each message as a single random
object and uses only mutual independence across files, identical distribution,
and independence from $(\theta,\mathcal{R},\A)$
(Remark~\ref{rem:app_nonuniform} in Appendix~\ref{app:kadhe}). Consequently,
under the hypotheses of Theorem~\ref{thm:converse} but allowing arbitrary
dependence among the coordinates of each innovation matrix, the full-rank
converse remains
\begin{equation}
  D \;\geq\; \frac{H(\Ntheta)}{\CPIR(N,P_c,L)}-o(m^2),
  \label{eq:beyond_memoryless}
\end{equation}
and the rank-deficient converse likewise remains
\begin{equation}
  D \;\geq\; \frac{H(\A^\top\btheta\mid\A)}{\CPIR(N,P_c,L)}-o(km).
  \label{eq:beyond_memoryless_rankk}
\end{equation}
These are distinct statements: for rank-deficient $\A$ the innovation is
generally not recoverable from the product, so~\eqref{eq:beyond_memoryless} is
not implied by~\eqref{eq:beyond_memoryless_rankk} and does not apply to
Theorem~\ref{prop:rankdef_lb}. If a support-refined Fano penalty is desired it
should be stated in its finite-error form; it becomes
$o(\log|\mathrm{supp}\,P_{\N,m}|)$ only when
$\log|\mathrm{supp}\,P_{\N,m}|\to\infty$, as for rank-$r$ structured updates
$\N=\mathbf{U}\mathbf{V}^\top$ with $\mathbf{U},\mathbf{V}\in\Fq^{m\times r}$
and $rm\to\infty$, and not when the support size stays bounded, where the
additive Fano constant does not vanish relative to it.
Likewise, the concealed-structure theorem's
common-conditional-support hypothesis is stated directly at library level
(Theorem~\ref{thm:rankk_impossible}(i)), so it too covers dependent-entry
innovations. Memorylessness enters solely on the achievability side, where linear
source coding of the innovation at rate $H(\Ntheta)+o(m^2)$ bits with vanishing
error is required: entrywise i.i.d.\ innovations supply this via random linear
binning, and the retrieve-then-compute architecture accommodates any innovation
family admitting such compression. Thus, for structured or dependent
innovations, the lower bounds stand, and any gap
to them is attributable to source coding, not to the privacy or security
constraints. The \emph{additive} form of the correlation is likewise a
convenience rather than a requirement of the converse: the per-realization
argument of Theorem~\ref{thm:converse} requires only that the files be
conditionally i.i.d.\ given $\A=\mathbf{a}$ (together with (M4)), yielding
$D\geq H(\btheta\mid\A)/\CPIR$ minus the corresponding slack for general
joint laws $P_{\A,\{\B\}}$ with that structure; additivity and full rank are
what furnish the \emph{achievability}-side innovation reduction and the
$\A$-independent effective library.
\end{remark}

\subsection{Finite-Length Accounting}
\label{subsec:finiteblock_thm}

Finite-length costs are quantified in Appendix~\ref{app:finiteblock}
(Proposition~\ref{thm:finiteblock}): a linear syndrome code composed with the
optimal-subpacketization Zhang--Xu colluding scheme~\cite{zhang2019optimal}
gives fixed-length downloaded communication exactly $\Lambda_n=\tilde{r}_n/\CPIR$ (an upper bound on the entropy download), error bounded by the source
decoder, and normalized download
$H_2(p)/\CPIR+O(n^{-1/2})+(s_{\mathrm{base}}/n)\,O(1)$~\eqref{eq:finiteblock_norm}; numerical
accounting is in Section~\ref{subsec:e2e}.

\section{The Rank-Deficient Regime: Functional Converses, Published Structure, and a Strengthened-Observation Concealed Converse}
\label{subsec:extension}

This section develops the paper's main converse results for the
\emph{functional} regime, where the operand is noninvertible and product
recovery no longer determines the requested file. Throughout this section,
the rank-profile axiom~(M1) is specialized to rank exactly $k$ with
$1\leq k<m$; (M2)--(M5) are unchanged. The lower bound $k\geq 1$ is assumed
throughout: at $k=0$ the operand is the zero map, $\A^\top\btheta=0$
identically, and the $o(km)$ scale is degenerate. Assumption~\ref{ass:fullrank}
and the innovation-equivalence
Lemma~\ref{lem:innovation} are full-rank tools and are not available here,
except for comparison with the full-rank base case. Table~\ref{tab:libraries}
summarizes the three per-index objects in play and the converse applicable to
each. The contrast between Theorems~\ref{thm:rankk_public}
and~\ref{thm:rankk_impossible} below is over two \emph{disjoint} admissible
classes, separated by whether the rank-$k$ structure of $\A$ is published as a
protocol parameter or withheld; the two theorems therefore do not contradict
each other, and Remark~\ref{rem:dichotomy_consistency} makes the separation
precise.

\begin{table}[t]
\caption{The three libraries; $\phi$ is a per-file coarsening
(e.g.\ $\mathbf{a}^\top(\cdot)$, $R_\A(\cdot)$, $((\cdot)-\mathbf{a})_U$).}
\label{tab:libraries}
\centering
\small
\renewcommand{\arraystretch}{1.3}
\begin{tabular}{@{}p{2.35cm}p{2.1cm}p{3.1cm}@{}}
\toprule
Object & vs.~$\A$ & Applicable converse \\
\midrule
$\B^{(\ell)}=\A+\Nell$ (stored file) & correlated, library-wide &
none directly; enters as the refinement $X_\ell$ of
Theorem~\ref{thm:refined_converse} \\
$\Nell=\B^{(\ell)}-\A$ (innovation) & independent &
Lemma~\ref{lem:kadhe} (full-rank reduction, Section~\ref{sec:results}) \\
$W_\ell(\mathbf{a})=\phi(\B^{(\ell)})$ (functional message) &
i.i.d.\ given $\A{=}\mathbf{a}$ &
Theorem~\ref{thm:refined_converse} with $X_\ell=\B^{(\ell)}$ \\
\bottomrule
\end{tabular}
\end{table}
When $\mathrm{rank}(\A)=k<m$, the computation $\Ctheta=\A^\top\btheta$
lives in a $k$-dimensional subspace, so less information about $\btheta$
is needed to reconstruct $\Ctheta$; the innovation reduction of the full-rank
case no longer applies, because $\Ctheta$ no longer determines $\btheta$.
The lower bound is subtle, because the servers store $\B^{(\ell)}$, not
$\C^{(\ell)}=\A^\top\B^{(\ell)}$, and~$\A$ is unknown to them. Two obstructions
would appear to force a conditional statement: privacy conditioned on the full
library need not transfer, as a generic matter, to privacy conditioned on a
coarsening of it; and the answers are deterministic functions of the stored
files, not of the functional messages, so the standard colluding converse does
not apply directly to the functional library. We show that under the model's
own definitions both obstructions dissolve---the first by a universal transfer
characterization, the second by the refined-storage extension of the colluding converse
(Theorem~\ref{thm:refined_converse})---so the functional lower bound is valid for every rank-$k$ operand, with no
structural hypothesis on the map.

\subsection{Two Tools: Universal Privacy Transfer and the Refined-Storage Converse}

\begin{proposition}[Functional-privacy transfer: universality and its boundary]
\label{thm:transfer}
Let $V_{\mathcal{T}}$ be the colluding view, $|\mathcal{T}|\leq P_c$.

\emph{(a)~Universality under the joint form.}
Assume side-information-robust index privacy (Definition~\ref{def:index}). Then
for $P_A$-a.e.\ $\mathbf{a}$ and \emph{every} measurable function
$W=\Phi_{\mathbf{a}}(\{\B^{(\ell)}\})$ of the stored library,
\begin{multline}
  I\!\left(\theta;V_{\mathcal{T}}\mid W,\A{=}\mathbf{a}\right)
  \;\leq\;
  I\!\left(\theta;V_{\mathcal{T}},\{\B^{(\ell)}\}\mid\A{=}\mathbf{a}\right)
  \;=\;0\\
  \text{for all }|\mathcal{T}|\leq P_c.
  \label{eq:transfer}
\end{multline}

\emph{(b)~The joint form is exactly the boundary (over arbitrary joint laws).}
The hypothesis cannot be weakened to the library-\emph{conditioned} form
$I(\theta;V_{\mathcal{T}}\mid\{\B^{(\ell)}\},\A{=}\mathbf{a})=0$ \emph{as a
statement about arbitrary joint laws of
$(\theta,V_{\mathcal{T}},\{\B^{(\ell)}\})$ given $\A=\mathbf{a}$}: whenever
$I(\theta;\{\B^{(\ell)}\}\mid\A{=}\mathbf{a})>0$, there exist a view satisfying
the library-conditioned form and a coarsening for which the transfer fails. An
explicit witness is
\begin{align}
  &\{\B\}=(W,U),\qquad \theta:=U,\qquad V=W\oplus\theta:\notag\\
  &\quad I(\theta;V\mid\{\B\})=0,\qquad I(\theta;V\mid W)=1~\text{bit}.
  \label{eq:transfer_witness}
\end{align}
Conversely, if $\theta\indep\{\B^{(\ell)}\}$ given $\A=\mathbf{a}$, the
conditioned and joint forms coincide and the universal transfer of part~(a)
applies. Thus, over arbitrary joint laws, the transfer holds for all views
satisfying the library-conditioned form and all coarsenings \emph{if and only
if} the demand is independent of the stored content given the side
information. Within the model, (M4) forbids the failure regime by axiom, which
is why Definition~\ref{def:index}'s joint form makes the transfer
universal there.
\end{proposition}
\begin{proof}
\emph{(a)} All alphabets are finite, so conditioning on $\A=\mathbf{a}$ (any
$\mathbf{a}$ in the support of $P_A$) is elementary, and $\Phi_{\mathbf{a}}$
may depend on~$\mathbf{a}$. By the chain rule,
$I(\theta;V_{\mathcal{T}},W\mid\A{=}\mathbf{a})
 =I(\theta;W\mid\A{=}\mathbf{a})+I(\theta;V_{\mathcal{T}}\mid W,\A{=}\mathbf{a})
 \geq I(\theta;V_{\mathcal{T}}\mid W,\A{=}\mathbf{a})$;
since $(V_{\mathcal{T}},W)$ is a function of $(V_{\mathcal{T}},\{\B^{(\ell)}\})$
given $\A=\mathbf{a}$, data processing gives
$I(\theta;V_{\mathcal{T}},W\mid\A{=}\mathbf{a})
 \leq I(\theta;V_{\mathcal{T}},\{\B^{(\ell)}\}\mid\A{=}\mathbf{a})$,
which is zero by the \emph{joint}, library-inclusive
form~\eqref{eq:index_privacy} of Definition~\ref{def:index}. Part~(a) holds
without change when queries depend on~$\A$: conditioning on $\A=\mathbf{a}$ renders
any such dependence a fixed function.

\emph{(b)} \emph{Failure direction.} Suppose
$I(\theta;\{\B^{(\ell)}\}\mid\A{=}\mathbf{a})>0$. The view
$V:=\{\B^{(\ell)}\}$ trivially satisfies the conditioned form,
$I(\theta;V\mid\{\B^{(\ell)}\},\A{=}\mathbf{a})=0$, yet with the constant
coarsening $W\equiv 0$ the transfer fails:
$I(\theta;V\mid W,\A{=}\mathbf{a})=I(\theta;\{\B^{(\ell)}\}\mid\A{=}\mathbf{a})>0$.
For a witness with a nondegenerate coarsening and complete failure
($I(\theta;V\mid W)$ equal to $H(\theta)$),
take~\eqref{eq:transfer_witness} with $W,U$ i.i.d.\ uniform bits: given
$\{\B\}=(W,U)$ the demand $\theta=U$ is determined, so
$I(\theta;V\mid\{\B\})=0$; given only $W=w$, the map $\theta\mapsto w\oplus\theta$
is a bijection and $\theta$ is uniform, so $I(\theta;V\mid W)=H(\theta)=1$~bit.
\emph{Boundary direction.} If $\theta\indep\{\B^{(\ell)}\}$ given
$\A=\mathbf{a}$, the chain rule gives
$I(\theta;V_{\mathcal{T}},\{\B^{(\ell)}\}\mid\A{=}\mathbf{a})
=I(\theta;\{\B^{(\ell)}\}\mid\A{=}\mathbf{a})
+I(\theta;V_{\mathcal{T}}\mid\{\B^{(\ell)}\},\A{=}\mathbf{a})
=I(\theta;V_{\mathcal{T}}\mid\{\B^{(\ell)}\},\A{=}\mathbf{a})$,
so the conditioned form implies the joint form and part~(a) applies.
\end{proof}

\begin{remark}[Interpretation and a randomized extension]
\label{rem:transfer_scope}
Conditional mutual information is not monotone in the conditioning
$\sigma$-algebra, so part~(a) is not a generic coarsening statement; part~(b)
locates the exact obstruction in demand--content correlation. The
witness~\eqref{eq:transfer_witness} sets $\theta:=U$, a function of the stored
library, and so violates (M4) \emph{by design}: that is the content of the
boundary claim---the failure regime lies entirely outside the model class, and
the joint form of Definition~\ref{def:index} encodes the demand-independence
(M4), so that $\theta$ carries no information about the library content whose
coarsening could unmask it. Finally, if the coarsening is
randomized by a public seed,
$W=\Phi_{\mathbf{a}}(\{\B^{(\ell)}\},\mathsf{P})$ with
$\mathsf{P}\indep(\theta,\{\B^{(\ell)}\})$ given $\A=\mathbf{a}$, the identical
argument with $\mathsf{P}$ adjoined to the conditioning yields
$I(\theta;V_{\mathcal{T}}\mid W,\A{=}\mathbf{a},\mathsf{P})=0$ whenever
Definition~\ref{def:index} holds conditioned on $\mathsf{P}$---automatic for
the schemes of this paper, whose public parameters are either deterministic
design objects or, in the published-structure regime, the announced $R_\A$ on
which Proposition~\ref{prop:sir_satisfiable} is instantiated.
\end{remark}

The second tool extends the colluding PIR converse to the situation created by
rank deficiency: the servers store more than the messages the user wants.
Dependent-message PIR~\cite{chen2020asymptotic} formulates answers directly in
terms of the complete dependent-message library, and the replicated
private-computation model of~\cite{sun2019capacity} contains a full-rank set of
candidate linear functions that determines the underlying datasets. By contrast,
the coded private-linear-computation model of~\cite{obead2022private} and the
general nonlinear-computation model of~\cite{obead2019nonlinear} already permit
answers to depend on underlying stored messages not necessarily determined by a
rank-deficient or arbitrary candidate-function family. Those results are
noncolluding and do not address decoder-side common-state conditioning, a
private file index under a decoder-private fixed map, conditionally independent
per-file functional messages, or arbitrary correlated refinements.
Theorem~\ref{thm:refined_converse} establishes a $P_c$-colluding refined-storage
converse under these latter conditions and supplies an explicit answer-locality
lemma that does not require message-level answer determinism.

\begin{theorem}[Colluding PIR converse with refined storage and functional
messages]
\label{thm:refined_converse}
Let $X_1,\ldots,X_L$ be \emph{refined storage} variables, let
$W_\ell=\phi(X_\ell)$, $\ell\in[L]$, be the \emph{functional messages} for a
fixed map $\phi$, and let $S$ be decoder side information. Assume:
\emph{(i)~storage--demand independence and full-support demand:}
$X_{1:L}\indep(\theta,\mathcal{R})$ given $S$, and
$\Pr[\theta=t\mid S]>0$ almost surely for every $t\in[L]$ (e.g., $\theta$
uniform on $[L]$ and independent of $S$, as (M4) supplies in every application);
\emph{(ii)~message structure:} $W_1,\ldots,W_L$ are mutually independent and
identically distributed given $S$, with $H:=H(W_\ell\mid S)$---the residual
content of $X_\ell$ beyond $W_\ell$ is otherwise arbitrary: residuals may be
correlated with their own message, with each other across files, and the
$X_\ell$ themselves need not be mutually independent;
\emph{(iii)~queries:} $Q_i=f_i(\theta,\mathcal{R},S)$;
\emph{(iv)~answers:} $\Yans_i=g_i(Q_i,X_{1:L},S)$, deterministic given the
\emph{refined} storage;
\emph{(v)~correctness:}
$H(W_\theta\mid\Yans_{1:N},Q_{1:N},W_J,S)\leq\varepsilon$ for every
$\theta\in[L]$ and $J\subseteq[L]\setminus\{\theta\}$; and
\emph{(vi)~$P_c$-colluding privacy conditioned on the functional library:}
$I(\theta;\{Q_i,\Yans_i\}_{i\in\mathcal{T}}\mid W_{1:L},S)=0$ for all
$|\mathcal{T}|\leq P_c$. (Privacy is required relative to $\{W_\ell\}$ only;
conditioning on $X_{1:L}$ instead is an incomparable hypothesis and is not
assumed. Under the full-support demand of hypothesis~(i), condition~(vi) is
equivalent to demand-invariance of the conditional law of
$\{Q_i,\Yans_i\}_{i\in\mathcal{T}}$ given $(W_{1:L},S)$ across all $t\in[L]$,
which is the form the proof uses; the bound therefore holds for every fixed
demand and for any demand distribution having full conditional support.) Then
\begin{equation}
  D_S \;=\; \sum_{i=1}^{N} H(\Yans_i\mid S)
  \;\geq\; \frac{H-\varepsilon}{\CPIR(N,P_c,L)}.
  \label{eq:refined_converse}
\end{equation}
\end{theorem}

\begin{proposition}[Refined storage is not reducible to message-level storage]
\label{prop:not_relabelling}
Fix the refinement map $\phi$ and set $W_\ell=\phi(X_\ell)$. Suppose the
refined storage carries nontrivial residual information beyond the complete
functional-message library,
\begin{equation}
  H\bigl(X_{1:L}\mid W_{1:L},S\bigr)>0,
  \label{eq:residual_entropy}
\end{equation}
equivalently that for some $s$ and $w_{1:L}$ of positive probability there are
distinct joint realizations
$x_{1:L},\tilde{x}_{1:L}\in\mathrm{supp}(X_{1:L}\mid W_{1:L}{=}w_{1:L},S{=}s)$.
If a converse
assumes that the answers are deterministic functions of the demanded messages,
$\Yans_i=g_i(Q_i,W_{1:L},S)$, then it does not apply to the class of
Theorem~\ref{thm:refined_converse}, in which
$\Yans_i=g_i(Q_i,X_{1:L},S)$: there exist correct, $P_c$-private schemes
admitted by that theorem whose answers are \emph{not}
$\sigma(Q_{1:N},W_{1:L},S)$-measurable. Thus a converse requiring answer
determinism with respect to $W_{1:L}$ cannot be applied merely by treating the
functional messages as the stored files.
\end{proposition}

\begin{proof}
Let $\pi$ be a standard $P_c$-private PIR scheme run directly on the
functional-message library: its queries are $Q_i=f_i(\theta,\mathcal{R},S)$ and
its answers are $\Yans_i=g_i(Q_i,W_{1:L},S)$, so its entire transcript is a
function of $(\theta,\mathcal{R},W_{1:L},S)$; it satisfies correctness~(v) at
some slack $\varepsilon$ and privacy~(vi). Form $\pi'$ by appending to
server~$1$'s answer the single symbol $Z\triangleq\mathbbm{1}\{X_{1:L}=x_{1:L}\}$,
so $\Yans_1'=(\Yans_1,Z)$ and $\Yans_i'=\Yans_i$ otherwise; this is permitted by
hypothesis~(iv), which lets each answer depend on the complete refined storage.

\emph{$\pi'$ is admissible.} Hypothesis~(iv) holds since $Z$ is a function of
$X_{1:L}$, so every $\Yans_i'$ remains deterministic in $(Q_i,X_{1:L},S)$. Correctness~(v)
holds at the same slack: the decoder discards $Z$, and extra conditioning does
not increase conditional entropy. For privacy~(vi), fix
$|\mathcal{T}|\leq P_c$ and expand
\begin{equation*}
  I\bigl(\theta;Q_{\mathcal{T}},\Yans'_{\mathcal{T}}\mid W_{1:L},S\bigr)
  =\underbrace{I\bigl(\theta;Q_{\mathcal{T}},\Yans_{\mathcal{T}}\mid W_{1:L},S\bigr)}_{=0\ \text{by (vi) for }\pi}
  +I\bigl(\theta;Z\mid Q_{\mathcal{T}},\Yans_{\mathcal{T}},W_{1:L},S\bigr).
\end{equation*}
For the second term: by hypothesis~(i),
$X_{1:L}\indep(\theta,\mathcal{R})$ given $S$, hence also given $(W_{1:L},S)$,
since $W_{1:L}$ is a function of $X_{1:L}$; thus
$Z\indep(\theta,\mathcal{R})$ given $(W_{1:L},S)$. The base view
$(Q_{\mathcal{T}},\Yans_{\mathcal{T}})$ is a deterministic function of
$(\theta,\mathcal{R},W_{1:L},S)$, so conditioning on it adds nothing beyond
those variables, and
$I(\theta;Z\mid Q_{\mathcal{T}},\Yans_{\mathcal{T}},W_{1:L},S)=0$.

\emph{Its answers are not $W$-measurable.} The realizations $x_{1:L}$ and
$\tilde{x}_{1:L}$ lie in the same fiber $\{W_{1:L}{=}w_{1:L},S{=}s\}$ of positive
probability, so they carry identical $W_{1:L}$ yet give $Z=1$ and $Z=0$. Hence
$\Yans_1'$ takes different values on two positive-probability realizations with
identical $(Q_{1:N},W_{1:L},S)$---the queries agree, being functions of
$(\theta,\mathcal{R},S)$ alone---so it is not a function of
$(Q_{1:N},W_{1:L},S)$.

Concretely, in the rank-deficient instances of
Theorem~\ref{prop:rankdef_lb} one has $X_\ell=\B^{(\ell)}$ and
$W_\ell=\mathbf{a}^\top\B^{(\ell)}$; the components of $\B^{(\ell)}$ orthogonal
to $\mathrm{row}(\mathbf{a}^\top)$ are residual content that the servers hold
and may answer from, but which the functional library does not determine. In
this application, condition~\eqref{eq:residual_entropy} holds whenever the
conditional support of $\B^{(1)}$ given $\A=\mathbf{a}$ contains two distinct
matrices $b\neq b'$ with $\mathbf{a}^\top b=\mathbf{a}^\top b'$: conditioned on
$W_1=\mathbf{a}^\top b$, the file $\B^{(1)}$ is then not determined, so
$H(\B^{(1)}\mid W_1,\A{=}\mathbf{a})>0$. Matrix-level full-support innovations
guarantee such a pair, because $\mathbf{a}^\top(\cdot)$ has a nontrivial kernel
when $k<m$. Cross-file independence under (M2) then permits the remaining
files to be held fixed: conditioning additionally on
$W_{2:L}$ does not alter the conditional law of $\B^{(1)}$, so
$H(\B^{(1)}\mid W_{1:L},\A{=}\mathbf{a})>0$. (We do not claim the residual
components are independent of the within-file message
$\mathbf{a}^\top\B^{(1)}$; for non-uniform innovations and dense
$\mathbf{a}^\top$ they generally are not.) The augmentation costs one symbol
of download, so it does not affect the first-order rate.
\end{proof}

\begin{corollary}[Strict extension of the message-level converse]
\label{cor:strict_extension}
Theorem~\ref{thm:refined_converse} strictly extends Lemma~\ref{lem:kadhe}.
Every instance covered by Lemma~\ref{lem:kadhe} is recovered by setting
$X_\ell=W_\ell$ and $\phi=\mathrm{id}$, under which answers deterministic in
the storage are deterministic in the messages. Conversely, whenever
$H(X_{1:L}\mid W_{1:L},S)>0$, Proposition~\ref{prop:not_relabelling} exhibits
correct, $P_c$-private refined-storage schemes whose answers are not
$\sigma(Q_{1:N},W_{1:L},S)$-measurable; such schemes lie outside the
message-level answer model of Lemma~\ref{lem:kadhe}, so the admissible class of
Theorem~\ref{thm:refined_converse} is strictly larger.
\end{corollary}

The structural condition that fails is measurability of the answers with
respect to the functional messages: the message-level model requires
$\Yans_i\in\sigma(Q_i,W_{1:L},S)$, whereas under refined storage only
$\Yans_i\in\sigma(Q_i,X_{1:L},S)$ holds, and by
Proposition~\ref{prop:not_relabelling} the former can fail whenever
$H(X_{1:L}\mid W_{1:L},S)>0$. This is the message-level \emph{locality}
property on which the standard peeling induction rests. Lemma~\ref{lem:app_locality_refined}
replaces it: the peeled entropy terms are controlled through a
refined-storage conditional-independence argument, using
$X_{1:L}\indep(\theta,\mathcal{R})$ given $S$ in place of answer measurability
in the messages.

\begin{remark}[Two design choices in Proposition~\ref{prop:not_relabelling}]
\label{rem:not_relabelling_design}
The construction fixes the base scheme to run on the functional library rather
than augmenting a generic admissible scheme, and it keys $Z$ on the joint
storage $X_{1:L}$ rather than on $X_1$. Both choices are necessary. Augmenting a
generic scheme can destroy privacy through a collider: if a base answer takes
the form $U\oplus\theta$ with $U$ a residual coordinate, it is private on its
own, yet $I(\theta;U\mid U\oplus\theta)=1$ bit, so appending $U$ couples the
demand to the residual. Keying $Z$ on $X_1$ alone can fail the
non-measurability step: under correlated refinements $W_{2:L}$ may determine the
residual of $X_1$, making every function of $X_1$ measurable with respect to the
full library. Condition~\eqref{eq:residual_entropy}, together with the joint
key, avoids both.
\end{remark}

\begin{remark}[Comparison with neighbouring converses]
\label{rem:prior_converse_comparison}
Table~\ref{tab:converse_comparison} records, for each neighbouring converse,
the hypothesis that the present model violates. The discriminating condition is
hypothesis~(iv) of Theorem~\ref{thm:refined_converse}: answers are
deterministic given the \emph{refined} storage $X_{1:L}$, not given the
functional messages $W_{1:L}=\phi(X_{1:L})$; Proposition~\ref{prop:not_relabelling}
shows this cannot be repaired by redefining the messages.
Colluding PIR~\cite{sun2018colluding} is the special case $X_\ell=W_\ell$,
$\phi=\mathrm{id}$, and its peeling induction relies on message-level
determinism directly. Private computation~\cite{sun2019capacity} lets the
private index select the \emph{function} from a \emph{public} candidate family
whose full-rank basis determines the underlying independent datasets, whereas
our $\phi=\phi_{\mathbf{a}}$ is fixed by the decoder-private $\A$ and never
available to the servers. The coded private-linear-computation
model~\cite{obead2022private} and the general nonlinear-computation
model~\cite{obead2019nonlinear} already permit a storage/function mismatch: when
the candidate coefficient matrix is rank-deficient (or the family arbitrary),
the demanded functions determine only a projection of the stored base messages,
and answers may depend on the residual. Those results are noncolluding, use
independent base messages, and carry no decoder-side common state $S=\A$ or
arbitrary cross-file-correlated refinements---which are exactly the conditions
Theorem~\ref{thm:refined_converse} addresses. The distinction from private
computation is thus not the presence of residual storage per se but the
combination of $P_c$-collusion, a private \emph{file} index under a
decoder-private fixed map, conditionally i.i.d.\ per-file functional messages,
and correlated refinements. Dependent-message
PIR~\cite{chen2020asymptotic} admits dependency \emph{among the demanded
messages}---more general than our hypothesis~(ii), under which the $W_\ell$ are
conditionally i.i.d.\ given $\A=\mathbf{a}$---but still computes answers from
those messages, so residual storage content has no counterpart. These
generalizations are transverse: no listed converse subsumes
Theorem~\ref{thm:refined_converse}.
A second, independent requirement is that privacy transfer from the raw
correlated library to its $\A$-dependent coarsening
(Proposition~\ref{thm:transfer}), which has no analogue in settings whose
stored objects coincide with the demanded ones.
\end{remark}

\begin{table}[t]
\caption{Which hypothesis of each neighbouring converse fails in the
refined-storage model of Theorem~\ref{thm:refined_converse}. ``Answers from''
records the variables the answer maps are assumed deterministic in.}
\label{tab:converse_comparison}
\centering
\renewcommand{\arraystretch}{1.25}
\begin{tabular}{p{0.15\textwidth}p{0.15\textwidth}p{0.15\textwidth}p{0.13\textwidth}p{0.24\textwidth}}
\toprule
converse & private object & answers from & residual storage? & distinction from Theorem~\ref{thm:refined_converse} \\
\midrule
Dependent-message PIR~\cite{chen2020asymptotic}
  & search value / message index
  & complete dependent-message library
  & no separate refinement
  & arbitrarily dependent messages, but message-level storage and noncollusion \\
Private computation~\cite{sun2019capacity}
  & public-family function / coefficient vector
  & replicated independent datasets
  & no, in its full-rank capacity formulation
  & public function family; noncolluding; function index, not file index \\
Private linear comp.~\cite{obead2022private}
  & linear function / coefficient vector
  & locally coded base-message storage
  & yes, when coefficient rank $r<f$
  & noncolluding; linear coded storage; independent base messages \\
Private nonlinear comp.~\cite{obead2019nonlinear}
  & nonlinear function
  & replicated base messages
  & yes in general; no in its identity-plus-nonlinear capacity case
  & noncolluding; no decoder-side state or correlated per-file refinements \\
\midrule
Theorem~\ref{thm:refined_converse}
  & file index under fixed $\phi$
  & arbitrary refined storage $X_{1:L}$
  & yes, including correlated residuals
  & $P_c$-collusion, decoder side information, conditional message structure, explicit locality proof \\
\bottomrule
\end{tabular}
\end{table}
\begin{proof}
The full proof, with every step displayed in refined-storage notation, is in
Appendix~\ref{app:kadhe}
(equations~\eqref{eq:ref_base}--\eqref{eq:ref_peel}): the base and peeling
steps use chain identities valid without answer determinism, the locality
lemma is Lemma~\ref{lem:app_locality_refined}, the privacy swap is carried out
with an explicit marginalization from $W_{1:L}$ to $W_{1:k}$, Shearer is
applied under the common conditioning $(W_{1:k},Q^{[k+1]}_{1:N},S)$, and the
final nonnegative residual is dropped on the lower-bound side. The refinement
could a priori break the classical argument at the locality step, repaired by
Lemma~\ref{lem:app_locality_refined}, and at the privacy swap, where
privacy~(vi) and the storage--demand independence and full-support demand
of~(i) are used together. Neither step uses any property of the residual
$X_\ell\mid W_\ell$, so refinements arbitrary within a file and correlated
across files are admissible.
\end{proof}

\subsection{The Functional Lower Bound for Every Fixed-Rank Operand}

\begin{theorem}[Rank-Deficient Functional Lower Bound]
\label{prop:rankdef_lb}
Consider the library-wide additive model~\eqref{eq:additive} under (M1)--(M5),
with $P_A$ supported on matrices of rank exactly $k$, where $1\leq k<m$. For
$P_A$-almost every realization $\mathbf{a}$,
define the induced \emph{functional messages}
\begin{equation}
  W_\ell(\mathbf{a}) \;\triangleq\; \mathbf{a}^\top\B^{(\ell)},\qquad \ell\in[L].
  \label{eq:functional_msgs}
\end{equation}
Any scheme satisfying correctness~\eqref{eq:correctness}, side-information-robust
index privacy (Definition~\ref{def:index}), and conditioned data security
(Definition~\ref{def:security})---the last again retained as part of the threat
model but not used by the proof
(Remark~\ref{rem:converse_security_unused})---obeys
\begin{equation}
  D \;\geq\; \frac{H(\Ctheta\mid\A)}{\CPIR(N,P_c,L)}-o(km),
  \label{eq:rankk_lb}
\end{equation}
with no assumption on the structure of the rank-$k$ map
(Remark~\ref{rem:unconditional_scope} delimits what the bound's universality does and
does not mean). The slack is stated at
the scale of the bound itself: $\Ctheta$ is a $k\times m$-dimensional object
(after removing the rank deficiency), so $H(\Ctheta\mid\A)=O(km\log q)$, and an
$o(m^2)$ slack would be vacuous for $k=o(m)$. Throughout, $L$, $N$, $P_c$, and
$q$ are held fixed as $m\to\infty$; the asymptotics are in $m$ alone. This
matters here because the factor $L$ enters the Fano bookkeeping (the
$\varepsilon_{\mathbf{a}}\leq 1+L\,P_e(\mathbf{a})\,m^2\log q$ accounting
of~\eqref{eq:eps_a_fullrank} and its rank-deficient analogue below), so the
penalty is $o(km)$ for fixed $L$ but not uniformly in~$L$.
\end{theorem}

\begin{remark}[Scope of the rank-$k$ bound]
\label{rem:unconditional_scope}
The bound's universality concerns the \emph{structure of the operand}: it holds
for every rank-$k$ map $\mathbf{a}^\top(\cdot)$, with no coordinate-selection,
sparsity, or genericity hypothesis, and no additional
functional-privacy-transfer assumption beyond the model definitions already
stated, since Proposition~\ref{thm:transfer} supplies that transfer from
Definition~\ref{def:index} itself. The theorem is \emph{not} assumption-free:
it retains the rank profile~(M1), the innovation
independence~(M2)--(M3), the demand and randomness axioms~(M4)--(M5),
replicated storage, correctness~\eqref{eq:correctness},
side-information-robust index privacy
(Definition~\ref{def:index}), conditioned data security
(Definition~\ref{def:security}), and fixed $L$ as $m\to\infty$.
\end{remark}

\begin{proof}
Fix a rank-deficient realization $\mathbf{a}$ and consider the induced instance
retrieving $W_\theta(\mathbf{a})$. We verify the hypotheses of
Theorem~\ref{thm:refined_converse} with refinement variables
$X_\ell:=\B^{(\ell)}$ and $\phi=\mathbf{a}^\top(\cdot)$.
\emph{(Refinement structure and demand.)} Given $\A=\mathbf{a}$, the files
$\B^{(\ell)}=\mathbf{a}+\N^{(\ell)}$ are mutually independent and independent of
$(\theta,\mathcal{R})$ jointly, by (M2) and (M4)--(M5); moreover $\theta$ is
uniform on $[L]$ and independent of $\A$ by (M4), so with $S$ the constant
$\A=\mathbf{a}$ the full-support demand condition of hypothesis~(i) holds. Next,
$W_\ell(\mathbf{a})=\mathbf{a}^\top\mathbf{a}+\mathbf{a}^\top\N^{(\ell)}$.
Since the shift $\mathbf{a}^\top\mathbf{a}$ is a constant given
$\A=\mathbf{a}$, the $W_\ell(\mathbf{a})$ are mutually independent and
identically distributed given $\A=\mathbf{a}$, so hypothesis~(ii) of
Theorem~\ref{thm:refined_converse} applies without modification with
$H=H(\mathbf{a}^\top\N^{(\theta)})$; the answers
$\Yans_i=g_i(Q_i,\{\B^{(\ell)}\})$ are deterministic in the refined library.
\emph{(Privacy.)} The functional-message privacy
\begin{equation}
  I\!\left(\theta;\{Q_i^{(\theta)},\Yans_i^{(\theta)}\}_{i\in\mathcal{T}}
  \;\big|\;\{W_\ell(\mathbf{a})\},\,\A=\mathbf{a}\right)=0,
  \qquad |\mathcal{T}|\leq P_c,
  \label{eq:func_privacy}
\end{equation}
holds for $P_A$-a.e.~$\mathbf{a}$ \emph{as a consequence of}
Definition~\ref{def:index}, by Proposition~\ref{thm:transfer} applied to the
coarsening $W=\{W_\ell(\mathbf{a})\}=\{\mathbf{a}^\top\B^{(\ell)}\}$; no
separate assumption is needed.
\emph{(Correctness.)} The decoder uses the user randomness $\mathcal{R}$,
which correlates with the transcript; it may nevertheless be dropped from the
Fano conditioning: writing $X=\{\B^{(\ell)}\}$,
$Z=(Q_{1:N},\theta,\A{=}\mathbf{a})$, and $\Yans_{1:N}=\psi(X,Z)$, the
independence $X\indep\mathcal{R}\mid Z$ (from (M5) and $Q=f(\theta,\A,\mathcal{R})$)
factorizes as
$P(x,r\mid y,z)=P(x\mid z)\mathbbm{1}\{\psi(x,z)=y\}P(r\mid z)/P(y\mid z)
=P(x\mid y,z)\,P(r\mid z)$, so $X\indep\mathcal{R}\mid(\Yans_{1:N},Z)$; since
$W_\theta(\mathbf{a})=\phi(X)$, this gives
$I(W_\theta(\mathbf{a});\mathcal{R}\mid\Yans_{1:N},Q_{1:N},\theta,\A{=}\mathbf{a})=0$.
The Fano alphabet is small: each column of
$W_\theta(\mathbf{a})=\mathbf{a}^\top\B^{(\theta)}$ lies in the column space of
$\mathbf{a}^\top$, a subspace of dimension~$k$, so $W_\theta(\mathbf{a})$ takes
at most $(q^{k})^{m}=q^{km}$ values. Define the uniform correctness penalty
\begin{equation}
  \varepsilon_{\mathbf{a}}
  \;:=\;
  \max_{t\in[L]}
  H\!\left(W_t(\mathbf{a})\,\middle|\,
  \Yans^{[t]}_{1:N},Q^{[t]}_{1:N},\A{=}\mathbf{a}\right).
  \label{eq:eps_a_def}
\end{equation}
Global correctness~\eqref{eq:correctness} bounds the $\theta$-averaged error
$P_e(\mathbf{a})$ with $\mathbb{E}_\A[P_e(\A)]\to 0$; since $\theta$ is uniform
on $[L]$ with $L$ fixed, the per-index error is at most $L\,P_e(\mathbf{a})$,
and Fano over the $q^{km}$-point alphabet gives
\begin{equation}
  \varepsilon_{\mathbf{a}}
  \;\leq\;
  1+L\,P_e(\mathbf{a})\,km\log q.
  \label{eq:eps_a_fano}
\end{equation}
Conditioning on any processed subset $W_J$ only decreases entropy, so
hypothesis~(v) of Theorem~\ref{thm:refined_converse} holds at level
$\varepsilon_{\mathbf{a}}$ uniformly over $(t,J)$. The messages
$W_\ell(\mathbf{a})$ need not have memoryless entries---for dense maps the
coordinates within a column of $\mathbf{a}^\top\N^{(\ell)}$ are dependent---but
the converse uses only the per-message entropy
(Remark~\ref{rem:app_nonuniform}, Appendix~\ref{app:kadhe}).
Theorem~\ref{thm:refined_converse} therefore gives, for $P_A$-a.e.\
$\mathbf{a}$,
\begin{equation}
  D_{\mathbf{a}}
  \;\geq\;
  \frac{H(W_\theta(\mathbf{a})\mid\A{=}\mathbf{a})-\varepsilon_{\mathbf{a}}}
       {\CPIR(N,P_c,L)},
  \label{eq:Da_bound}
\end{equation}
with $W_\theta(\mathbf{a})=\Ctheta$ when $\A=\mathbf{a}$. Finally,
by~\eqref{eq:eps_a_fano},
$\mathbb{E}_\A[\varepsilon_\A]\leq 1+L\,km\log q\;\mathbb{E}_\A[P_e(\A)]=o(km)$,
so the averaging transfer $D\geq\mathbb{E}_\A[D_\A]$
(Lemma~\ref{lem:averaging}, which uses no rank hypothesis) applied
to~\eqref{eq:Da_bound}, together with
$\mathbb{E}_\A[H(W_\theta(\A)\mid\A)]=H(\Ctheta\mid\A)$ and
$\varepsilon_\A/\CPIR\leq L\varepsilon_\A$, yields~\eqref{eq:rankk_lb}.
\end{proof}

\begin{remark}[From conditional to unconditional: the two obstructions]
\label{rem:rankdef_gap}
An earlier version of this result assumed~\eqref{eq:func_privacy} as an extra
hypothesis. Two apparent obstructions motivated the caution, and both dissolve
under the model's own definitions. \emph{First}, privacy conditioned on the full
library does not, as a generic matter of conditional mutual information,
transfer to privacy conditioned on the coarser functional library
$\sigma(\{W_\ell\})\subseteq\sigma(\{\B^{(\ell)}\})$
(Proposition~\ref{thm:transfer}(b)); but part~(a) shows the
transfer is universal here, because Definition~\ref{def:index}'s joint form
encodes the demand-independence~(M4). \emph{Second}, the servers store the
larger files $\B^{(\ell)}$, so the answers are \emph{not} deterministic
functions of the functional library and the colluding converse of
Lemma~\ref{lem:kadhe} does not apply directly to $\{W_\ell\}$; this is resolved
by the refined-storage extension (Theorem~\ref{thm:refined_converse}), whose proof
shows the determinism hypothesis is needed only at the refined level. The
resulting statement---a colluding converse for private retrieval of a fixed
per-file linear function when servers store the larger independent files---is a
fixed-function, private-index counterpart of private
computation~\cite{sun2019capacity}, where instead the function index is private.
\end{remark}

The upper bound, by contrast, requires no structural hypothesis either:
\begin{equation}
  D^* \;\leq\; \frac{H(\btheta\mid\A)}{\CPIR(N,P_c,L)}+o(m^2),
  \label{eq:rankk_ub}
\end{equation}
obtained by retrieving enough to reconstruct $\btheta$ (the achievability
construction of Theorem~\ref{thm:achievability} applied with the innovation
library---a construction that nowhere uses invertibility of~$\A$ in the
retrieval itself) and computing $\Ctheta=\A^\top\btheta$ locally. The functional lower bound~\eqref{eq:rankk_lb} is valid for every rank-$k$
operand (Theorem~\ref{prop:rankdef_lb}); what remains open in the general rank-$k$ case
is \emph{not} the converse but the matching achievability at the functional
lower endpoint, the gap between~\eqref{eq:rankk_lb} and~\eqref{eq:rankk_ub}
reflecting the private functional-compression problem. For the
coordinate-selection class, the published branch closes this gap
(Theorem~\ref{thm:rankk_public}), while the concealed branch bounds it from the
other side under the strengthened hypotheses of
Theorem~\ref{thm:rankk_impossible}.

Combining the lower bound~\eqref{eq:rankk_lb} with the
upper bound~\eqref{eq:rankk_ub}, both free of structural hypotheses on the
map, the optimal rank-$k$ cost satisfies
\begin{equation}
  \frac{H(\Ctheta|\A)}{\CPIR}\;-\;o(km) \;\leq\; D^* \;\leq\; \frac{H(\btheta|\A)}{\CPIR}
  \;+\;o(m^2).
  \label{eq:rankk_interval}
\end{equation}
The bounds may be strictly separated
(Example~\ref{ex:rankdeficient}), and what remains open is which endpoint is
tight in general: a matching private functional-compression achievability at
the lower endpoint, addressed by the results below for coordinate-selection
structure. Classical functional
compression~\cite{orlitsky2001coding,doshi2010functional} exploits the
graph-coloring structure of $f(\A,\cdot)$, but in a PIR setting the server cannot
evaluate $f$ without leaking $\theta$. Closing the gap requires a new architecture
combining coded computation with PIR, at the intersection of functional
compression and private computation.
Example~\ref{ex:rankdeficient} below shows the gap is not vacuous.

\subsection{Specialization: The Coordinate-Selection Class}

The class on which the contrast between Theorems~\ref{thm:rankk_public}
and~\ref{thm:rankk_impossible} is resolved is the following structured family,
for which the functional messages are lossless sub-blocks of the innovations.

\begin{definition}[Coordinate-selection class $\mathcal{A}_k$]
\label{def:selection_class}
$\A\in\mathcal{A}_k$ if $\A$ has rank $k$ and $\A^\top$ acts as a
\emph{coordinate row-selector}: there is a size-$k$ index set
$\mathcal{S}(\A)\subseteq[m]$ such that $\mathbf{a}^\top\N$ reproduces the rows of
$\N$ indexed by $\mathcal{S}(\A)$ (up to an $\A$-determined invertible relabeling
within the selected block), and discards the rest. Precisely: for each
$\mathbf{a}$ there is a bijection $\psi_{\mathbf{a}}$, determined by
$\mathbf{a}$, with
$\psi_{\mathbf{a}}(\mathbf{a}^\top\N)=\N_{\mathcal{S}(\mathbf{a})}$ for every
$\N$, an invertible relabeling of the selected block, so that
\begin{equation}
  H\!\left(\mathbf{a}^\top\N\,\middle|\,\A{=}\mathbf{a}\right)
  =H\!\left(\N_{\mathcal{S}(\mathbf{a})}\right)
  \label{eq:selection_equiv}
\end{equation}
and exact recovery of $\mathbf{a}^\top\N^{(\ell)}$ given $\A=\mathbf{a}$ is
exact recovery of $\N^{(\ell)}_{\mathcal{S}(\mathbf{a})}$, and conversely. We
refer to this property as \emph{informational equivalence} of
$W_\ell(\mathbf{a})=\mathbf{a}^\top\N^{(\ell)}$ and
$\N^{(\ell)}_{\mathcal{S}(\mathbf{a})}$.
\end{definition}

\begin{corollary}[Rank-$k$ functional converse for $\mathcal{A}_k$]
\label{thm:rankk_selection}
Consider the library-wide model~\eqref{eq:additive} with replicated storage,
$P_c$-collusion, side-information-robust index privacy
(Definition~\ref{def:index}), and conditioned data security
(Definition~\ref{def:security}). If $\A\in\mathcal{A}_k$ almost surely, with
$k=\alpha m$,
$0<\alpha<1$, then
\begin{equation}
  D^* \;\geq\; \frac{H(\A^\top\btheta\mid\A)}{\CPIR(N,P_c,L)}-o(km).
  \label{eq:rankk_selection_lb}
\end{equation}
For binary innovations $N_{i,j}\sim\mathrm{Bern}(p)$ this is
$D^*\geq km\,\Hb(p)/\CPIR-o(km)$.
\end{corollary}
\begin{proof}
Immediate from Theorem~\ref{prop:rankdef_lb}. For the binary value: given
$\A=\mathbf{a}\in\mathcal{A}_k$ with selected-row set
$\mathcal{S}=\mathcal{S}(\mathbf{a})$, Definition~\ref{def:selection_class}
makes $W_\theta(\mathbf{a})$ informationally equivalent to the selected
sub-block $\N^{(\theta)}_{\mathcal{S}}$, so by~\eqref{eq:selection_equiv},
$H(\A^\top\btheta\mid\A=\mathbf{a})=H(\N_{\mathcal{S}})=km\,\Hb(p)$. We note
for later use (Theorem~\ref{thm:rankk_impossible}, Step~4) that for this class
the functional library is the lossless sub-block library
$\{\B^{(\ell)}_{\mathcal{S}}\}$, so Proposition~\ref{thm:transfer} yields privacy
conditioned on any sub-block library directly.
\end{proof}

\begin{remark}[The functional regime is not a disguised retrieval problem]
\label{rem:rankk_significance}
The functional lower bound is valid for every rank-$k$
operand (Theorem~\ref{prop:rankdef_lb}); the coordinate-selection class is
where its \emph{tightness} can be resolved. In this regime $\A^\top\btheta$ does not determine $\btheta$, so the result
is not an artifact of full-rank invertibility.
For the branches studied here, tightness of the functional lower bound turns on
whether the selected-coordinate pattern $\mathcal{S}(\A)$ is public or private:
we characterize the published case, and derive a concealed-structure penalty
under the strengthened hypotheses of Theorem~\ref{thm:rankk_impossible}; the
general concealed capacity is not characterized (Remark~\ref{rem:dichotomy}).
\end{remark}

\smallskip
\subsection{Published Functional Structure: A Functional-Rate Capacity Theorem}

The coordinate-selection converse extends to \emph{any} published
rank-$k$ row-space structure, and the resulting bound is tight, the matching
achievability resting on the uniform projected-source coding of
Lemma~\ref{lem:uniform_source}. Care is needed
about \emph{what} is published: announcing the map $F_\A(\cdot)=\A^\top(\cdot)$
itself would reveal $\A$ entirely and render data security vacuous. What is
published is a \emph{canonical representative} $R_\A\in\Fq^{k\times m}$ of the
rank-$k$ row space of $\A^\top$ (for the coordinate-selection subclass, the
selector onto the pattern $\mathcal{S}(\A)$), fixed by a public canonicalization
rule (e.g.\ reduced row-echelon form). Given the published $R_\A$, the unique
coefficient matrix $M_\A\in\Fq^{m\times k}$, $\mathrm{rank}(M_\A)=k$, determined
by $\A^\top=M_\A R_\A$ (uniqueness because $R_\A$ has full row rank), completely
determines $\A^\top$, and hence $\A$; $M_\A$ remains private. Full column rank
furnishes a left inverse $L_\A\in\Fq^{k\times m}$
with $L_\A M_\A=I_k$. The
servers form the \emph{canonical functional messages}
\begin{equation}
  W_\ell \;=\; R_\A\,\B^{(\ell)} \;=\; R_\A\,\A+R_\A\,\N^{(\ell)},
  \label{eq:pub_functional_msgs}
\end{equation}
from which the user recovers
$\A^\top\B^{(\ell)}=M_\A W_\ell$ using its private $M_\A$; conversely
$W_\ell=L_\A\,\A^\top\B^{(\ell)}$, so given $\A=\mathbf{a}$ the two are
deterministic functions of each other---informationally equivalent at the
decoder---and $H(\A^\top\B^{(\theta)}\mid\A)=H(W_\theta\mid\A)$. The retrieval
index $\theta$ remains private. Since the innovations are i.i.d.\ across~$\ell$,
the $W_\ell$ are conditionally i.i.d.\ given $\A=\mathbf{a}$ (up to the common
shift $R_\A\A$).

Because $R_\A$ is published before the protocol runs---known to user and
servers alike---the operative download measure in this subsection conditions
on it:
\begin{equation}
  D_{\mathrm{pub}}\;\triangleq\;\sum_{i=1}^{N}
  H\!\bigl(\Yans_i^{(\theta)}\,\big|\,R_\A\bigr)
  \;=\;\mathbb{E}_{R_\A}\!\left[\sum_{i=1}^{N}
  H\!\bigl(\Yans_i^{(\theta)}\,\big|\,R_\A{=}\mathsf{P}\bigr)\right].
  \label{eq:Dpub}
\end{equation}
Conditioning removes only the mixture uncertainty of the already-public
parameter: the unconditional $\sum_i H(\Yans_i^{(\theta)})$ would charge the
protocol for the entropy of $R_\A$ itself, which every party already holds.
Every per-realization converse below transfers to $D_{\mathrm{pub}}$
unchanged: since $\A$ determines $R_\A$, conditioning on $\A$ is finer, so
$D_{\mathrm{pub}}\geq\sum_i H(\Yans_i^{(\theta)}\mid\A)
=\mathbb{E}_\A[D_\A]$ (Lemma~\ref{lem:averaging}).

\begin{lemma}[Uniform linear compression of projected innovations]
\label{lem:uniform_source}
Let $\N_m\in\Fq^{m\times m}$ have i.i.d.\ entries following an arbitrary fixed
distribution on $\Fq$, and let $1\leq k<m$. Write $\mathcal{P}_{m,k}$ for the
family of full-row-rank $\mathsf{P}\in\Fq^{k\times m}$. There exist public
linear encoders $h_{m,\mathsf{P}}\colon\Fq^{km}\to\Fq^{r_m(\mathsf{P})}$ and
decoders such that
\begin{equation}
  \sup_{\mathsf{P}\in\mathcal{P}_{m,k}}
  P_e(m,\mathsf{P})
  \;\leq\;\frac{C_0}{\sqrt m}+2^{-(k\log_2 q+1)m^{3/4}}
  \;\longrightarrow\;0,
  \label{eq:uniform_source}
\end{equation}
where $P_e(m,\mathsf{P})=\Pr[\widehat{\mathsf{P}\N_m}\neq\mathsf{P}\N_m]$ and
$C_0$ is a universal constant, and
\begin{equation}
  \sup_{\mathsf{P}\in\mathcal{P}_{m,k}}
  \bigl[\,r_m(\mathsf{P})\log_2 q-H(\mathsf{P}\N_m)\,\bigr]
  \;\leq\;2\bigl(k\log_2 q+1\bigr)m^{3/4}+\log_2 q
  \;=\;o(km).
  \label{eq:uniform_rate}
\end{equation}
\end{lemma}

\begin{remark}[Scope of the uniformity]
\label{rem:uniform_source_scope}
``Uniform'' refers to performance across the published representatives
$\mathsf{P}$, not to the innovation law: $P_N$ may be Bernoulli-$p$, sparse
$q$-ary, or any other fixed i.i.d.\ law, and need not have full support. Two
features of the argument (Appendix~\ref{app:uniform_source}) are worth noting.
First, the variance bound~\eqref{eq:iota_var} on the information density
follows from the alphabet size $q^{k}$ alone, so it is free of $\mathsf{P}$ and
of the induced law. Second, no union bound over $\mathcal{P}_{m,k}$ is
required: since $\mathsf{P}=R_\A$ is published before retrieval, a separate
encoder may be fixed for each representative, and the Gaussian-binomial
cardinality $\binom{m}{k}_q$ never enters. The bound~\eqref{eq:uniform_source}
is an asymptotic existence statement, not a finite-blocklength guarantee.
\end{remark}

\begin{theorem}[Published rank-$k$ functional-structure capacity]
\label{thm:rankk_public}
Suppose $P_A$ is supported on matrices of rank exactly $k$, with $1\leq k<m$,
under (M1)--(M5), and the canonical row-space representative $R_\A$ is
published
as a protocol parameter, while the coefficient matrix $M_\A$ stays private.
Given $R_\A$, the unique full-column-rank $M_\A$ in $\A^\top=M_\A R_\A$
completely determines $\A^\top$, and hence $\A$. Under side-information-robust index privacy (Definition~\ref{def:index})
and conditioned data security relative to the published structure
(Definition~\ref{def:security_public} with $\mathsf{P}=R_\A$), and for
innovations with i.i.d.\ entries following an arbitrary fixed distribution
on~$\Fq$, the optimal
\emph{public-setup} download~\eqref{eq:Dpub} for computing
$\Ctheta=\A^\top\B^{(\theta)}$ is
\begin{equation}
  D^*_{\mathrm{pub},k}
  \;=\; \frac{H(\A^\top\B^{(\theta)}\mid\A)}{\CPIR(N,P_c,L)}+o(km).
  \label{eq:rankk_public_cap}
\end{equation}
For i.i.d.\ $q$-ary innovations with $\A^\top$ of rank $k$,
$H(\A^\top\B^{(\theta)}\mid\A)=H(\A^\top\N^{(\theta)}\mid\A)\leq km\log q$; for the
binary coordinate-selection subclass with $N_{i,j}\sim\mathrm{Bern}(p)$ it equals
$km\,\Hb(p)$, giving $D^*_{\mathrm{pub},k}=km\,\Hb(p)/\CPIR+o(km)$.
\end{theorem}
\begin{proof}
\emph{Achievability.} For the published representative
$\mathsf{P}=R_\A$, let
$H_{\mathsf{P}}:=H(\mathsf{P}\,\Ntheta)$ denote the entropy (in bits) of the
functional innovation---a function of $\mathsf{P}$ alone, since
$\Ntheta\indep\A$ by~(M2); note
$H(W_\theta\mid\A{=}\mathbf{a})=H(R_\A\mathbf{a}+R_\A\Ntheta\mid\A{=}\mathbf{a})
=H_{R_\A(\mathbf{a})}$, so
$H(W_\theta\mid\A)=\mathbb{E}_{R_\A}[H_{R_\A}]$. By
Lemma~\ref{lem:uniform_source}, for each realizable $\mathsf{P}$ select the
public encoder $h_{m,\mathsf{P}}\colon\Fq^{k\times m}\to\Fq^{r_m(\mathsf{P})}$
supplied by that lemma, whose length obeys the \emph{uniform} rate
margin~\eqref{eq:uniform_rate},
$r_m(\mathsf{P})\log q\leq H_{\mathsf{P}}+\gamma_m$ with
$\gamma_m:=\sup_{\mathsf{P}}[r_m(\mathsf{P})\log q-H_{\mathsf{P}}]=o(km)$, and
whose decoding error vanishes uniformly over $\mathcal{P}_{m,k}$
per~\eqref{eq:uniform_source}.
Each server computes the canonical functional messages
$W_\ell=R_\A\B^{(\ell)}$ (a $k\times m$ object, computable from the published
$R_\A$ and its stored raw files alone---no private content of $\mathbf{a}$ is
needed) and stores the sketches
$T_\ell=h_{m,\mathsf{P}}(W_\ell)$ with $\mathsf{P}=R_\A$. The user privately retrieves the \emph{sketch}
$T_\theta$---not $W_\theta$ itself, whose direct retrieval would cost
$km\log q/\CPIR$ bits---by a capacity-achieving $P_c$-colluding PIR scheme over
the sketch library $\{T_\ell\}$ of $L$ files of $r_m(R_\A)$ $\Fq$-symbols;
subtracts the locally computable $h_{m,\mathsf{P}}(R_\A\mathbf{a})$, obtaining by
linearity the innovation sketch
$h_{m,\mathsf{P}}(R_\A\Ntheta)=T_\theta-h_{m,\mathsf{P}}(R_\A\mathbf{a})$; and source-decodes
$R_\A\Ntheta$ with the decoder of $h_{m,\mathsf{P}}$, whose error is at most
$\sup_{\mathsf{P}}\Pr[\widehat{\mathsf{P}\Ntheta}\neq\mathsf{P}\Ntheta]\to0$
by~\eqref{eq:uniform_source}.

\emph{Uniform reliability across published parameters.} Averaging over the
published $R_\A$ therefore preserves vanishing error,
$\mathbb{E}_{R_\A}[P_e(R_\A)]\leq\sup_{\mathsf{P}\in\mathcal{P}_{m,k}}
P_e(\mathsf{P})\to0$, which is what
correctness~\eqref{eq:correctness} requires; this is exactly the content
of~\eqref{eq:uniform_source} in Lemma~\ref{lem:uniform_source}. The user then forms
$W_\theta=R_\A\mathbf{a}+R_\A\Ntheta$ and
$\Ctheta=M_\A W_\theta$ locally using its private $M_\A$, the equality because
$W\mapsto M_\A W$ is injective, inverted on its image by $L_\A$, given
$\A=\mathbf{a}$. Conditioned on $R_\A=\mathsf{P}$ the download is at most
$r_m(\mathsf{P})\log q/\CPIR+o(km)\leq(H_{\mathsf{P}}+\gamma_m)/\CPIR+o(km)$
bits, with $\gamma_m=o(km)$ \emph{uniform} in $\mathsf{P}$
by~\eqref{eq:uniform_rate}; averaging over the published
$R_\A$ as in~\eqref{eq:Dpub}---the uniform margin lets the $\gamma_m/\CPIR$ term
pass through the expectation as a single $o(km)$ quantity---
\begin{equation*}
  D_{\mathrm{pub}}\;\leq\;\frac{\mathbb{E}_{R_\A}[r_m(R_\A)\log q]}{\CPIR}+o(km)
  \;\leq\;\frac{\mathbb{E}_{R_\A}[H_{R_\A}]+\gamma_m}{\CPIR}+o(km)
  \;=\;\frac{H(W_\theta\mid\A)}{\CPIR}+o(km)
  \;=\;\frac{H(\A^\top\B^{(\theta)}\mid\A)}{\CPIR}+o(km).
\end{equation*}
\emph{Converse.} An admissible protocol's answers are deterministic in the
stored raw files $\B^{(\ell)}$---a strict refinement of the functional messages
$W_\ell=R_\A\B^{(\ell)}$---so Lemma~\ref{lem:kadhe} cannot be applied without modification; we
instead invoke Theorem~\ref{thm:refined_converse} with $X_\ell=\B^{(\ell)}$ and
$\phi=R_\A(\cdot)$. Hypothesis~(vi)---privacy conditioned on
$\{W_\ell\}$---follows from Definition~\ref{def:index} by
Proposition~\ref{thm:transfer} applied to the coarsening
$\Phi_{\mathbf{a}}(\{\B^{(\ell)}\})=\{R_\A\B^{(\ell)}\}$
(Remark~\ref{rem:transfer_scope} covers the deterministic public parameter
$R_\A$, a function of $\mathbf{a}$). Hypothesis~(i) holds as in the proof of
Theorem~\ref{prop:rankdef_lb}: given $\A=\mathbf{a}$ the files are mutually
independent and independent of $(\theta,\mathcal{R})$ by (M2) and (M4)--(M5),
and $\theta$ is uniform on $[L]$ by (M4), supplying the full-support demand
condition; the remaining hypotheses likewise hold as in that proof, and the
averaging transfer
(Lemma~\ref{lem:averaging}) gives
$\mathbb{E}_\A[D_\A]\geq H(W_\theta\mid\A)/\CPIR-o(km)$; since $\A$ determines
$R_\A$, $D_{\mathrm{pub}}\geq\mathbb{E}_\A[D_\A]$
(see~\eqref{eq:Dpub}), completing the matching lower bound on retrieving
$W_\theta$ in the public-setup measure; any correct scheme
for $\Ctheta$ retrieves $W_\theta$ up to the user-known correspondence
$W_\theta\mapsto M_\A W_\theta$, inverted on its image by $L_\A$. Index privacy holds
by Proposition~\ref{prop:sir_satisfiable} in its public-parameter form: the
queries of the scheme above depend on $(\theta,\mathcal{R})$ and on $\A$ only
through the declared public $\mathsf{P}=R_\A$ (which also fixes the public
encoder $h_{m,\mathsf{P}}$), and conditioning on $\A=\mathbf{a}$ fixes both.
Conditioned data security
holds \emph{relative to the published structure}: given
$(\{\B^{(\ell)}\},R_\A)$, the canonical library $\{W_\ell\}=\{R_\A\B^{(\ell)}\}$,
the public encoder $h_{m,\mathsf{P}}$, and hence the sketch library $\{T_\ell\}$
are all determined, and the transcript is a function of $(\theta,\mathcal{R})$ and
these determined objects; since $(\theta,\mathcal{R})\indep(\A,\{\B^{(\ell)}\})$, this gives
$I(\A;V_{\mathcal{T}}\mid\{\B^{(\ell)}\},R_\A)=0$, i.e.\
Definition~\ref{def:security_public} with $\mathsf{P}=R_\A$. The guarantee can
remain non-vacuous whenever $H(\A\mid\{\B^{(\ell)}\},R_\A)>0$; the published
representative need not reveal the coefficient matrix $M_\A$.
\end{proof}

\begin{remark}[Setup cost and amortized communication]
\label{rem:setup_cost}
The metric $D_{\mathrm{pub}}$ of~\eqref{eq:Dpub} measures retrieval-phase
download after the canonical representative $R_\A$ has been announced. The user
announces $R_\A$ at setup, as part of the public parameter tuple $\mathsf{P}$
of Definition~\ref{def:security_public}; setup and upload lie outside the
download criterion, which is standard in the PIR-capacity literature
(Remark~\ref{rem:upload}). Let $\ell(R_\A)$ denote the description length of
the representative; the canonical row-reduced form gives
$\ell(R_\A)\leq km\log_2 q$ bits. For a one-shot retrieval the total
communication attributable to publication and retrieval is
$D_{\mathrm{one\text{-}shot}}=\ell(R_\A)+D_{\mathrm{pub}}$; if the same
representative serves $T$ retrievals, the amortized cost per retrieval is
$D_{\mathrm{amortized}}(T)=D_{\mathrm{pub}}+\ell(R_\A)/T$. Publication is
worthwhile once
\begin{equation*}
  T\;>\;\frac{\ell(R_\A)}{D_{\mathrm{full}}-D_{\mathrm{pub}}},
\end{equation*}
where $D_{\mathrm{full}}$ is the full-file download and
$D_{\mathrm{full}}>D_{\mathrm{pub}}$. For the binary coordinate-selection
subclass, the leading-order per-retrieval saving is
\begin{equation*}
  D_{\mathrm{full}}-D_{\mathrm{pub}}
  \;=\;\frac{m(m-k)\Hb(p)}{\CPIR(N,P_c,L)}+o(m^2).
\end{equation*}
Using the general description-length bound $\ell(R_\A)\leq km\log_2 q$, the
break-even ratio satisfies
\begin{equation*}
  \frac{\ell(R_\A)}{D_{\mathrm{full}}-D_{\mathrm{pub}}}
  \;\lesssim\;\frac{k\,\CPIR(N,P_c,L)\log_2 q}{(m-k)\Hb(p)}.
\end{equation*}
This ratio need not be below one, particularly when $k$ is close to $m$ or the
innovation entropy is small. Publication is most advantageous when the
row-space description is compact, the codimension $m-k$ is appreciable, or the
same representative is reused across multiple retrievals. For coordinate-selector
row spaces, describing the selected index set rather than a complete
$k\times m$ matrix can further reduce the setup cost. Reuse over $T>1$
retrievals requires only that the \emph{row space} of $\A$ stay fixed---the
private coefficient matrix $M_\A$ and the demands may vary---which matches the
motivating applications, where a user holds one operand $\A$ and retrieves
products against successive files. Theorem~\ref{thm:rankk_public}
therefore characterizes the retrieval phase; the one-shot total exceeds it by
the representation cost of $R_\A$, and the amortized cost approaches the
functional rate as the published structure is reused.
\end{remark}

\begin{remark}[Why Definition~\ref{def:security_public} is the right guarantee
here, and consistency of the two results]
\label{rem:dichotomy_consistency}
In the public-structure model the colluding view contains the announced
representative $R_\A$ itself, a deterministic function of $\A$, so the
library-only form of conditioned data security (Definition~\ref{def:security})
is \emph{infeasible by the joint distribution} whenever the structure is not
determined by the library: if $H(R_\A\mid\{\B^{(\ell)}\})>0$, then
$I(\A;R_\A\mid\{\B^{(\ell)}\})=H(R_\A\mid\{\B^{(\ell)}\})>0$ already,
independently of any protocol, as in Proposition~\ref{prop:infeasible}
applied to the published parameter. Publishing an $\A$-dependent parameter imposes
a protocol-independent floor of leakage relative to the library alone, and the
meaningful relative guarantee conditions on the published object as well
(Remark~\ref{rem:security_public}(ii)); this is what
Theorem~\ref{thm:rankk_public} certifies: once $R_\A$ is fixed, the unique
coefficient matrix $M_\A$ carries all remaining information in $\A^\top$, and it
stays protected. Conversely,
Theorem~\ref{thm:rankk_impossible} below concerns protocols whose public
parameters are \emph{$\A$-independent}---the functional structure is
concealed---so that Definitions~\ref{def:security}
and~\ref{def:security_public} coincide there
(Remark~\ref{rem:security_public}(i)). The two theorems therefore occupy
\emph{disjoint} admissible classes, distinguished by whether the functional
structure is published as a protocol parameter or withheld as private content
of~$\A$; the contrast is between \emph{published} and \emph{concealed}
functional structure, not two readings of one definition.
\end{remark}

\begin{corollary}[Published coordinate-selection capacity]
\label{cor:rankk_public_selection}
For the coordinate-selection subclass $\mathcal{A}_k^{\mathcal{S}}$ with a fixed,
published row set $\mathcal{S}$ (so $R_\A$ is the public selector onto rows
$\mathcal{S}$), the induced functional innovation $R_\A\Ntheta$ consists of $km$
i.i.d.\ symbols. For a fixed published selector $R_\A$, the set of realizable
public parameters is the singleton $\{R_\A\}$; consequently, the suprema
in~\eqref{eq:uniform_source}--\eqref{eq:uniform_rate} may be restricted to that
singleton. Theorem~\ref{thm:rankk_public} yields
$D^*_{\mathcal{A}_k^{\mathcal{S}}}=km\,\Hb(p)/\CPIR+o(km)$ in the binary case;
here $R_\A$ is deterministic, so the public-setup
measure~\eqref{eq:Dpub} coincides with the unconditional download.
\end{corollary}

\smallskip
\subsection{Concealed Functional Structure: The Join-Subspace Penalty}

The published-structure capacity relied on announcing $R_\A$, which let the
achievability scheme compress against a known row space. Suppose instead that the
row space is itself private---part of the secret content of~$\A$. Then no single
row space is available to compress against. A correct scheme must decode
$\mathbf{a}^\top\B^{(\theta)}$ whichever $\mathbf{a}$ the user holds, and, under
the exactness, common-support, and full-transcript hypotheses below, a
transcript decodable for one supported row space becomes simultaneously
decodable across the row-space support. The user then pays not for the
$k$-dimensional map it
needs, but for the linear \emph{join} of every row space in the support of~$\A$.
The functional rate is recovered only when that join is itself $k$-dimensional
(a fixed, public row space), and degrades smoothly as the join grows.

\begin{theorem}[Concealed row-space join converse]
\label{thm:rankk_impossible}
Consider the library-wide model~\eqref{eq:additive} under (M1)--(M5), with
$\mathrm{rank}(\A)=k$ almost surely for a fixed $1\leq k<m$. For each
$\mathbf{a}\in\mathrm{supp}(P_A)$
let $R_{\mathbf{a}}\in\Fq^{k\times m}$ be a canonical full-row-rank
representative of $\mathrm{row}(\mathbf{a}^\top)$, and define the
\emph{join} of the concealed row spaces,
\begin{equation}
  \mathcal{V}_m\;\triangleq\!\!\sum_{\mathbf{a}\in\mathrm{supp}(P_A)}\!\!
  \mathrm{row}(\mathbf{a}^\top),
  \qquad r_m\triangleq\dim\mathcal{V}_m,
  \label{eq:join_subspace}
\end{equation}
with $G_m\in\Fq^{r_m\times m}$ any full-row-rank basis matrix for
$\mathcal{V}_m$. Assume:
\emph{(i)}~\emph{common conditional library support}:
$\mathrm{supp}(\{\B^{(\ell)}\}\mid\A{=}\mathbf{a})=\mathcal{B}_m$ for every
$\mathbf{a}\in\mathrm{supp}(P_A)$;
\emph{(ii)}~exact correctness at each blocklength,
$\Pr[\hat{\C}^{(\theta)}\neq\A^\top\btheta]=0$;
\emph{(iii)}~side-information-robust index privacy (Definition~\ref{def:index});
and \emph{(iv)}~the \emph{$\theta$-conditioned full-transcript form} of
Definition~\ref{def:security} (see
Remark~\ref{rem:rankk_impossible_scope}(a) for its exact relation to
Definition~\ref{def:security}), with all public parameters independent of~$\A$
(so Definitions~\ref{def:security} and~\ref{def:security_public} coincide,
Remark~\ref{rem:security_public}(i)):
\begin{equation}
  I\!\left(\A;\,Q_{1:N}^{(\theta)},\Yans_{1:N}^{(\theta)}
  \;\Big|\;\{\B^{(\ell)}\},\theta\right)=0.
  \label{eq:full_transcript_security}
\end{equation}
Then, writing
\begin{equation}
  D^*_{\mathrm{priv},k}\;\triangleq\;
  \inf_{\pi\in\Pi_{\mathrm{priv},k}}D(\pi),
  \label{eq:dpriv_class}
\end{equation}
where $\Pi_{\mathrm{priv},k}$ is the class of protocols satisfying
hypotheses~(i)--(iv) above---exact correctness, common conditional library
support, side-information-robust index privacy, and $\theta$-conditioned
full-transcript security with $\A$-independent public parameters---we have
\begin{equation}
  D^*_{\mathrm{priv},k} \;\geq\;
  \frac{H\!\big(G_m\N^{(\theta)}\big)}{\CPIR(N,P_c,L)}.
  \label{eq:rankk_impossible}
\end{equation}
The infimum is over this class and not over the wider class admitted by
Definitions~\ref{def:index} and~\ref{def:security} alone; the bound is
therefore a statement about the strengthened observational model, not about
$D^*$ as used elsewhere in the paper.
The bound depends only on the subspace $\mathcal{V}_m$, not on the chosen
basis: any other basis is $G_m'=TG_m$ with $T$ invertible, so
$H(G_m'\N^{(\theta)})=H(G_m\N^{(\theta)})$.
Concealing several possible row spaces therefore forces retrieval over their
linear join, and the apparent rank-$k$ functional gain disappears whenever
$r_m>k$, even though the user needs only the $km$ functional symbols.
\end{theorem}

\begin{remark}[Support domination suffices]
\label{rem:support_domination}
Hypothesis~(i) can be weakened to a one-sided condition. Let
$\mathbf{a}_1,\dots,\mathbf{a}_J$ be the spanning realizations selected in
Step~3 of the proof. It suffices that
$\mathrm{supp}(\{\B^{(\ell)}\}\mid\A{=}\mathbf{a})\subseteq
\mathrm{supp}(\{\B^{(\ell)}\}\mid\A{=}\mathbf{a}_j)$ for every
$\mathbf{a}\in\mathrm{supp}(P_A)$ and every $j$: exactness under
$\mathbf{a}_j$ then covers every library realization arising under any actual
$\mathbf{a}$. Matrix-level full-support innovations
($P_{\N}(x)>0$ for all $x\in\Fq^{m\times m}$) imply hypothesis~(i), but are
strictly stronger than what the proof uses.
\end{remark}

\begin{corollary}[Coordinate-selection specialization]
\label{cor:join_selector}
If $\A\in\mathcal{A}_k$ is a coordinate selector with pattern support $\Sigma$,
then $\mathcal{V}_m=\mathrm{span}\{e_i:i\in U(\Sigma)\}$ with
$U(\Sigma)=\bigcup_{S\in\Sigma}S$, and $G_m\N^{(\theta)}$ is the union block
$\N^{(\theta)}_{U(\Sigma)}$, so~\eqref{eq:rankk_impossible} reads
$D^*_{\mathrm{priv},k}\geq H(\N^{(\theta)}_{U(\Sigma)})/\CPIR$. For binary
$\mathrm{Bern}(p)$ innovations with $|U(\Sigma)|=r$ this is $rm\,\Hb(p)/\CPIR$,
versus the $km\,\Hb(p)/\CPIR$ of the published-structure rate
(Theorem~\ref{thm:rankk_public}).
\end{corollary}

\begin{corollary}[Full-join specialization]
\label{cor:join_full}
If $\mathcal{V}_m=\Fq^m$ then $G_m$ is invertible and
$H(G_m\N^{(\theta)})=H(\Ntheta)$, so
$D^*_{\mathrm{priv},k}\geq H(\Ntheta)/\CPIR$: the full-file rate, with no
functional gain.
\end{corollary}

\begin{corollary}[$\A$-oblivious queries]
\label{cor:a_oblivious}
Suppose the queries are generated as $Q_{1:N}^{(\theta)}=f(\theta,\mathcal{R},
\mathsf{P})$ with a public parameter $\mathsf{P}\indep(\A,\{\B^{(\ell)}\},\theta)$,
and the answers are deterministic functions of the queries and the stored
library. Then, conditioned on $(\{\B^{(\ell)}\},\theta)$, the transcript is a
function of $(\mathcal{R},\mathsf{P})$ alone, which is independent of~$\A$;
hence~\eqref{eq:full_transcript_security} holds automatically and
hypothesis~(iv) need not be assumed separately. This covers the concealed-structure
regime with $\A$-independent public parameters and the full-rank
retrieve-then-compute construction, but \emph{not} the published-structure
scheme of Theorem~\ref{thm:rankk_public}, whose public parameter $R_\A$ is a
function of~$\A$. The corollary identifies a subclass for which the strongest
hypothesis is automatic; it does not strengthen
Theorem~\ref{thm:rankk_impossible} over all admissible schemes.
\end{corollary}

\begin{remark}[Dimension does not determine the bound]
\label{rem:join_not_dimension}
For non-coordinate joins, $H(G_m\N^{(\theta)})\neq r_m m\,\Hb(p)$ in general:
the equality holds for coordinate selectors under binary $\mathrm{Bern}(p)$
innovations, but dense linear combinations drive the projected law toward
uniformity and raise the entropy. At $m=4$, $q=2$, $p=0.15$, concealed row
spaces $\mathrm{span}\{e_1{+}e_2,e_3\}$ and $\mathrm{span}\{e_1{+}e_2,e_4\}$
give a join of dimension $r_m=3$ with
$H(G_m\N^{(\theta)})=8.155$ bits, whereas a \emph{coordinate} join of the same
dimension gives $r_m m\,\Hb(0.15)=7.318$ bits. The
theorem therefore retains the exact expression $H(G_m\N^{(\theta)})$ rather
than a dimension count.
\end{remark}
\begin{proof}
Write $V=(Q_{1:N}^{(\theta)},\Yans_{1:N}^{(\theta)})$ for the full transcript.

\emph{Step 1 (User randomness is redundant for decoding).}
Since $Q_{1:N}=f(\theta,\A,\mathcal{R})$ and
$\{\N^{(\ell)}\}\indep(\theta,\A,\mathcal{R})$, the conditional law factorizes,
$P(\{\mathbf{n}\},r\mid\theta,\mathbf{a},Q_{1:N}=q)
 =P(\{\mathbf{n}\})\,P(r\mid\theta,\mathbf{a},q)$,
so $\{\N^{(\ell)}\}\indep\mathcal{R}\mid(\theta,\A,Q_{1:N})$. The answers
$\Yans_{1:N}=g(Q_{1:N},\{\mathbf{a}+\N^{(\ell)}\})$ are a deterministic function
of $\{\N^{(\ell)}\}$ and the conditioning variables; for
$X\indep\mathcal{R}\mid Z$ and $Y=\psi(X,Z)$ one has
$I(X;\mathcal{R}\mid Y,Z)
 =H(\mathcal{R}\mid Y,Z)-H(\mathcal{R}\mid X,Z)
 \leq H(\mathcal{R}\mid Z)-H(\mathcal{R}\mid X,Z)=0$,
so conditioning additionally on the answers preserves the independence. By exact
correctness, recovering $\Ctheta$ given $\A=\mathbf{a}$ recovers the canonical
projection: writing $\mathbf{a}^\top=M_{\mathbf{a}}R_{\mathbf{a}}$ with
$M_{\mathbf{a}}\in\Fq^{m\times k}$ of full column rank and left inverse
$L_{\mathbf{a}}$ ($L_{\mathbf{a}}M_{\mathbf{a}}=I_k$), we have
$R_{\mathbf{a}}\btheta=L_{\mathbf{a}}\,\mathbf{a}^\top\btheta$. The decoder is
a function of $(\theta,\mathbf{a},\mathcal{R},\Yans_{1:N})$, hence
$H(R_{\mathbf{a}}\btheta\mid V,\theta,\mathcal{R},\A=\mathbf{a})=0$,
and dropping $\mathcal{R}$ by the independence just shown,
\begin{equation*}
  H\!\left(R_{\mathbf{a}}\btheta\,\middle|\,
  V,\theta,\A=\mathbf{a}\right)=0
  \qquad\text{for $P_A$-a.e.\ }\mathbf{a}.
\end{equation*}

\emph{Step 2 (Common answer channel).}
Hypothesis~\eqref{eq:full_transcript_security} states
$\A\indep V\mid(\{\B^{(\ell)}\},\theta)$, and (M4) gives
$\theta\indep(\A,\{\B^{(\ell)}\})$; together the joint law factorizes as
\begin{equation*}
  P(\mathbf{a},\{\mathbf{b}\},t,v)
  =P(\mathbf{a})\,P(\{\mathbf{b}\}\mid\mathbf{a})\,P(t)\,
   P(v\mid\{\mathbf{b}\},t),
\end{equation*}
i.e.\ given $\theta=t$ the chain
$\A\to\{\B^{(\ell)}\}\to V$ is Markov and the transcript channel
$P(v\mid\{\mathbf{b}\},t)$ is \emph{common to all realizations of}~$\A$. In
particular, since $\mathrm{row}(\A^\top)$ is a function of $\A$, data processing
gives $I(\mathrm{row}(\A^\top);V\mid\{\B^{(\ell)}\},\theta)=0$: conditioned on
the stored library, the transcript law does not depend on the concealed row
space.

\emph{Step 3 (Simultaneous decodability across row spaces).}
Choose finitely many $\mathbf{a}_1,\dots,\mathbf{a}_J\in\mathrm{supp}(P_A)$ with
$\sum_{j=1}^{J}\mathrm{row}(\mathbf{a}_j^\top)=\mathcal{V}_m$; such a collection
exists with $J\leq r_m$ since each realization contributes a $k$-dimensional
subspace and $\mathcal{V}_m$ is finite-dimensional. Fix $t$ and $j$. By Step~1
there is a measurable map $\Psi_j(\cdot,t)$ with
$\Pr[\Psi_j(V,t)=R_{\mathbf{a}_j}\btheta\mid\A=\mathbf{a}_j,\theta=t]=1$.
Expanding over the common channel of Step~2,
\begin{multline*}
  1=\sum_{\{\mathbf{b}\}}P(\{\mathbf{b}\}\mid\mathbf{a}_j)\\
  \times\Pr_{v\sim P(\cdot\mid\{\mathbf{b}\},t)}
  \!\left[\Psi_j(v,t)=R_{\mathbf{a}_j}\mathbf{b}^{(t)}\right].
\end{multline*}
Every $\{\mathbf{b}\}\in\mathcal{B}_m$ has $P(\{\mathbf{b}\}\mid\mathbf{a}_j)>0$
by hypothesis~(i), and the alphabet is finite, so the inner probability equals
$1$ for every $\{\mathbf{b}\}\in\mathcal{B}_m$: common conditional support makes
every realizable library positively weighted under $\mathbf{a}_j$, and exact
correctness forces the inner conditional success probability to equal one for
each of them. Under vanishing error neither step holds, and
Corollary~\ref{cor:rankk_vanishing} assumes~\eqref{eq:SD} directly. For any
other realization $\A=\mathbf{a}$ the transcript is generated by the \emph{same}
channel from a library in the same support $\mathcal{B}_m$, so
\begin{equation*}
  \Pr[\Psi_j(V,t)=R_{\mathbf{a}_j}\btheta\mid\A=\mathbf{a},\theta=t]
  =\sum_{\{\mathbf{b}\}}P(\{\mathbf{b}\}\mid\mathbf{a})\cdot 1=1 .
\end{equation*}
Hence, $t$ being arbitrary,
$H(R_{\mathbf{a}_j}\btheta\mid V,\theta,\A{=}\mathbf{a})=0$ for every $j$
simultaneously. Stacking $R_{\mathbf{a}_1},\dots,R_{\mathbf{a}_J}$, whose rows
span $\mathcal{V}_m$, there is a deterministic linear map carrying the stacked
outputs to $G_m\btheta$; subtracting the known $G_m\mathbf{a}$,
\begin{equation}
  H\!\left(G_m\N^{(\theta)}\,\middle|\,V,\theta,\A=\mathbf{a}\right)=0
  \qquad\text{for $P_A$-a.e.\ }\mathbf{a}:
  \label{eq:simultaneous_decodability}
\end{equation}
the transcript of any admissible protocol determines the innovation projection
onto the \emph{join}, regardless of which row space the user actually holds.

\emph{Step 4 (Privacy transfer to the join projection).}
The projected library $\{G_m\B^{(\ell)}\}$ is a measurable function of the
stored library, so Proposition~\ref{thm:transfer} gives
$I(\theta;V_{\mathcal{T}}\mid\{G_m\B^{(\ell)}\},\A{=}\mathbf{a})=0$ for all
$|\mathcal{T}|\leq P_c$: side-information-robust index privacy conditioned on
the full library implies $P_c$-privacy conditioned on the projected
library---equivalently on the messages $\{G_m\N^{(\ell)}\}$, since
$G_m\B^{(\ell)}$ and $G_m\N^{(\ell)}$ determine each other given
$\A=\mathbf{a}$.

\emph{Step 5 (Induced converse and averaging).}
For $P_A$-a.e.\ $\mathbf{a}$, the answers are deterministic in the stored raw
files $\B^{(\ell)}$, a strict refinement of the projected messages
$G_m\N^{(\ell)}=\phi_{\mathbf{a}}(\B^{(\ell)})$ with
$\phi_{\mathbf{a}}(x)=G_m(x-\mathbf{a})$ (non-invertible for $r_m<m$), so
Lemma~\ref{lem:kadhe} does not apply verbatim; we invoke
Theorem~\ref{thm:refined_converse} with $X_\ell=\B^{(\ell)}$ and
$\phi=\phi_{\mathbf{a}}$. The messages $\{G_m\N^{(\ell)}\}$ are independent
across~$\ell$, independent of $(\theta,\mathcal{R})$ jointly (by (M2) and
(M4)--(M5)), and identically distributed, and the demand is uniform on $[L]$
and independent of $\A$ by (M4), so hypothesis~(i) holds in full, including its
full-support demand condition; exact
correctness~\eqref{eq:simultaneous_decodability} gives hypothesis~(v) with
$\varepsilon=0$ uniformly over $(\theta,J)$ (extra conditioning on $W_J$ only
decreases entropy); and hypothesis~(vi) is precisely the $P_c$-privacy of
Step~4. Theorem~\ref{thm:refined_converse} gives
$D_{\mathbf{a}}\geq H(G_m\N^{(\theta)})/\CPIR$, and the averaging transfer
(Lemma~\ref{lem:averaging}) yields~\eqref{eq:rankk_impossible};
$\mathcal{V}_m=\Fq^m$ yields the full-file rate.
\end{proof}

\begin{lemma}[Simultaneous decodability under exactness]
\label{lem:simdec}
Under hypotheses (i)--(iv) of Theorem~\ref{thm:rankk_impossible} and
(M2), (M4)--(M5), every admissible exact-correctness protocol satisfies
\begin{equation}
  H\!\left(G_m\N^{(\theta)}\,\middle|\,
  Q_{1:N}^{(\theta)},\Yans_{1:N}^{(\theta)},\theta,\A\right)=0:
  \label{eq:simdec}
\end{equation}
the transcript determines the innovation projection onto the \emph{join}
$\mathcal{V}_m$ of all row spaces in the support, not merely onto the realized
one.
\end{lemma}
\begin{proof}
Steps~1--3 of the proof of Theorem~\ref{thm:rankk_impossible} above, whose only
exactness-critical ingredient is the average-to-worst-case sweep of Step~3.
\end{proof}

\noindent Condition~\eqref{eq:SD} of the corollary below is precisely the
$o(r_m m)$-relaxation of~\eqref{eq:simdec}; deriving it from the security
definitions alone under vanishing error is the open problem of
Remark~\ref{rem:rankk_impossible_scope}(b).

\begin{corollary}[Vanishing-error extension under simultaneous decodability]
\label{cor:rankk_vanishing}
Let $r_m=\dim\mathcal{V}_m$. Replace exact correctness (hypothesis~(ii) of
Theorem~\ref{thm:rankk_impossible}) by vanishing-error
correctness~\eqref{eq:correctness}, keeping hypotheses (i), (iii), (iv). If in
addition the scheme satisfies the \emph{simultaneous-decodability condition}
\begin{equation}
  H\!\left(G_m\N^{(\theta)}\,\middle|\,
  Q_{1:N}^{(\theta)},\Yans_{1:N}^{(\theta)},\theta,\A\right)\;\leq\;o(r_m m),
  \tag{SD}
  \label{eq:SD}
\end{equation}
then the bound holds with an $o(r_m m)$ penalty at the join-projection scale:
\begin{equation}
  D^*_{\mathrm{priv},k}\;\geq\;
  \frac{H\!\left(G_m\N^{(\theta)}\right)}{\CPIR(N,P_c,L)}-o(r_m m).
  \label{eq:rankk_impossible_vanishing}
\end{equation}
When the simultaneous-decodability conditional entropy is exactly zero, the
same argument recovers the exact bound~\eqref{eq:rankk_impossible} without an
asymptotic penalty. Here
$D^*_{\mathrm{priv},k}$ denotes the infimum over the class
of~\eqref{eq:dpriv_class} with exact correctness replaced by vanishing-error
correctness together with~\eqref{eq:SD}; this class contains
$\Pi_{\mathrm{priv},k}$, so the corollary is not implied by
Theorem~\ref{thm:rankk_impossible}.
\end{corollary}
\begin{proof}
Steps~2 and~4 of the proof of Theorem~\ref{thm:rankk_impossible} use only
hypotheses (iv) and (iii) respectively and are unchanged; exactness entered
only through Steps~1 and~3, whose sole role is to \emph{derive} the zero-error
form of~\eqref{eq:SD}. Assuming~\eqref{eq:SD} directly, consider the induced
instance of Step~5---retrieving $G_m\N^{(\theta)}$ from the projected library
$\{G_m\N^{(\ell)}\}$. Hypothesis~(v) of Theorem~\ref{thm:refined_converse}
requires a correctness penalty that is \emph{uniform over the demand}, whereas
\eqref{eq:SD} controls the $\theta$-\emph{average}: writing the per-demand
penalty as
$\delta_{\mathbf{a},t}:=H(G_m\N^{(t)}\mid
Q^{[t]}_{1:N},\Yans^{[t]}_{1:N},\theta{=}t,\A{=}\mathbf{a})$
and
$\delta_{\mathbf{a}}:=H(G_m\N^{(\theta)}\mid
Q_{1:N},\Yans_{1:N},\theta,\A{=}\mathbf{a})$, uniformity of $\theta$ on $[L]$
(M4) gives $\delta_{\mathbf{a}}=\frac{1}{L}\sum_{t\in[L]}\delta_{\mathbf{a},t}$,
hence, exactly as in the $L$-factor bookkeeping of
Theorem~\ref{prop:rankdef_lb},
\begin{equation*}
  \varepsilon_{\mathbf{a}}
  \;:=\;\max_{t\in[L]}\delta_{\mathbf{a},t}
  \;\leq\; L\,\delta_{\mathbf{a}}.
\end{equation*}
Further conditioning on any processed subset $W_J$, as required by
hypothesis~(v), only decreases each term, so hypothesis~(v) holds at level
$\varepsilon_{\mathbf{a}}$ uniformly over $(t,J)$. The refined-storage converse
(Theorem~\ref{thm:refined_converse}, applied as in Step~5) then gives
$D_{\mathbf{a}}\geq\bigl(H(G_m\N^{(\theta)})-\varepsilon_{\mathbf{a}}\bigr)/\CPIR$,
and since $L$ is fixed,
$\mathbb{E}_\A[\varepsilon_\A]\leq L\,\mathbb{E}_\A[\delta_\A]\leq
L\cdot o(r_m m)=o(r_m m)$ by~\eqref{eq:SD}, with
$\varepsilon_\A/\CPIR\leq L\,\varepsilon_\A$ as in
Theorem~\ref{prop:rankdef_lb}. The averaging transfer
(Lemma~\ref{lem:averaging}) completes the proof, with slack $o(r_m m)$.
\end{proof}

\begin{remark}[Scope of Theorem~\ref{thm:rankk_impossible}, and what remains
open]
\label{rem:rankk_impossible_scope}
\emph{(a)~The $\theta$-conditioned full-transcript hypothesis.}
Hypothesis~\eqref{eq:full_transcript_security} strengthens
Definition~\ref{def:security} in \emph{two} independent directions. First, from
colluding views $|\mathcal{T}|\leq P_c$ to the whole transcript. Second, it
conditions on~$\theta$: writing $V:=(Q_{1:N},\Yans_{1:N})$, the chain rule and
(M4) (which gives $I(\A;\theta\mid\{\B^{(\ell)}\})=0$) yield
\begin{equation*}
  I(\A;V\mid\{\B^{(\ell)}\},\theta)
  = I(\A;V\mid\{\B^{(\ell)}\})
  + I(\A;\theta\mid\{\B^{(\ell)}\},V),
\end{equation*}
so \eqref{eq:full_transcript_security} is equivalent to full-transcript
Definition~\ref{def:security} \emph{together with} the extra condition
$I(\A;\theta\mid\{\B^{(\ell)}\},V)=0$: transcript and library must not leak
about $\A$ jointly with the demand. The extra condition is not implied by
Definition~\ref{def:security}---a query $\A_{1,1}\oplus\theta$ (admissible,
since queries may depend on $(\theta,\A,\mathcal{R})$) has
$I(\A;V\mid\{\B^{(\ell)}\})=0$ yet
$I(\A;\theta\mid\{\B^{(\ell)}\},V)=H(\A_{1,1}\mid\{\B^{(\ell)}\})>0$ whenever the
library does not determine $\A_{1,1}$ (attaining $1$~bit exactly when that
conditional law is uniform, e.g.\ for $\mathrm{Bern}(1/2)$ innovations). Both strengthenings are cost-free on
the achievability side: the full-rank and concealed-structure schemes have
queries depending only on $(\theta,\mathcal{R})$ and $\A$-independent public
parameters, so given $(\{\B^{(\ell)}\},\theta)$ their transcript is a function of
$\mathcal{R}$ alone, and $\mathcal{R}\indep\A$ by (M5) gives
$I(\A;V\mid\{\B^{(\ell)}\},\theta)=0$. The full-transcript hypothesis prevents
the concealed row space from being encoded solely in correlations across server
views. Under only the $|\mathcal{T}|\leq P_c$ constraints, such cross-server
encodings are not excluded, and the join-subspace conclusion is not
established. Whether such a scheme exists remains open.
\emph{(b)~Exact correctness.} Capacity-achieving PIR layers are exact, and in the
finite-length scheme of Proposition~\ref{thm:finiteblock} (Appendix~\ref{app:finiteblock}) the only error source is
the source decoder, so exactness is the natural regime for this statement.
Corollary~\ref{cor:rankk_vanishing} extends the conclusion to vanishing-error
schemes under the single displayed condition~\eqref{eq:SD}, which
Lemma~\ref{lem:simdec} establishes when correctness is exact. What remains open is deriving
\eqref{eq:SD} from the security definitions alone under vanishing error: the
per-pattern Fano bounds are averages over pattern-dependent library laws, and
stitching them requires quantitative control of the posterior ambiguity of the
pattern.
\emph{(c)~Common conditional support.} Hypothesis~\emph{(i)} yields the strong
form of ``the library does not determine the row space'': every library
realization in $\mathcal{B}_m$ is consistent with every
$\mathbf{a}\in\mathrm{supp}(P_A)$, so the posterior support of
$\mathrm{row}(\A^\top)$ given $\{\B^{(\ell)}\}$ is the whole family for every
library realization. It is implied by matrix-level full-support innovations,
and holds for the binary additive model with $p\in(0,1)$ and the
sparse additive model with $\rho\in(0,1)$; the weaker one-sided form of
Remark~\ref{rem:support_domination} suffices. Note that~\eqref{eq:rankk_impossible}
carries \emph{no} asymptotic slack: exact
correctness~\eqref{eq:simultaneous_decodability} instantiates hypothesis~(v) of
Theorem~\ref{thm:refined_converse} with $\varepsilon=0$, and that theorem's
conclusion is the exact inequality $D\geq(H-\varepsilon)/\CPIR$, so Step~5
delivers $D_{\mathbf{a}}\geq H(G_m\N^{(\theta)})/\CPIR$ pointwise and
Lemma~\ref{lem:averaging} transfers it without penalty. Under the vanishing-error
relaxation of Corollary~\ref{cor:rankk_vanishing} the bound reacquires an
$o(r_m m)$ term, at the scale of the join-projection alphabet $q^{r_m m}$.
\emph{(d)~Summary of the hypothesis hierarchy.} The concealed-structure
converse is therefore a \emph{stronger-observation} converse: it is proved
under the $\theta$-conditioned full-transcript security
condition~\eqref{eq:full_transcript_security}, exact correctness, and the
common-conditional-support hypothesis~\emph{(i)} (implied by matrix-level
full-support innovations), with the vanishing-error extension
requiring the simultaneous-decodability condition of
Corollary~\ref{cor:rankk_vanishing}. These observational hypotheses have no
counterpart in the published-structure capacity theorem
(Theorem~\ref{thm:rankk_public}), whose achievability holds under
Definitions~\ref{def:index} and~\ref{def:security_public} as stated.
\end{remark}

\begin{remark}[Interpretation]
\label{rem:dichotomy}
Theorems~\ref{thm:rankk_public} and~\ref{thm:rankk_impossible} exhibit the
rank-deficient case as a \emph{contrast} over two disjoint admissible classes
(Remark~\ref{rem:dichotomy_consistency}): \emph{published} structure admits a
capacity theorem at the functional rate, for a non-invertible
functional-computation problem not reducible to full-file retrieval, while
\emph{concealed} structure, under the strengthened observational
hypotheses of Theorem~\ref{thm:rankk_impossible}, forces download at the join-subspace rate
$H(G_m\N^{(\theta)})/\CPIR$, which degrades the $m/k$ separation to $m/r_m$ and
collapses it entirely when $\mathcal{V}_m=\Fq^m$. Under the
respective hypotheses of Theorems~\ref{thm:rankk_public}
and~\ref{thm:rankk_impossible}, the communication requirement depends on whether
the functional structure is published or concealed, not solely on the rank
of~$\A$.

The two branches use different hypotheses. Theorem~\ref{thm:rankk_public}
applies to published rank-$k$ row spaces under Definitions~\ref{def:index}
and~\ref{def:security_public}, with achievability supplied by
Lemma~\ref{lem:uniform_source} under the paper's i.i.d.\ innovation model.
Theorem~\ref{thm:rankk_impossible} applies to
concealed row spaces of arbitrary rank $k$ under exact correctness, common
conditional library support, and the stronger full-transcript security
condition; Corollary~\ref{cor:a_oblivious} shows the last is automatic for
$\A$-oblivious schemes. What holds across the whole regime, for every fixed-rank operand, is the
functional converse of Theorem~\ref{prop:rankdef_lb}. The general concealed case under $P_c$-server
observations remains open.
\end{remark}

\begin{example}[Rank-Deficient Case: A Non-Vacuous Functional Gap]
\label{ex:rankdeficient}
Let $m=2$, $q=2$, $N=8$, $P_c=2$, $L\geq 5$, with $\btheta=\A+\N$ and
$N_{i,j}\overset{\mathrm{i.i.d.}}{\sim}\mathrm{Bern}(0.15)$. Take $P_A$ supported
on coordinate-selector rank-one matrices, assigning positive probability to
both selector patterns $\mathcal{S}(\A)=\{1\}$ and $\mathcal{S}(\A)=\{2\}$; then
$\mathrm{rank}(\A)=1$ almost surely.

The full-file conditional entropy is exact:
$H(\btheta\mid\A)=H(\N)=4\Hb(0.15)\approx 2.44$ bits.
For the functional target, since $\A$ has rank~$1$,
$\Ctheta=\A^\top\btheta=\A^\top\A+\A^\top\N$ lies in the column span of
$\A^\top$, a one-dimensional subspace; the $2\times2$ matrix $\A^\top\N$ thus has
at most $km=2$ free $\mathbb{F}_2$-symbols, giving the dimension bound
$H(\Ctheta\mid\A=\mathbf{a})=H(\A^\top\N\mid\A=\mathbf{a})\leq km\log_2 q=2$ bits.

The exact value of $H(\A^\top\N\mid\A=\mathbf{a})$ depends on the structure of
$\mathbf{a}$: over $\mathbb{F}_2$ a row of $\mathbf{a}^\top$ with a single
nonzero entry \emph{selects} one innovation row, whereas a row with two nonzero
entries \emph{XORs} two independent rows, driving the output toward uniform.
For every support point of $P_A$ the selected block carries $km=2$ free
$\mathbb{F}_2$-symbols, so
$H(\Ctheta\mid\A=\mathbf{a})=2\Hb(0.15)=1.220$ bits, the full $m/k$ separation.
For comparison, enumerating all nine rank-$1$ matrices in
$\mathbb{F}_2^{2\times2}$ (with $H(\btheta\mid\A)=4\Hb(0.15)=2.44$ bits fixed)
yields two possible conditional functional entropies: $2\Hb(0.15)=1.220$ bits
for the six coordinate-selector matrices (ratio $2.000$) and $1.638$ bits for
the three XOR-type matrices (ratio $1.489$), which lie outside the support of
$P_A$ chosen here.

\begin{table}[t]
\caption{$H(\A^\top\N\mid\A{=}\mathbf{a})$ over rank-$k$
$\mathbf{a}\in\mathbb{F}_2^{3\times3}$,
$N_{i,j}\overset{\mathrm{i.i.d.}}{\sim}\mathrm{Bern}(p)$, $H(\N)=9\Hb(p)$.
Ratio $=H(\N)/H(\A^\top\N\mid\A)$. For the enumerated dimensions and
$p\in\{0.15,0.30\}$, the maximal $m/k$ separation is attained only by selector
or selector-equivalent structures (reproducibility script).}
\label{tab:rankdef_enum3}
\centering
\renewcommand{\arraystretch}{1.2}
\begin{tabular}{cclcc}
\toprule
$p$ & $k$ & type & $H(\A^\top\N\mid\A)$ & ratio \\
\midrule
$0.15$ & $1$ & selector   & $1.830$ ($=3\Hb(p)$) & $3.000$ \\
$0.15$ & $1$ & dense/XOR  & $2.457$--$2.740$      & $2.00$--$2.23$ \\
$0.15$ & $2$ & selector-equiv. & $3.659$ ($=6\Hb(p)$) & $1.500$ \\
$0.15$ & $2$ & dense/XOR  & $4.287$--$4.698$      & $1.17$--$1.28$ \\
\midrule
$0.30$ & $1$ & selector   & $2.644$ ($=3\Hb(p)$) & $3.000$ \\
$0.30$ & $1$ & dense/XOR  & $2.944$--$2.991$      & $2.65$--$2.69$ \\
$0.30$ & $2$ & selector-equiv. & $5.288$ ($=6\Hb(p)$) & $1.500$ \\
$0.30$ & $2$ & dense/XOR  & $5.588$--$5.848$      & $1.36$--$1.42$ \\
\bottomrule
\end{tabular}
\end{table}

To confirm this is not an artifact of the $m=2$ toy case, Table~\ref{tab:rankdef_enum3}
reports the analogous enumeration for $m=3$, $k\in\{1,2\}$, grouping the rank-$k$
matrices by structure type. The pattern is robust across the tabulated cases:
the maximal $m/k$ separation
(ratio $H(\N)/H(\A^\top\N\mid\A)=m/k$, i.e.\ $3.0$ for $k{=}1$ and $1.5$ for $k{=}2$)
is attained \emph{in these enumerations} only by coordinate-selection-type
structure, for which
$H(\A^\top\N\mid\A{=}\mathbf{a})=km\,\Hb(p)$; dense/XOR-type maps have strictly larger
functional entropy (ratio $<m/k$), because XOR-ing $\mathrm{Bern}(p)$ innovation rows
drives the output toward uniformity. (For $k{=}2$ some non-selector row spaces are
selector-\emph{equivalent} and also attain $km\,\Hb(p)$; the strict increase is for
genuinely dense maps.) This is an enumerated observation at
$p\in\{0.15,0.30\}$, not a characterization: at $p=\tfrac12$ the separation
degenerates and every rank-$k$ map attains the same value
(Remark~\ref{rem:rankdef_interp}).

Consequently the functional lower bound
$D^*\geq H(\Ctheta\mid\A)/\CPIR(N,P_c,L)-o(km)$ (Theorem~\ref{prop:rankdef_lb})
and the full-file upper bound
$D^*\leq H(\btheta\mid\A)/\CPIR(N,P_c,L)+o(m^2)$ can differ by as much as the factor
$m/k=2$, \emph{attained in the coordinate-selector case}. At $L=5$ the exact
inverse capacity is
\begin{equation*}
  \frac{1}{\CPIR(8,2,5)}=\sum_{j=0}^{4}\left(\frac{P_c}{N}\right)^{\!j}
  =1+\tfrac14+\tfrac1{16}+\tfrac1{64}+\tfrac1{256}
  =\tfrac{341}{256}\approx 1.3320,
\end{equation*}
so with $H(\Ctheta\mid\A)=2\Hb(0.15)$ the bounds are, to leading order,
$D^*\gtrsim 1.3320\cdot 1.220\approx 1.62$ bits and
$D^*\lesssim 1.3320\cdot 2.439\approx 3.25$ bits.
(The large-library limit replaces $1/\CPIR$ by $N/(N-P_c)=4/3$, giving
$1.63$ and $3.25$ bits; the ratio $m/k=2$ is unaffected, since the same factor
multiplies both bounds.) The coordinate-selector matrices
here are exactly the class $\mathcal{A}_k$ of
Corollary~\ref{thm:rankk_selection}; the lower bound $1.62$ bits---like the
functional lower bound for the XOR-type $\mathbf{a}$, where the separation is
smaller---holds for every rank-$k$ operand (Theorem~\ref{prop:rankdef_lb}). The example therefore exhibits a
\emph{potential factor-$m/k$ separation}---attained under structured
(coordinate-selecting) rank-$k$ instances, not for every operand---and shows the
rank-deficient gap is not vacuous. Within the two protocol classes analyzed
here, Theorem~\ref{thm:rankk_public} attains the functional rate when the
pattern is published, whereas Theorem~\ref{thm:rankk_impossible} forces the
full-join rate for concealed patterns under its strengthened observational
assumptions; the general concealed case remains open. Under the
exactness, common-support, and full-transcript-security assumptions of
Theorem~\ref{thm:rankk_impossible}, concealment forces the full-join rate: the
chosen support has patterns $\{1\}$ and $\{2\}$, so the join is
$U(\Sigma)=\{1,2\}=[m]$ and the bound is the full-file rate.
\end{example}

\begin{example}[An asymptotic separating family]
\label{ex:asymptotic_family}
The enumeration above is at $m=3$; the same separation persists at first order
along a growing family. Fix $0<\alpha\leq\beta\leq 1$ with $\alpha<1$, let
$k=\alpha m$, and let $U\subseteq[m]$ be a fixed coordinate pool of size
$r_m=\beta m$. Take $\mathrm{supp}(P_\A)$ to consist of coordinate selectors
whose supports are $k$-subsets of $U$ whose union is all of $U$ (for instance
the $\beta/\alpha$ disjoint blocks of size $k$ partitioning $U$, when
$\beta/\alpha$ is an integer). Every realization then has rank exactly $k$, and
the join of the row spaces is $U(\Sigma)=U$, of dimension $r_m$. For a selector,
$\mathbf{a}^\top\N^{(\ell)}$ consists of $km$ i.i.d.\ $\mathrm{Bern}(p)$
entries, so $H(\mathbf{a}^\top\N^{(\ell)})=km\,\Hb(p)$, and likewise
$H(G_m\N^{(\ell)})=r_m m\,\Hb(p)$ for a basis $G_m$ of the join. Hence
Theorem~\ref{thm:rankk_public} gives the published rate
\begin{equation*}
  D^*_{\mathrm{pub}}
  \;=\;\frac{\alpha\, m^2\,\Hb(p)}{\CPIR(N,P_c,L)}+o(m^2),
\end{equation*}
while under the hypotheses of Theorem~\ref{thm:rankk_impossible} concealment
forces the join-scale bound
\begin{equation*}
  D
  \;\geq\;\frac{\beta\, m^2\,\Hb(p)}{\CPIR(N,P_c,L)}-o(m^2).
\end{equation*}
The two differ by the factor $\beta/\alpha$, which is independent of $m$ and
of $p$: a first-order separation, not an artifact of small blocklength. Taking
$\beta=1$ (supports whose union is all of $[m]$) gives the full-file rate
$m^2\Hb(p)/\CPIR$ and a separation factor $1/\alpha=m/k$, matching
Example~\ref{ex:rankdeficient}; taking $\beta=\alpha$ (a single fixed support,
publicly determined by the join) collapses the two rates, consistently with the
functional rate being recovered when the join is $k$-dimensional.
\end{example}

\begin{remark}[Interpretation of Example~\ref{ex:rankdeficient}]
\label{rem:rankdef_interp}
The example shows only that rank deficiency can create a non-vacuous gap; it
does not assert a universal $m/k$ ratio. Two distinct quantities should be kept
apart. For coordinate-selection maps under binary $\mathrm{Bern}(p)$
innovations,
\begin{equation*}
  H(\mathbf{a}^\top\N\mid\A{=}\mathbf{a})=km\,\Hb(p),
\end{equation*}
which is the value used in Example~\ref{ex:rankdeficient}. For a general
rank-$k$ map,
\begin{equation*}
  H(\mathbf{a}^\top\N\mid\A{=}\mathbf{a})\leq km\log_2 q,
\end{equation*}
and its value depends on the algebraic structure of the map, so the ratio must
be computed from the law of~$\A$. Coordinate selection is not the only
structure attaining $km\,\Hb(p)$: at $p=\tfrac12$ every rank-$k$ map
over $\mathbb{F}_2$ yields uniform output, so all of them attain
$km\,\Hb(1/2)=km\log_2 q$ simultaneously.
\end{remark}

\begin{remark}[Upper bound is achievable]
\label{rem:ub_achievable}
The bound $D^*\leq H(\btheta|\A)/\CPIR$ is achievable by running SW
compression on $\btheta$ and retrieving the sketch via any PIR scheme.
Entropy flatness ($H(\Ntheta\mid\A=\mathbf{a})$ constant in $\mathbf{a}$) is the
sufficient condition used by our single-rate construction. Without it, the
average conditional entropy $\HcondBA$ need not be achievable by that
construction, and a more general source-coding analysis would be required.
Whether $H(\Ctheta|\A)/\CPIR$ is achievable---which would require private
functional compression of $\Ctheta$ without revealing $\btheta$ or
$\theta$---remains open.
\end{remark}

\section{Correlation Models and Finite-Length Accounting}
\label{sec:examples}

Within the model class of Corollary~\ref{cor:capacity}, download depends on the
correlation structure only through the scalar $\HcondBA$; for entropy-flat
sources (Remark~\ref{rem:ub_achievable}) the algebraic form of the noise model drops
out, and only the residual uncertainty of $\btheta$ given $\A$ matters. Three
canonical families illustrate the range: the \emph{additive} model
$\btheta=\A+\N$ with i.i.d.\ entries gives $\HcondBA=H(\N)=m^2H(N_{1,1})$, hence
$D^*/m^2=\Hb(p)/\CPIR(N,P_c,L)+o(1)$ at fixed $L$ for binary
$\mathrm{Bern}(p)$ innovations, becoming $(N/(N-P_c))\Hb(p)$ in the
large-library limit ($m\to\infty$ at each fixed $L$, then $L\to\infty$;
Remark~\ref{rem:iterated_limit}); the
\emph{erasure/refresh} model (each entry independently replaced by a uniform
$\Fq$-symbol with probability $\varepsilon$) gives, per entry,
\begin{equation*}
  H(B_{i,j}\mid A_{i,j})
  =-\pi_0\log\pi_0-(q-1)\tfrac{\varepsilon}{q}\log\tfrac{\varepsilon}{q},
  \qquad \pi_0=1-\varepsilon+\tfrac{\varepsilon}{q},
\end{equation*}
so $H(\B\mid\A)=m^2\bigl[-\pi_0\log\pi_0-(q-1)\tfrac{\varepsilon}{q}\log\tfrac{\varepsilon}{q}\bigr]$;
here $\pi_0$ collects the two ways an entry can remain unchanged (no refresh, or
a refresh that happens to redraw the same symbol). The simpler quantity
$\varepsilon\log q$ equals $H(B_{i,j}\mid A_{i,j},E_{i,j})$, the conditional
entropy when the refresh mask $E$ is also revealed to the decoder; without the
mask the conditional entropy increases by $I(E_{i,j};B_{i,j}\mid A_{i,j})$
(e.g.\ $0.169$ versus $0.050$ bits per entry at $q=2$, $\varepsilon=0.05$, a
gap of $0.119$ bits). If the refresh mask is revealed, the relevant conditional
entropy becomes $H(\B\mid\A,E)$; extending the present operational reduction to
that decoder-side-information model requires an appropriately conditioned PIR
converse, since the innovations would no longer be independent of the decoder's
side information as hypothesis~(iii) of Lemma~\ref{lem:kadhe} requires, and is
not claimed here. The entropy formula in this paragraph concerns the
unrevealed-mask variant.
And the \emph{sparse additive} model
($N_{i,j}=Z_{i,j}U_{i,j}$, where $Z_{i,j}\sim\mathrm{Bern}(\rho)$ and
$U_{i,j}$ is independent of $Z_{i,j}$ and uniform over
$\Fq\setminus\{0\}$) gives
$H(\B\mid\A)=m^2[\Hb(\rho)+\rho\log(q-1)]$, much smaller than $m^2\log q$ for
small~$\rho$; the formula depends on this choice of $U_{i,j}$---taking
$U_{i,j}$ uniform over all of $\Fq$ instead merges a mass $\rho/q$ back onto
the zero symbol and yields a strictly smaller entropy. In the large-library limit, the normalized download
$(N/(N-P_c))\Hb(p)$ measures the saving relative to the $m^2\log_2 q$
independence baseline: for $q=2$ it equals $0.33$ at
$(p,N,P_c)=(0.05,8,1)$ (a $67\%$ saving) and $0.62$ at $(0.11,10,2)$ (a
$38\%$ saving), the gain vanishing as $p\to\tfrac12$ or $P_c\to N$; at fixed
$L$ the corresponding normalized download is $\Hb(p)/\CPIR(N,P_c,L)$.

\subsection{Finite-Length Download Accounting}
\label{subsec:e2e}
Proposition~\ref{thm:finiteblock} (Appendix~\ref{app:finiteblock}) composes a
finite-length source code with an exact colluding-PIR layer: under its
hypotheses (P1)--(P3), a source code compressing $\Ntheta$ to length
$r_n$ padded to a multiple of the base-field padding unit $s_{\mathrm{base}}$
gives fixed-length downloaded communication of exactly
\begin{equation}
  \Lambda_n\;=\;\tilde{r}_n/\CPIR,\qquad
  \tilde{r}_n=\min\bigl\{s_{\mathrm{base}}\lceil r_n/s_{\mathrm{base}}\rceil,\;
  s_{\mathrm{base}}\lceil n/s_{\mathrm{base}}\rceil\bigr\},
  \label{eq:e2e_download}
\end{equation}
with $r_n$ the lossless-source normal-approximation length. We size the code by
this design rule; second-order attainability by the chosen linear syndrome
ensemble is \emph{not} established here (Appendix~\ref{app:finiteblock}).
Table~\ref{tab:e2e} gives a representative accounting at three $m=128$ operating points
via~\eqref{eq:e2e_download} at $(N,P_c,L)=(4,1,4)$ with a binary source and
an $\mathbb{F}_4$ PIR layer grouping two bits per extension-field symbol (so
$a=2$ and $s_{\mathrm{base}}=a\,s^*=128$ bits), $\CPIR=64/85$; as $m$ grows the normalized download converges to the
dispersion prediction and, once the syndrome branch is active
($r_n=n\Hb(p)+O(\sqrt{n})$ with $\Hb(p)<1$), the padding overhead vanishes. The full accounting
across all twelve combinations of $p\in\{0.05,0.15,0.30\}$ and
$m\in\{16,32,64,128\}$ (with the selected branch---syndrome, raw, or
tie---identified), the parameter instantiation and arithmetic accounting under
(P1)--(P3), and the padding calculations are in the supplementary material,
together with scripts reproducing every tabulated value. No empirical
decoding-error claim is made: $\epsilon_s$ is the target error entering the
design length, and exhaustive maximum-likelihood decoding is infeasible at the
tabulated blocklengths.

\begin{table}[!ht]
\caption{Representative design-level finite-length download accounting
(lossless-source normal approximation, Zhang--Xu colluding
scheme~\cite{zhang2019optimal,xu2018building} with an $\mathbb{F}_4$ PIR layer over the binary source ($a=2$), $(N,P_c,L)=(4,1,4)$, $\CPIR=64/85$, $s_{\mathrm{base}}=128$ bits,
$\epsilon_s=10^{-3}$). $\Lambda_n=\tilde{r}_n/\CPIR$ is the fixed-length downloaded communication, an upper bound on the entropy download. Entries are \emph{design} values from the
normal-approximation length, not measured decoding errors. The full table is in
the supplement.}
\label{tab:e2e}
\centering
\renewcommand{\arraystretch}{1.2}
\setlength{\tabcolsep}{4pt}
\begin{tabular}{ccccccc}
\toprule
$p$ & $m$ & $r_n$ & $\tilde{r}_n$ & $\Lambda_n/m^2$ & disp. & asym. \\
\midrule
$0.05$ & $128$ & $5066$  & $5120$  & $0.415$ & $0.410$ & $0.380$ \\
$0.15$ & $128$ & $10353$ & $10368$ & $0.840$ & $0.839$ & $0.810$ \\
$0.30$ & $128$ & $14668$ & $14720$ & $1.193$ & $1.188$ & $1.170$ \\
\bottomrule
\end{tabular}
\end{table}

\section{Discussion}
\label{sec:discussion}

\subsection{On the Relationship to PIR}
\label{subsec:nontrivial}

The full-rank characterization $D^*=H(\Ntheta)/\CPIR+o(m^2)$ is the natural
formula for private retrieval of the innovation; what the full-rank part
establishes is where the formula comes from. First, the problem is not
well-posed under the classical definitions: the library-inclusive unconditional
security condition is infeasible under correlation
(Proposition~\ref{prop:infeasible}), and unconditional index privacy need not
survive conditioning on $\A=\mathbf{a}$ (Remark~\ref{rem:index_equiv}), so
Definitions~\ref{def:index} and~\ref{def:security} must be in place before any
per-realization converse can be run. Second, the retrieve-then-compute
architecture is a consequence, not a modeling choice:
Theorem~\ref{thm:equivalence} shows every admissible protocol induces, after
conditioning on $\A=\mathbf{a}$, a colluding PIR instance over the innovation
library with matched error, privacy, and averaged download---whereas
in PIR with side information~\cite{kadhe2017private,chen2018private} the
retrieval structure is a modeling assumption. The rank-deficient regime, where
the reduction fails, is where the paper's main results lie
(Section~\ref{subsec:extension}).

\subsection{Extensions and Open Problems}
\label{subsec:open}

The full-rank and rank-deficient regimes together suggest a change in the
communication mechanism, from retrieval-dominated to
functional-compression-dominated behavior,
which we leave for future work.
Further open directions include: (i)~MDS-coded storage, which may reduce
storage while changing the PIR capacity factor; (ii)~partial server-side
correlation, where servers have a degraded version of the correlation;
(iii)~multi-round or adaptive queries; and (iv)~finite-blocklength (finite
$m$) analysis providing practically relevant bounds.

\begin{remark}[Scope of Corollary~\ref{cor:capacity}]
\label{rem:scope_limit}
The converse $D\geq\HcondBA/\CPIR$ holds without entropy-flatness, as
argued in Remark~\ref{rem:singleton}.
The achievability additionally requires a SW code at rate $\HcondBA+\varepsilon$,
guaranteed by the i.i.d.\ assumption or, more broadly, by the AEP for
stationary ergodic sources.
Relaxing replicated storage, full rank of~$\A$, the additive correlation
structure, or innovation independence each requires new techniques and is left
open.
\end{remark}


\section{Conclusion}
\label{sec:conclusion}

The paper's central contributions are converse-side. For rank-deficient
operands, the functional lower bound
$D^*\geq H(\A^\top\btheta\mid\A)/\CPIR-o(km)$ is valid for every fixed-rank
operand under the proposed correlated-library model
(Theorem~\ref{prop:rankdef_lb}). It rests on two tools of independent interest.
The first characterizes when index privacy survives coarsening of the
conditioning (Proposition~\ref{thm:transfer}); it is universal under the
model's joint privacy form, admits an explicit failure witness outside that form,
and is used by every converse of the rank-deficient section. The second is
a colluding PIR converse for per-file function retrieval with refined storage
(Theorem~\ref{thm:refined_converse}), a fixed-function, private-index
counterpart of private computation~\cite{sun2019capacity,obead2022private}.
Under its per-file refined-storage structure the colluding
converse holds even when every server stores the full refined library, whose
residual content need not be recoverable from the functional messages. On top of these, the regime exhibits a published/concealed contrast:
published functional structure admits a capacity theorem at the functional
rate (Theorem~\ref{thm:rankk_public}), while for concealed structure a
strengthened-observation converse forces download at the row-space-join scale
(Theorem~\ref{thm:rankk_impossible}, relaxed to vanishing error in
Corollary~\ref{cor:rankk_vanishing}). One branch is a capacity theorem, the
other a conditional converse under stronger observational hypotheses, stated
with those results.

In the full-rank base case, the problem reduces---after repairing the security
definitions, whose library-inclusive unconditional form is infeasible under
correlation (Proposition~\ref{prop:infeasible})---to colluding PIR over the
$\A$-independent innovation library, yielding the tight characterization
$D^*=\HcondBA/\CPIR+o(m^2)$ with a matching linear Slepian--Wolf plus PIR
scheme. The converses require no entrywise memorylessness
(Remark~\ref{rem:beyond_memoryless}); memorylessness enters only through the
achievability's source code.

What remains open: deriving condition~\eqref{eq:SD} from the security
definitions alone under vanishing error; the $P_c$-view-only version of
Theorem~\ref{thm:rankk_impossible}; tightness of the functional rate for
concealed non-selector structure; explicit source codes for structured
innovations; cross-file innovation dependence beyond the common state~$\A$
(PIR over correlated messages); and coded storage. We believe the transfer characterization and
the refined-storage converse will find use beyond this model, wherever privacy
constraints must survive a change of the conditioning library.

\appendices

\section{Proof of Theorem~\ref{thm:converse}: Full-Rank Converse}
\label{app:converse}
\begin{proof}[Proof of Theorem~\ref{thm:converse}]
\textbf{Step~0 (Reduction to innovation retrieval).}
By Lemma~\ref{lem:innovation}, under Assumption~\ref{ass:fullrank} recovering
$\Ctheta$ with vanishing error is equivalent to recovering the innovation
$\Ntheta=\btheta-\A$ with vanishing error; the user's local post-processing
$\Ctheta=\A^\top\A+\A^\top\Ntheta$ is free, so $D$ is unchanged. It therefore
suffices to lower-bound the download of any scheme that privately retrieves
$\Ntheta$ from the replicated \emph{innovation library}
$\{\N^{(1)},\ldots,\N^{(L)}\}$, where the decoder side information $\A=\mathbf{a}$
appears only as a fixed parameter of the induced query and answer maps. By~(M2),
$\Nell\indep\A$ for every~$\ell$, so the applicable bound is the standard
$P_c$-colluding PIR converse, whereas the raw files $\B^{(\ell)}=\A+\Nell$ are
all $\A$-correlated and violate the message-independence hypothesis
(Remark~\ref{rem:standard_converse}).

Fix any full-rank realization $\mathbf{a}$ of $\A$.
Define the \emph{induced PIR instance at $\mathbf{a}$}, denoted~$\pi_{\mathbf{a}}$:
the protocol $\pi$ run with $\A$ fixed to $\mathbf{a}$, retrieving $\Ntheta$ from
$\{\Nell\}$, producing queries
$Q_i^{(\theta)}=f_i(\theta,\mathbf{a},\mathcal{R})$ and answers
$\Yans_i^{(\theta)}=g_i(Q_i^{(\theta)},\{\B^{(\ell)}\})$.
The original answer function $g_i$ acts on the stored files $\B^{(\ell)}$, not on
the innovations; to make the induced instance act on $\{\Nell\}$ with
$\mathbf{a}$ a fixed parameter, define the \emph{induced answer function}
\begin{equation}
  g_{i,\mathbf{a}}\bigl(q,\N^{(1)},\ldots,\N^{(L)}\bigr)
  \;\triangleq\;
  g_i\bigl(q,\,\mathbf{a}+\N^{(1)},\ldots,\mathbf{a}+\N^{(L)}\bigr),
  \label{eq:induced_answer}
\end{equation}
so that $\Yans_i^{(\theta)}=g_{i,\mathbf{a}}(Q_i^{(\theta)},\{\Nell\})$ exactly
reproduces the original answers when $\A=\mathbf{a}$. This makes $\pi_{\mathbf{a}}$
a bona fide colluding PIR instance over the innovation library $\{\Nell\}$ with
$\mathbf{a}$ a fixed parameter of the answer maps, removing any objection that the
servers act on $\B$ rather than $\N$.
Its download is $D_{\mathbf{a}}\triangleq\sum_{i=1}^N H(\Yans_i^{(\theta)}\mid\A=\mathbf{a})$,
with entropy over the remaining randomness $(\mathcal{R},\{\Nell\})$.
The queries of $\pi_{\mathbf{a}}$ may depend on $\mathbf{a}$ through $f_i$; as in
PIR with side information, user queries are permitted to depend on the fixed
side-information realization, but here the effect is only parametric since the
messages $\{\Nell\}$ are independent of~$\A$.

\textbf{Step~1 (Chain-rule transfer: $D\geq\mathbb{E}_\A[D_\A]$).}
By the chain rule of entropy applied to each server:
$H(\Yans_i^{(\theta)})\geq H(\Yans_i^{(\theta)}\mid\A)$ for every $i$.
Summing and using $\sum_i H(\Yans_i^{(\theta)}\mid\A)=\mathbb{E}_\A[D_\A]$
(linearity of expectation):
\begin{equation}
  D\;=\;\sum_i H(\Yans_i^{(\theta)})\;\geq\;\sum_i H(\Yans_i^{(\theta)}\mid\A)
  \;=\;\mathbb{E}_\A[D_\A].
  \label{eq:step1_chain}
\end{equation}
Any lower bound on $\mathbb{E}_\A[D_\A]$ is therefore a lower bound on $D$.
Note: $H(Y_i\mid\A=\mathbf{a})$ may exceed $H(Y_i)$ for specific $\mathbf{a}$
(see Remark~\ref{rem:singleton}); the pointwise inequality $D\geq D_{\mathbf{a}}$
is \emph{not} used.

\textbf{Step~2 (Verify the colluding PIR converse conditions for
$\pi_{\mathbf{a}}$).}
We verify, for $P_A$-a.e.\ full-rank $\mathbf{a}$, the hypotheses of the
$P_c$-colluding PIR converse applied to retrieval of
$\Ntheta$ from the innovation library $\{\Nell\}$ (with $\mathbf{a}$ a fixed
parameter of the query/answer maps):
\begin{enumerate}[(a)]
  \item \emph{Correctness (per realization, transcript-conditioned, with
        penalty).} The decoder of the original protocol is a function of
        $(\theta,\A,\mathcal{R},Q_{1:N}^{(\theta)},\Yans_{1:N}^{(\theta)})$, so
        Fano's inequality directly bounds only the entropy conditioned on
        everything the decoder observes; two reductions align this with
        hypothesis~(i) of Lemma~\ref{lem:kadhe}. \emph{First (private
        randomness is removable).} Fix a demand $t$ and $\A=\mathbf{a}$. Given
        $(\theta{=}t,\A{=}\mathbf{a})$ we have
        $\{\Nell\}\indep\mathcal{R}$ by (M2) and (M4)--(M5); conditioning
        additionally on $Q_{1:N}^{(t)}=f(t,\mathbf{a},\mathcal{R})$, a function
        of $\mathcal{R}$ alone, preserves the product structure of the joint
        law, and the answers $\Yans_{1:N}^{(t)}$ are deterministic in
        $(\{\Nell\},Q_{1:N}^{(t)},\mathbf{a})$, so by the factorization
        identity used in the proof of Theorem~\ref{prop:rankdef_lb},
        $\{\Nell\}\indep\mathcal{R}$ given
        $(\Yans_{1:N}^{(t)},Q_{1:N}^{(t)},\theta{=}t,\A{=}\mathbf{a})$: the
        private randomness may be added to the conditioning at no entropy cost.
        \emph{Second (per-demand Fano with queries retained).} With
        $P_{e,t}(\mathbf{a})$ the error probability at demand~$t$ under
        $\A=\mathbf{a}$, Fano's inequality for the decoder---which uses
        $(t,\mathbf{a},\mathcal{R},Q,\Yans)$---gives
        \begin{align*}
          H\!\left(\N^{(t)}\,\middle|\,\Yans_{1:N}^{(t)},Q_{1:N}^{(t)},
          \A{=}\mathbf{a}\right)
          &= H\!\left(\N^{(t)}\,\middle|\,\Yans_{1:N}^{(t)},Q_{1:N}^{(t)},
          \mathcal{R},\A{=}\mathbf{a}\right)\\
          &\leq 1+P_{e,t}(\mathbf{a})\,m^2\log q.
        \end{align*}
        Since $\theta$ is uniform on $[L]$ and independent of $\A$ by~(M4),
        $P_{e,t}(\mathbf{a})\leq L\,P_e(\mathbf{a})$, so the \emph{uniform
        per-demand penalty}
        \begin{equation}
          \varepsilon_{\mathbf{a}}\;\triangleq\;\max_{t\in[L]}
          H\!\left(\N^{(t)}\,\middle|\,\Yans_{1:N}^{(t)},Q_{1:N}^{(t)},
          \A{=}\mathbf{a}\right)
          \label{eq:eps_a_fullrank}
        \end{equation}
        satisfies
        $\varepsilon_{\mathbf{a}}\leq 1+L\,P_e(\mathbf{a})\,m^2\log q$, and
        hypothesis~(i) of Lemma~\ref{lem:kadhe} holds for $\pi_{\mathbf{a}}$
        with $\varepsilon=\varepsilon_{\mathbf{a}}$, where
        $P_e(\mathbf{a})=\Pr[\text{error}\mid\A=\mathbf{a}]$. Pointwise convergence
        of $P_e(\mathbf{a})$ is not required; the penalty is retained as
        $\varepsilon_{\mathbf{a}}$ and controlled \emph{after} averaging
        over~$\A$ (Step~3), using $\mathbb{E}_\A[P_e(\A)]=P_e\to 0$ from global
        correctness~\eqref{eq:correctness}; no pointwise-convergence assumption
        is used.
  \item \emph{$P_c$-privacy after conditioning.} By side-information-robust index
        privacy (Definition~\ref{def:index}),
        $I(\theta;\{Q_i^{(\theta)},\Yans_i^{(\theta)}\}_{i\in\mathcal{T}},
        \{\B^{(\ell)}\}\mid\A=\mathbf{a})=0$ holds for $P_A$-a.e.\ $\mathbf{a}$
        \emph{by hypothesis}, hence in particular
        $I(\theta;\{Q_i^{(\theta)},\Yans_i^{(\theta)}\}_{i\in\mathcal{T}}\mid
        \{\B^{(\ell)}\},\A=\mathbf{a})=0$. Since $\{\Nell\}$ is a deterministic
        shift of $\{\B^{(\ell)}\}$ by~$\mathbf{a}$, privacy transfers to the
        innovation library. Merely unconditional privacy would not suffice here
        (Remark~\ref{rem:index_equiv}).
  \item \emph{Message independence of the side information.} By~(M2),
        $\Nell\indep\A$ for \emph{every}~$\ell$, so $\Nell\indep\mathbf{a}$. The
        independence holds library-wide, not target-only.
  \item \emph{Effective file size.} $H(\Ntheta\mid\A=\mathbf{a})=H(\Ntheta)$ for
        every full-rank $\mathbf{a}$, since $\Ntheta\indep\A$ by~(M2); flatness
        is automatic here and no further hypothesis is needed.
  \item \emph{Full-support demand.} By~(M4), $\theta$ is uniform on $[L]$ and
        independent of $\A$, so conditioned on $\A=\mathbf{a}$ every demand
        $t\in[L]$ has probability $1/L>0$; the full-support demand hypothesis of
        Lemma~\ref{lem:kadhe} holds with $S$ the constant $\A=\mathbf{a}$.
\end{enumerate}
Therefore, for $P_A$-a.e.\ full-rank $\mathbf{a}$, the induced
protocol $\pi_{\mathbf{a}}$ is a valid $P_c$-colluding PIR
instance on the innovation library---with $\mathbf{a}$ entering only
parametrically and the messages independent of~$\A$---and the standard converse
applies to $D_{\mathbf{a}}$.

\textbf{Step~3 (Per-realization bound + expectation).}
Applying the \emph{finite-error} form~\eqref{eq:kadhe_finite} of
Lemma~\ref{lem:kadhe} to the induced instance
$\pi_{\mathbf{a}}$ for $P_A$-a.e.\ full-rank $\mathbf{a}$, with $S$ the constant
$\A=\mathbf{a}$ and correctness slack
$\varepsilon=\varepsilon_{\mathbf{a}}$ from~\eqref{eq:eps_a_fullrank}---valid
for arbitrary $\varepsilon_{\mathbf{a}}\geq0$, so no pointwise convergence is
needed here:
\begin{equation}
  D_{\mathbf{a}}\;\geq\;
  \frac{H(\Ntheta\mid\A=\mathbf{a})}{\CPIR(N,P_c,L)} - \delta_{\mathbf{a}},
  \label{eq:pointwise_bound}
\end{equation}
where the penalty term is
\begin{equation}
  \delta_{\mathbf{a}} \;\triangleq\;
  \frac{\varepsilon_{\mathbf{a}}}{\CPIR(N,P_c,L)}.
  \label{eq:delta_a}
\end{equation}
By~(M2), $H(\Ntheta\mid\A=\mathbf{a})=H(\Ntheta)$ for every full-rank
$\mathbf{a}$ (entropy flatness is automatic since $\Ntheta\indep\A$).
Taking expectations over $\A$ on both sides of~\eqref{eq:pointwise_bound}:
\begin{equation}
  \mathbb{E}_\A[D_\A]\;\geq\;
  \frac{H(\Ntheta)}{\CPIR(N,P_c,L)}
  - \mathbb{E}_\A[\delta_\A],
  \qquad
  \mathbb{E}_\A[\delta_\A]=\frac{\mathbb{E}_\A[\varepsilon_\A]}{\CPIR},
  \label{eq:error_expectation}
\end{equation}
the equality by linearity of expectation and~\eqref{eq:delta_a}.
Note also that
$0\leq\varepsilon_{\mathbf{a}}\leq m^2\log q$ uniformly in $\mathbf{a}$
(a conditional entropy of a variable on an alphabet of size $q^{m^2}$), so
$\delta_\A$ is bounded and hence integrable (no regularity condition on the
expectation is required beyond this uniform bound).
By Step~2(a),
$\varepsilon_{\mathbf{a}}\leq 1+L\,P_e(\mathbf{a})\,m^2\log q$; taking
expectations and using $\mathbb{E}_\A[P_e(\A)]=P_e$ from the global correctness
condition~\eqref{eq:correctness},
$\mathbb{E}_\A[\varepsilon_\A]\leq 1+L\,P_e\,m^2\log q=o(m^2)$ for fixed
$L$ and $q$, since $P_e\to 0$---note that $P_e\to 0$ does \emph{not} imply
$P_e\,m^2\to 0$, so no vanishing of the penalty itself is claimed, only its
sublinearity at the $m^2$ scale of the bound.
Therefore $\mathbb{E}_\A[\delta_\A]=o(m^2)$, since $\CPIR$ is a constant
independent of~$m$, and using
$H(\Ntheta)=\HcondBA$ (Lemma~\ref{lem:innovation}):
\begin{equation}
  \mathbb{E}_\A[D_\A]\;\geq\;
  \frac{\HcondBA}{\CPIR(N,P_c,L)}-o(m^2).
  \label{eq:step3_expect}
\end{equation}

\textbf{Step~4 (Conclusion).}
Combining~\eqref{eq:step1_chain} and~\eqref{eq:step3_expect}:
$D\geq\mathbb{E}_\A[D_\A]\geq\HcondBA/\CPIR-o(m^2)$,
which is~\eqref{eq:converse}.
The large-library form~\eqref{eq:converse_largeL} follows by letting
$L\to\infty$ \emph{after} the limit in $m$, so that $\CPIR\to 1-P_c/N$
(Remark~\ref{rem:iterated_limit}).
\end{proof}

\section{Uniform Linear Compression of Projected Innovations}
\label{app:uniform_source}

This appendix proves Lemma~\ref{lem:uniform_source}, which supplies
Theorem~\ref{thm:rankk_public} with a family of public encoders whose
redundancy and error are controlled \emph{uniformly} over the published
row-space representatives. No uniformity of the innovation law is assumed: the
entries of $\N_m$ follow an arbitrary fixed distribution on $\Fq$, possibly
without full support.

\begin{proof}[Proof of Lemma~\ref{lem:uniform_source}]
Write $X_\mathsf{P}=\mathsf{P}\N_m\in\Fq^{k\times m}$ and let
$Z_\mathsf{P}\in\Fq^{k}$ denote one projected column. Since the columns of
$\N_m$ are i.i.d., so are the columns of $X_\mathsf{P}$, for every
$\mathsf{P}$; their common law may depend on $\mathsf{P}$ and be strongly
nonuniform.

\emph{Step 1 (A $\mathsf{P}$-free variance bound).}
Let $\imath_\mathsf{P}(z)=-\log_2 P_{Z_\mathsf{P}}(z)$ and $M=q^{k}$, so the
alphabet of $Z_\mathsf{P}$ has at most $M$ elements. At most $M$ outcomes carry
probability at most $2^{-t}$, whence
$\Pr[\imath_\mathsf{P}(Z_\mathsf{P})\geq t]\leq\min\{1,M2^{-t}\}$. Therefore
\begin{align*}
  \mathbb{E}\bigl[\imath_\mathsf{P}(Z_\mathsf{P})^2\bigr]
  &=\int_0^\infty 2t\,\Pr[\imath_\mathsf{P}(Z_\mathsf{P})\geq t]\,dt\\
  &\leq\int_0^{\log_2 M}\!\!2t\,dt+\int_{\log_2 M}^\infty\!\!2tM2^{-t}\,dt\\
  &=(\log_2 M)^2+\frac{2\log_2 M}{\ln 2}+\frac{2}{(\ln 2)^2},
\end{align*}
so that, with $C_0$ a universal constant,
\begin{equation}
  \mathrm{Var}\bigl[\imath_\mathsf{P}(Z_\mathsf{P})\bigr]
  \;\leq\;C_0\bigl(k\log_2 q+1\bigr)^2
  \label{eq:iota_var}
\end{equation}
uniformly over $\mathsf{P}$ and over the induced law. This is an analytic
consequence of the alphabet bound alone; no minimum-mass or full-support
hypothesis is used.

\emph{Step 2 (Uniform information-spectrum tail).}
By independence across the $m$ columns,
$\mathrm{Var}[\imath_\mathsf{P}(X_\mathsf{P})]\leq
C_0m(k\log_2 q+1)^2$ and
$\mathbb{E}[\imath_\mathsf{P}(X_\mathsf{P})]=H(X_\mathsf{P})$. Setting
\begin{equation}
  \tau_m:=\bigl(k\log_2 q+1\bigr)m^{3/4},
  \label{eq:tau_m}
\end{equation}
Chebyshev's inequality gives the $\mathsf{P}$-free tail bound
\begin{equation}
  \sup_{\mathsf{P}}\;
  \Pr\bigl[\imath_\mathsf{P}(X_\mathsf{P})>H(X_\mathsf{P})+\tau_m\bigr]
  \;\leq\;\frac{C_0}{\sqrt{m}}.
  \label{eq:uniform_tail}
\end{equation}

\emph{Step 3 (Encoder, in two branches).}
Define
\begin{equation}
  r_m(\mathsf{P})=\min\left\{km,\;
  \left\lceil\frac{H(X_\mathsf{P})+2\tau_m}{\log_2 q}\right\rceil\right\}.
  \label{eq:rm_branched}
\end{equation}
The cap is needed: for projected sources close to uniform on $\Fq^{k}$ the
unbranched ceiling can exceed the input dimension $km$, and no scheme should
transmit more than the uncompressed source.

\emph{Compression branch, $H(X_\mathsf{P})+2\tau_m<km\log_2 q$.} Draw
$h_{m,\mathsf{P}}\colon\Fq^{km}\to\Fq^{r_m(\mathsf{P})}$ uniformly at random
from the linear maps, so that $\Pr[h_{m,\mathsf{P}}(x)=h_{m,\mathsf{P}}(x')]
=q^{-r_m(\mathsf{P})}$ for $x\neq x'$. On the information-spectrum set
$\mathcal{T}_{m,\mathsf{P}}=\{x:-\log_2 P_{X_\mathsf{P}}(x)\leq
H(X_\mathsf{P})+\tau_m\}$, which satisfies
$|\mathcal{T}_{m,\mathsf{P}}|\leq 2^{H(X_\mathsf{P})+\tau_m}$, a unique-bin
decoder errs only if $X_\mathsf{P}\notin\mathcal{T}_{m,\mathsf{P}}$ or some
other typical sequence shares its bin. Averaging over the ensemble,
\begin{equation}
  \mathbb{E}_h\bigl[P_e(h,\mathsf{P})\bigr]
  \;\leq\;\frac{C_0}{\sqrt m}
  +|\mathcal{T}_{m,\mathsf{P}}|\,q^{-r_m(\mathsf{P})}
  \;\leq\;\frac{C_0}{\sqrt m}+2^{-\tau_m}.
  \label{eq:ensemble_error}
\end{equation}

\emph{Identity branch, $H(X_\mathsf{P})+2\tau_m\geq km\log_2 q$.} Take
$h_{m,\mathsf{P}}=\mathrm{id}$, with $r_m(\mathsf{P})=km$ and zero error; the
branch condition gives $km\log_2 q-H(X_\mathsf{P})\leq 2\tau_m$.

\emph{Step 4 (Selection, without a union bound).}
The right-hand side of~\eqref{eq:ensemble_error} does not depend on
$\mathsf{P}$. Hence for \emph{each} $\mathsf{P}$ separately there exists at
least one linear map attaining it; fix one such $h_{m,\mathsf{P}}$ for every
$\mathsf{P}\in\mathcal{P}_{m,k}$. The resulting family is deterministic, and
\begin{equation*}
  \sup_{\mathsf{P}\in\mathcal{P}_{m,k}}P_e(m,\mathsf{P})
  \;\leq\;\frac{C_0}{\sqrt m}+2^{-\tau_m}\;\longrightarrow\;0 .
\end{equation*}
No claim is made that one draw of the ensemble succeeds simultaneously for all
$\mathsf{P}$, so the cardinality
$|\mathcal{P}_{m,k}|=\binom{m}{k}_q$ never enters. This is legitimate because
$\mathsf{P}=R_\A$ is published before retrieval begins, so the protocol may fix
a different encoder for each representative.

\emph{Step 5 (Uniform redundancy).}
Both branches satisfy
$r_m(\mathsf{P})\log_2 q-H(X_\mathsf{P})\leq 2\tau_m+\log_2 q$. Since
\begin{equation*}
  \frac{\tau_m}{km}=\frac{k\log_2 q+1}{k\,m^{1/4}}=O\bigl(m^{-1/4}\bigr)
\end{equation*}
uniformly for every $1\leq k<m$ (the ratio is decreasing in $k$ and bounded by
$(\log_2 q+1)m^{-1/4}$ at $k=1$),
$\sup_{\mathsf{P}}[r_m(\mathsf{P})\log_2 q-H(\mathsf{P}\N_m)]=o(km)$.
\end{proof}

\section{Proof of Theorem~\ref{thm:achievability}: Full-Rank Achievability}
\label{app:achievability}
\begin{proof}[Proof of Theorem~\ref{thm:achievability}]
\textbf{Step~1 (Linear sketching with server-side computation).}
Fix $\varepsilon>0$. Entropies are in bits (Section~\ref{sec:formulation});
one $\Fq$-symbol carries $\log q$ bits, so set the sketch length, in
$\Fq$-symbols, to
\begin{equation*}
  r_q \;=\; \left\lceil
  \frac{H(\Ntheta)+\varepsilon m^2}{\log q}\right\rceil
  \;=\;\left\lceil\frac{\HcondBA+\varepsilon m^2}{\log q}\right\rceil
\end{equation*}
(for $q=2$ this is $H(\Ntheta)+\varepsilon m^2$ bits, up to rounding).
Choose a public \emph{linear} map $h\colon\Fq^{m\times m}\to\Fq^{r_q}$, fixed once at
setup, independent of $\A$ and $\theta$. Each server stores the sketches
\begin{equation*}
  T_\ell \;=\; h(\B^{(\ell)}),\qquad \ell\in[L],
\end{equation*}
which are computable from the stored library alone---the servers do not
know~$\A$. By linearity of~$h$,
\begin{equation}
  h(\Nell) \;=\; h(\B^{(\ell)}-\A) \;=\; T_\ell - h(\A),
  \label{eq:sketch_cancel}
\end{equation}
so the user, holding~$\A$, can recover the innovation sketch $h(\Nell)$ from
$T_\ell$ by subtracting the locally computed $h(\A)$.

\emph{Linearity is essential to this sketch-cancellation mechanism}: it is what
allows the user to cancel the shared component $h(\A)$ after retrieval. A general
(nonlinear) binning map would require the servers to evaluate $h$ on the
innovation $\Nell=\B^{(\ell)}-\A$, which they cannot form without~$\A$. This establishes the need for linearity only in the proposed cancellation
construction; it does not rule out other nonlinear schemes.

\textbf{Step~2 (Private retrieval of the target sketch).}
Treat $\{T_1,\ldots,T_L\}$ as $L$ files of length $r_q$ $\Fq$-symbols
replicated at $N$ servers.
By~(M4), $\theta$ is independent of the stored library and therefore
independent of its public sketch library, which is a deterministic function of
the stored library and the public map. The sketches
$T_\ell=h(\A)+h(\Nell)$ also have identical marginals across file indices
by~(M2). Run a capacity-achieving
$P_c$-colluding PIR protocol~\cite{sun2018colluding} to retrieve $T_\theta$ with
download, in bits,
\begin{equation*}
  D_{\mathrm{PIR}}\;\leq\;\frac{r_q\log q}{\CPIR}+o(m^2)
  \;\leq\;\frac{H(\Ntheta)+\varepsilon m^2}{\CPIR}+o(m^2),
\end{equation*}
the second inequality since $r_q\log q\leq H(\Ntheta)+\varepsilon m^2+\log q$
and $\log q=O(1)$ is absorbed in the $o(m^2)$ term.

\textbf{Step~3 (Innovation decoding and computation).}
After retrieving $T_\theta=h(\B^{(\theta)})$, the user subtracts the locally known
$h(\A)$ to obtain the innovation sketch $h(\Ntheta)=T_\theta-h(\A)$
via~\eqref{eq:sketch_cancel}. The user then recovers $\Ntheta$ from this linear
sketch by \emph{linear lossless source coding of the innovation}: since
$r_q\log q>H(\Ntheta)$ and the entries of $\Ntheta$ are i.i.d., a syndrome-based decoder
$\varphi$ succeeds with $\Pr[\varphi(h(\Ntheta))\neq\Ntheta]\to 0$~\cite{slepian1973noiseless}.
For finite-alphabet memoryless innovations the required linear source code may be
realized through random linear binning with typical-set decoding, which achieves
any rate above $H(\Ntheta)$ with vanishing error~\cite{csiszar1982linear}; for the
binary additive case this specializes to syndrome coding over $\mathbb{F}_2$. In either case the map
$h$ is linear, which is what the sketch-cancellation step $h(\Ntheta)=T_\theta-h(\A)$
requires.
This is \emph{equivalently} Slepian--Wolf coding of the raw matrix~$\btheta$ with
decoder side information~$\A$, but that is only an equivalent interpretation: the
role of~$\A$ is operational---it cancels the shared component~$h(\A)$ so that the
residual to be decoded is exactly the innovation sketch, and the entropy of
$\Ntheta$ is not itself reduced by~$\A$, since $\Ntheta\indep\A$. The user then
forms $\btheta=\A+\Ntheta$ and computes $\Ctheta=\A^\top\btheta$ locally.

\textbf{Step~4 (Privacy and security).}
\emph{Side-information-robust index privacy (Definition~\ref{def:index}).}
We verify the two hypotheses of Proposition~\ref{prop:sir_satisfiable} for the
composed source-coding/PIR scheme. (a)~The queries in Step~2 depend only on
$(\theta,\mathcal{R})$ and the public linear map~$h$, never on~$\A$, since the
sketches $\{T_\ell\}$ are formed by the servers without knowledge of~$\A$ and the
PIR queries address the sketch library. (b)~The underlying capacity-achieving
$P_c$-colluding PIR protocol has, by construction, a colluding-set query
distribution that is independent of~$\theta$ for every fixed realization of the
sketch library. \emph{Therefore, by Proposition~\ref{prop:sir_satisfiable}, the
composed scheme satisfies side-information-robust index privacy
(Definition~\ref{def:index}),} i.e.~\eqref{eq:index_privacy} holds.
Definition~\ref{def:index} is what legitimizes the per-$\mathbf{a}$ induced
instances used in the converse, and Proposition~\ref{prop:sir_satisfiable}
certifies that the composed scheme meets it at no additional cost.

\emph{Conditioned data security (Definition~\ref{def:security}).}
Given $\{\B^{(\ell)}\}$, the server view $\{Q_i^{(\theta)},\Yans_i^{(\theta)}\}$ is
a deterministic function of $(\theta,\mathcal{R},h)$, hence independent of~$\A$;
thus $I(\A;\{Q_i^{(\theta)},\Yans_i^{(\theta)}\}\mid\{\B^{(\ell)}\})=0$.
That $h$ is public introduces no $\theta$-dependence, since $h$ maps all matrices
identically and the PIR guarantee is with respect to the sketch
library $\{T_\ell\}$.
\emph{Diagonalization.} The construction above is indexed by a fixed
$\varepsilon>0$: for each $m$ it yields a scheme with download
$(\HcondBA+\varepsilon m^2)/\CPIR+o(m^2)$ whose source-decoding error
$\eta_m(\varepsilon)\to0$ as $m\to\infty$, since the sketch length exceeds
$H(\Ntheta)$ by the margin $\varepsilon m^2\to\infty$. Fix a sequence
$\varepsilon_j\downarrow 0$. For each $j$, choose $M_j$ large enough that, for
every $m\geq M_j$,
\begin{equation*}
  \eta_m(\varepsilon_j)\leq\tfrac1j
  \qquad\text{and}\qquad
  \varepsilon_j m^2\geq j,
\end{equation*}
which is possible because $\eta_m(\varepsilon_j)\to0$ and
$\varepsilon_j m^2\to\infty$ as $m\to\infty$ at fixed $j$. Taking
$M_{j+1}>M_j$ and setting $\varepsilon_m=\varepsilon_j$ for
$M_j\leq m<M_{j+1}$ gives $\varepsilon_m\to0$,
$\eta_m(\varepsilon_m)\to0$, and $\varepsilon_m m^2\to\infty$. The resulting
scheme sequence has vanishing error and redundancy
$\varepsilon_m m^2+O(1)=o(m^2)$, which gives~\eqref{eq:achievability}.
\end{proof}

\section{Finite-Length Accounting: Statement and Proof}
\label{app:finiteblock}

\begin{proposition}[Finite-length innovation-code plus finite-PIR accounting]
\label{thm:finiteblock}
Let $\Ntheta\in\{0,1\}^n$, $n=m^2$, have i.i.d.\ $\mathrm{Bern}(p)$ entries,
$p\in(0,\tfrac12)$, under (M1)--(M5) with $r=m$. Compose a binary linear syndrome source
code with decoding error at most $\epsilon_s$ and a $P_c$-colluding PIR scheme
satisfying the exactness, library-independent-query, and deterministic-answer
hypotheses (P1)--(P3) below, instantiated by the
optimal-subpacketization capacity-achieving colluding scheme of Zhang and
Xu~\cite{zhang2019optimal};
the user retrieves the shorter of the padded syndrome and the padded raw file,
of padded length $\tilde{r}_n=\min\{s_{\mathrm{base}}\lceil r_n/s_{\mathrm{base}}\rceil, s_{\mathrm{base}}\lceil n/s_{\mathrm{base}}\rceil\}$
(the source-code and uncompressed branches; $r_n$ is the syndrome length and
$s_{\mathrm{base}}=a\,s^*$ the subpacketization in base-field symbols, both
fixed in~(P1)--(P3) below). The composition
achieves fixed-length downloaded communication of exactly
$\Lambda_n=\tilde{r}_n/\CPIR(N,P_c,L)$ bits, with total error at most
$\epsilon_s$; the entropy download of the main text satisfies
$\sum_i H(\Yans_i^{(\theta)})\leq\Lambda_n$, the entropy of a fixed-length
transmission being at most its length. The scheme
satisfies side-information-robust index privacy (Definition~\ref{def:index})
and conditioned data security (Definition~\ref{def:security}). If the syndrome
length is set to the optimal-lossless-source design benchmark
$r_n=nH_2(p)+\sqrt{nV(p)}\,Q^{-1}(\epsilon_s)+O(\log n)$~\cite{kontoyiannis2014optimal}---which
we do not establish is attained at second order by the selected linear syndrome
ensemble---then
\begin{equation}
  \frac{\Lambda_n}{n}=\frac{H_2(p)}{\CPIR(N,P_c,L)}+O(n^{-1/2})+\frac{s_{\mathrm{base}}}{n}\,O(1).
  \label{eq:finiteblock_norm}
\end{equation}
\end{proposition}

\noindent The hypotheses are as follows.

\emph{(P1) Source code.} $\mathcal{C}_n$
is a binary linear syndrome source code with parity-check matrix
$H_n\in\mathbb{F}_2^{r_n\times n}$ and decoder $\varphi_n$ satisfying
\begin{equation}
  \Pr[\varphi_n(H_n\Ntheta)\neq\Ntheta]\leq\epsilon_s.
  \label{eq:sourcecode_guarantee}
\end{equation}

\emph{(P2) Exact colluding PIR.} $\Pi_{\mathrm{PIR}}$ is a $P_c$-colluding PIR
scheme for $L$ replicated files with the following three properties:
\emph{(a)~exactness}: there is a finite subpacketization $s_{\mathrm{base}}$,
measured in base-field symbols, such that any file
whose length $\bar{r}$ is a multiple of $s_{\mathrm{base}}$ is retrieved with zero error and
download exactly $D_{\mathrm{PIR}}(\bar{r};N,P_c,L)=\bar{r}/\CPIR(N,P_c,L)$, at
query upload $U_{\mathrm{PIR}}(\bar{r};N,P_c,L)$;
\emph{(b)~library-independent queries}: the queries are generated from
$(\theta,\mathcal{R})$ alone, with the same distribution for every library
realization; and
\emph{(c)~deterministic answers}: each answer is a deterministic function of the
received query and the stored files.

\emph{(P3) Instantiation.} Assume $1\leq P_c<N$. The linear capacity-achieving
$P_c$-colluding
scheme of Zhang and Xu~\cite{zhang2019optimal} satisfies (P2)(a)--(c) at the
\emph{optimal} subpacketization
$s^*(N,P_c,L)=d\,(N/d)^{L-1}$, $d=\gcd(N,P_c)$, which is also a lower bound for
all linear capacity-achieving schemes~\cite{zhang2019optimal}; the original
Sun--Jafar scheme~\cite{sun2018colluding} satisfies the same properties at
subpacketization $N^{L}$; both values are in extension-field symbols, and the
base-field padding unit of (P2)(a) is $s_{\mathrm{base}}=a\,s^*$. The small-field optimal-subpacketization construction~\cite{xu2018building}
achieves the same $s^*$ over any field with $q\geq N$; over a base field with
$q<N$ one groups symbols into extension-field symbols over $\Fq^a$ with
$a=\min\{j\geq 1:q^j\geq N\}$, and the extension degree $a$ enters the
finite-length padding through $s_{\mathrm{base}}=a\,s^*$
(Appendix~\ref{app:finiteblock}). At $P_c=0$ (no
privacy against any single server) the
formula would give $s^*=N$, but retrieval is trivial---the user reads the
desired file directly---so (P3) is stated for $1\leq P_c<N$ and the $P_c=0$ case
is handled by the trivial scheme. The asymptotic converse and capacity theorems
retain the full range $0\leq P_c<N$. Only the \emph{existence} of a finite
$s_{\mathrm{base}}$ is used below, not its value.

The scheme chooses between two retrieval modes: (i) \emph{syndrome
mode}, using $H_n\in\mathbb{F}_2^{r_n\times n}$; and (ii) \emph{uncompressed mode},
using $H_n=I_n$ (retrieve $\B^{(\theta)}$ itself, then form
$\Ntheta=\B^{(\theta)}-\A$ losslessly). Since padding can inflate the compressed
length beyond the raw innovation length $n$ (at high $p$, small $m$, or coarse
$s$), the scheme uses whichever mode has the shorter padded length. When the
subpacketization is realized over an extension field $\Fq^a$ with
$a=\min\{j\geq 1:q^j\geq N\}$ (needed when $q<N$; $a=1$ when $q\geq N$), one
extension-field PIR symbol comprises $a$ base-field symbols, so lengths measured
in base-field symbols must be multiples of $s_{\mathrm{base}}=a\,s^*$:
\begin{equation}
  \tilde{r}_n \;=\; \min\Bigl\{\,s_{\mathrm{base}}\bigl\lceil r_n/s_{\mathrm{base}}\bigr\rceil,\;\;s_{\mathrm{base}}\bigl\lceil n/s_{\mathrm{base}}\bigr\rceil\,\Bigr\}.
  \label{eq:padded_len}
\end{equation}
Then the composed innovation-retrieval scheme---servers store the public syndrome
files $H_n\B^{(\ell)}$ (or the raw files in uncompressed mode), the user retrieves
$H_n\B^{(\theta)}$ by $\Pi_{\mathrm{PIR}}$,
cancels $H_n\A$, and decodes $\Ntheta$ by $\varphi_n$---achieves, in $\mathbb{F}_2$
symbols,
\begin{align}
  \Lambda_n&=D_{\mathrm{PIR}}(\tilde{r}_n;N,P_c,L),\qquad
  P_e^{\mathrm{total}}\leq\epsilon_s,\notag\\
  U_n&=U_{\mathrm{PIR}}(\tilde{r}_n;N,P_c,L)+U_H,
  \label{eq:finiteblock_exact}
\end{align}
and satisfies side-information-robust index privacy (Definition~\ref{def:index})
and conditioned data security (Definition~\ref{def:security}). By (P2)(a) the
download is exact with \emph{no} overhead at padded lengths:
\begin{equation}
  D_{\mathrm{PIR}}(\tilde{r}_n;N,P_c,L)=\frac{\tilde{r}_n}{\CPIR(N,P_c,L)},
  \label{eq:dpir_exact}
\end{equation}
with $\Delta_{\mathrm{PIR}}=0$ since $s_{\mathrm{base}}\mid\tilde{r}_n$ and $s^*\mid s_{\mathrm{base}}$.
If moreover $\mathcal{C}_n$ is drawn from a family whose syndrome length is set
by the \emph{design normal approximation}
\begin{equation}
  r_n=nH_2(p)+\sqrt{nV(p)}\,Q^{-1}(\epsilon_s)+O(\log n),
  \label{eq:syndrome_len}
\end{equation}
where
$V(p)=p(1-p)\bigl[\log_2\frac{1-p}{p}\bigr]^2$ is the source dispersion (the
variance of the information density $-\log_2 P_N(N_{i,j})$) and $Q^{-1}$ is the
inverse of the Gaussian tail function
$Q(x)=\int_x^\infty\frac{1}{\sqrt{2\pi}}e^{-u^2/2}\,du$, then the syndrome branch
of $\tilde{r}_n$ is active for all large~$n$, and
the normalized download~\eqref{eq:finiteblock_norm} of the main text holds.
The numerical values of Table~\ref{tab:e2e} are generated from the explicit
instantiation of~\eqref{eq:syndrome_len} obtained by taking the $O(\log n)$
term to be $\tfrac12\log_2 n$ with zero additive constant, and rounding up:
\begin{equation}
  r_n=\left\lceil\, nH_2(p)+\sqrt{nV(p)}\,Q^{-1}(\epsilon_s)
  +\tfrac12\log_2 n \,\right\rceil,
  \qquad n=m^2,
  \label{eq:syndrome_len_design}
\end{equation}
with $\epsilon_s=10^{-3}$, entropies in bits, and the subpacketization padding
$\tilde{r}_n$ of~\eqref{eq:padded_len} applied afterwards. All entries of
Table~\ref{tab:e2e} are reproduced exactly
by~\eqref{eq:syndrome_len_design}; the choice of $\tfrac12\log_2 n$ is a design
convention, not a claim about the true third-order term.
Achievability by linear codes is first-order standard~\cite{csiszar1982linear};
the second-order term is used only to \emph{size} the code, not proved
achievable by the selected ensemble (Appendix~\ref{app:finiteblock}).

\begin{remark}[Subpacketization blow-up for large $L$]
\label{rem:subpacket}
Exactness~\eqref{eq:dpir_exact} costs subpacketization
$s^*=\gcd(N,P_c)(N/\gcd(N,P_c))^{L-1}$~\cite{zhang2019optimal}: modest for small
configurations ($64$ at $(4,1,4)$) but exponential in $L$, where a
smaller-subpacketization scheme with genuine $\Delta_{\mathrm{PIR}}>0$ would be
used instead; we report exact figures only where an explicit scheme makes them exact
(Table~\ref{tab:e2e}).
\end{remark}

\begin{remark}[Upload/query cost]
\label{rem:upload}
$U_n=U_{\mathrm{PIR}}(\tilde{r}_n)+U_H$; the parity-check matrix $H_n$ is a
public fixed design object (or public-seed generated), so $U_H$ is a one-time
cost and per-retrieval upload reduces to the PIR query cost. Download is the
performance criterion throughout, as standard in the PIR-capacity literature.
\end{remark}

\begin{remark}[Privacy, security, and correctness of the finite scheme]
\label{rem:finite_verification}
The three guarantees hold at finite length, and each rests on an explicit
hypothesis of Proposition~\ref{thm:finiteblock} rather than on folklore
properties of a particular construction. \emph{Index privacy}
$I(\theta;V_{\mathcal{T}},\{\B^{(\ell)}\}\mid\A)=0$: by (P2)(b) the queries
address the public syndrome library $\{H_n\B^{(\ell)}\}$ and
are generated from $(\theta,\mathcal{R})$ with the same distribution for every
library realization---in particular independently of the private entries
of~$\A$---so Proposition~\ref{prop:sir_satisfiable} applies without modification:
conditioning on $\A=\mathbf{a}$ does not affect the colluding query
distribution, which is $\theta$-independent for every $\leq P_c$ subset. The
Sun--Jafar construction satisfies (P2)(b) because its queries are drawn from a
fixed distribution determined by $(\theta,\mathcal{R})$
alone~\cite{sun2018colluding}. \emph{Conditioned data security}
$I(\A;V_{\mathcal{T}}\mid\{\B^{(\ell)}\})=0$: the public matrix $H_n$, the syndrome
files $H_n\B^{(\ell)}$, the mode selection (a deterministic function of the public
constants $(r_n,n,s_{\mathrm{base}})$), and the zero-padding are all deterministic functions of the
library and public parameters; by (P2)(b)--(c) the transcript is a function of
$(\theta,\mathcal{R},H_n)$ and the library, hence carries no $\A$-dependence beyond
the library. \emph{Correctness}: by (P2)(a) the PIR layer returns
$H_n\B^{(\theta)}$ exactly; the user computes
$H_n\Ntheta=H_n\B^{(\theta)}-H_n\A$ and decodes, and full-rank invertibility (M1)
makes $\widehat{\C}^{(\theta)}\neq\Ctheta$ equivalent to
$\widehat{\N}\neq\Ntheta$, so
$\Pr[\widehat{\C}^{(\theta)}\neq\A^\top\B^{(\theta)}]=\Pr[\varphi_n(H_n\Ntheta)\neq\Ntheta]\leq\epsilon_s$.
The only error source is the syndrome decoder.
\end{remark}


\begin{proof}[Proof of Proposition~\ref{thm:finiteblock}]
\emph{Source stage.} By (P1), $\mathcal{C}_n$ compresses $\Ntheta$ to the
$r_n$-symbol syndrome $H_n\Ntheta$ with decoding error $\leq\epsilon_s$;
(P1) is a hypothesis on the code family, not a claim proved here. First-order
achievability of the lossless source-coding rate by linear codes is
standard~\cite{csiszar1982linear}, and~\eqref{eq:syndrome_len} is the optimal
lossless-source normal approximation~\cite{kontoyiannis2014optimal} used as a
\emph{design} length: we do not claim that the selected linear syndrome
ensemble provably attains that second-order expansion, and the $O(\log n)$ term
represents the usual third-order/design margin and is not optimized here.

\emph{PIR stage.} The servers store the public syndrome files
$H_n\B^{(\ell)}=H_n\A+H_n\N^{(\ell)}$, each of length $r_n$. The user retrieves,
whichever is cheaper, either the padded syndrome
($s_{\mathrm{base}}\lceil r_n/s_{\mathrm{base}}\rceil$ symbols) or
the padded uncompressed file ($s_{\mathrm{base}}\lceil n/s_{\mathrm{base}}\rceil$
symbols, i.e.\ $H_n=I_n$); both
branches are zero-padded to a multiple of~$s_{\mathrm{base}}$, the mode choice is a deterministic
function of the public constants $(r_n,n,s_{\mathrm{base}})$ and reveals nothing, and
$\tilde{r}_n$ is the smaller padded length. By (P2)(a) the padded file is
retrieved with zero error and download exactly $\tilde{r}_n/\CPIR$,
giving~\eqref{eq:dpir_exact}; the exchange is lossless, so it adds no error.

\emph{Decoding and correctness.} The user forms
$H_n\Ntheta=H_n\B^{(\theta)}-H_n\A$ (linearity of $H_n$) and decodes by
$\varphi_n$ (the uncompressed branch decodes trivially), obtaining
$\widehat{\N}$ and outputting
$\widehat{\C}^{(\theta)}=\A^\top(\A+\widehat{\N})$. Since $\A$ is invertible
almost surely (M1), $\A^\top(\widehat{\N}-\Ntheta)\neq\mathbf{0}$ whenever
$\widehat{\N}\neq\Ntheta$, so
$\Pr[\widehat{\C}^{(\theta)}\neq\Ctheta]
=\Pr[\varphi_n(H_n\Ntheta)\neq\Ntheta]\leq\epsilon_s$,
giving~\eqref{eq:finiteblock_exact}. Index privacy and conditioned data security
follow from (P2)(b)--(c) exactly as verified in
Remark~\ref{rem:finite_verification}.

\emph{Second-order rate.} Under~\eqref{eq:syndrome_len}, $H_2(p)<1$ for
$p\in(0,\tfrac12)$ gives $r_n\leq n$ for all large $n$, hence
$\lceil r_n/s_{\mathrm{base}}\rceil\leq\lceil n/s_{\mathrm{base}}\rceil$ and the syndrome branch is active with
$\tilde{r}_n=r_n+O(s_{\mathrm{base}})$; dividing by $n$ and absorbing $O(\log n)/n$ into
$O(n^{-1/2})$ yields~\eqref{eq:finiteblock_norm}.
\end{proof}

\section{Proof of Lemma~\ref{lem:kadhe}: the Colluding Converse for Independent
Non-Uniform Messages}
\label{app:kadhe}

The standard $P_c$-colluding PIR converse~\cite{sun2018colluding} is formulated
for independent uniform messages. We prove the required finite-error extension
for independent, identically distributed, potentially non-uniform messages and
an independent decoder parameter $S$. The library to which this paper applies
the bound---the innovation library $\{\Nell\}$---is non-uniform, so we give a
self-contained proof under the hypotheses of Lemma~\ref{lem:kadhe}, tracking
exactly which properties of the message law are used. The answer is: mutual
independence, identical entropies, and independence from
$(\theta,\mathcal{R},S)$, together
with full conditional support of the demand given $S$ (used only in the
privacy-swap step~\eqref{eq:app_privacy_used});
\emph{uniformity of the message law is never used}.

Throughout, $W_{1:L}$ are mutually independent and identically distributed with
common entropy $H:=H(W_\ell)$; by hypothesis~(iii), $H(W_\ell\mid S)=H$. We
write $Q^{[\theta]}_{1:N}$ and $\Yans^{[\theta]}_{1:N}$ for the queries and
answers when the retrieval index is $\theta$, so
$Q^{[\theta]}_i=f_i(\theta,\mathcal{R},S)$ and
$\Yans^{[\theta]}_i=\varphi_i(Q^{[\theta]}_i,W_{1:L},S)$, with
$(\theta,\mathcal{R},S)\indep W_{1:L}$. Let $\varepsilon\geq 0$ denote the
finite-error correctness slack of hypothesis~(i); after deriving the
finite-error inequality we specialize to $\varepsilon=o(\log|\mathcal{W}|)$ for
the asymptotic statement. Hypothesis~(i) is the
transcript-conditioned bound
$H(W_\theta\mid\Yans^{[\theta]}_{1:N},Q^{[\theta]}_{1:N},S)\leq\varepsilon$ for
each demand, and since extra conditioning on $W_J$ cannot increase entropy,
(i) implies
\begin{multline}
  H\!\left(W_{\theta}\mid \Yans^{[\theta]}_{1:N},Q^{[\theta]}_{1:N},
  W_{J},S\right)\leq\varepsilon\\
  \text{for every }\theta\in[L],\ J\subseteq[L]\setminus\{\theta\}.
  \label{eq:app_correctness}
\end{multline}

\begin{lemma}[Answer locality]
\label{lem:app_locality}
For every $\theta\in[L]$, every $\mathcal{T}\subseteq[N]$, and every
$J\subseteq[L]$,
\begin{equation*}
  I\!\left(\Yans^{[\theta]}_{\mathcal{T}};\,Q^{[\theta]}_{\mathcal{T}^c}
  \,\middle|\,Q^{[\theta]}_{\mathcal{T}},W_J,S\right)=0.
\end{equation*}
\end{lemma}
\begin{proof}
Since $W_{1:L}\indep(\theta,\mathcal{R},S)$ and the queries are functions of
$(\theta,\mathcal{R},S)$, we have $W_{1:L}\indep(Q^{[\theta]}_{1:N},S)$; with
mutual independence of the messages this gives
$P(w_{J^c}\mid q_{1:N},s,w_J)=P(w_{J^c})$, i.e.\
$W_{J^c}\indep Q^{[\theta]}_{\mathcal{T}^c}$ given
$(Q^{[\theta]}_{\mathcal{T}},W_J,S)$. Conditioned on
$(Q^{[\theta]}_{\mathcal{T}},W_J,S)$, the answers
$\Yans^{[\theta]}_{\mathcal{T}}
=\varphi_{\mathcal{T}}(Q^{[\theta]}_{\mathcal{T}},W_J,W_{J^c},S)$ are a
deterministic function of $W_{J^c}$ alone, so they inherit this conditional
independence.
\end{proof}

\begin{lemma}[Conditional Shearer inequality]
\label{lem:app_shearer}
Let $X_{1:N}$ and $C$ be jointly distributed, and let $\mathcal{F}$ be a family
of subsets of $[N]$ in which every index appears in at least $t$ members. Then
$\sum_{\mathcal{T}\in\mathcal{F}}H(X_{\mathcal{T}}\mid C)\geq t\,H(X_{1:N}\mid C)$.
\end{lemma}
\begin{proof}
By the chain rule and the fact that conditioning cannot increase entropy,
$H(X_{\mathcal{T}}\mid C)=\sum_{i\in\mathcal{T}}
H(X_i\mid X_{\mathcal{T}\cap[i-1]},C)
\geq\sum_{i\in\mathcal{T}}H(X_i\mid X_{[i-1]},C)$. Summing over
$\mathcal{T}\in\mathcal{F}$, each index $i$ contributes at least $t$ times, so
the sum is at least
$t\sum_{i=1}^{N}H(X_i\mid X_{[i-1]},C)=t\,H(X_{1:N}\mid C)$~\cite{chung1986some}.
\end{proof}

\begin{proposition}[Colluding converse, non-uniform messages]
\label{prop:app_converse}
Under the hypotheses of Lemma~\ref{lem:kadhe}, for every fixed $\theta\in[L]$,
\begin{equation*}
  \sum_{i=1}^{N}H\!\left(\Yans^{[\theta]}_i\mid S\right)
  \;\geq\;(H-\varepsilon)\sum_{k=0}^{L-1}\left(\frac{P_c}{N}\right)^{\!k}
  \;=\;\frac{H-\varepsilon}{\CPIR(N,P_c,L)}.
\end{equation*}
Consequently $D_S\geq H(W_\theta\mid S)/\CPIR-o(\log|\mathcal{W}|)$ for fixed
$L$, both for fixed $\theta$ and for random $\theta$ (since
$H(\Yans_i^{(\theta)}\mid S)\geq H(\Yans_i^{(\theta)}\mid S,\theta)$ and each
per-$\theta$ instance obeys the same bound).
\end{proposition}
\begin{proof}
Relabel the messages so that the fixed retrieval index is $\theta=1$ and the
remaining messages are visited in the order $2,\ldots,L$; privacy~(ii) holds for
every pair of indices, so the labeling is immaterial.

\emph{Base step.} Using subadditivity, then the deterministic-answer property
$H(\Yans^{[1]}_{1:N}\mid Q^{[1]}_{1:N},W_{1:L},S)=0$ and
$H(W_1\mid Q^{[1]}_{1:N},S)=H$ (messages independent of queries and side
information):
\begin{align}
  \sum_{i}H(\Yans^{[1]}_i\mid S)
  &\geq H\!\left(\Yans^{[1]}_{1:N}\mid Q^{[1]}_{1:N},S\right)\notag\\
  &= I\!\left(W_{1:L};\Yans^{[1]}_{1:N}\mid Q^{[1]}_{1:N},S\right)\notag\\
  &= I\!\left(W_1;\Yans^{[1]}_{1:N}\mid Q^{[1]}_{1:N},S\right)\notag\\
  &\quad + H\!\left(\Yans^{[1]}_{1:N}\mid W_1,Q^{[1]}_{1:N},S\right)\notag\\
  &\geq (H-\varepsilon)\notag\\
  &\quad + H\!\left(\Yans^{[1]}_{1:N}\mid W_1,Q^{[1]}_{1:N},S\right),
  \label{eq:app_base}
\end{align}
where the last line uses~\eqref{eq:app_correctness} with $\theta=1$.

\emph{Induction step.} Fix $k\in[L-1]$ and let
$\mathcal{T}_n:=\{n,n+1,\ldots,n+P_c-1\}\ (\mathrm{mod}\ N)$, $n\in[N]$, be the
cyclic windows of size $P_c$; every server lies in exactly $P_c$ of them. For
each $n$,
\begin{align}
  H\!\left(\Yans^{[k]}_{1:N}\mid W_{1:k},Q^{[k]}_{1:N},S\right)
  &\geq H\!\left(\Yans^{[k]}_{\mathcal{T}_n}\mid W_{1:k},Q^{[k]}_{1:N},S\right)
  \label{eq:app_subset}\\
  &= H\!\left(\Yans^{[k]}_{\mathcal{T}_n}\mid W_{1:k},Q^{[k]}_{\mathcal{T}_n},S\right)
  \label{eq:app_locality_used}\\
  &= H\!\left(\Yans^{[k+1]}_{\mathcal{T}_n}\mid W_{1:k},Q^{[k+1]}_{\mathcal{T}_n},S\right)
  \label{eq:app_privacy_used}\\
  &\geq H\!\left(\Yans^{[k+1]}_{\mathcal{T}_n}\mid W_{1:k},Q^{[k+1]}_{1:N},S\right),
  \label{eq:app_recondition}
\end{align}
where \eqref{eq:app_subset} drops answers, \eqref{eq:app_locality_used} is
Lemma~\ref{lem:app_locality} with $J=[k]$,
\eqref{eq:app_privacy_used} is $P_c$-privacy~(ii) \emph{together with} message
independence: privacy gives
$P(q_{\mathcal{T}_n},y_{\mathcal{T}_n}\mid w_{1:L},s,\theta=u)
=P(q_{\mathcal{T}_n},y_{\mathcal{T}_n}\mid w_{1:L},s)$ for every demand $u$ in
the conditional support of $\theta$ given $S=s$---all of $[L]$, by the
full-support demand hypothesis of Lemma~\ref{lem:kadhe}, and this is the one
step where that hypothesis is used: a demand of conditional probability
zero would be unconstrained by the mutual-information form of~(ii)---while
$(\theta,\mathcal{R},S)\indep W_{1:L}$ gives
$P(w_{1:L}\mid s,\theta=u)=P(w_{1:L})$ for every such $u$. Hence, for every
$s$, the conditional law of
$(Q^{[u]}_{\mathcal{T}_n},\Yans^{[u]}_{\mathcal{T}_n},W_{1:L})$ given $S=s$ is
the same under demand $u=k$ and demand $u=k+1$; so is its $(W_{1:k})$-marginal,
and the conditional entropies---averages over $S$ of per-$s$ quantities
computed from this common law---coincide; and \eqref{eq:app_recondition} adds conditioning.
Averaging over $n\in[N]$ and applying Lemma~\ref{lem:app_shearer} with
$t=P_c$, $|\mathcal{F}|=N$, and $C=(W_{1:k},Q^{[k+1]}_{1:N},S)$:
\begin{multline}
  H\!\left(\Yans^{[k]}_{1:N}\mid W_{1:k},Q^{[k]}_{1:N},S\right)\\
  \geq\frac{P_c}{N}\,
  H\!\left(\Yans^{[k+1]}_{1:N}\mid W_{1:k},Q^{[k+1]}_{1:N},S\right).
  \label{eq:app_shrink}
\end{multline}
Exactly as in the base step, peeling off message $W_{k+1}$ with
correctness~\eqref{eq:app_correctness} at $\theta=k+1$ and
$H(W_{k+1}\mid W_{1:k},Q^{[k+1]}_{1:N},S)=H$ (mutual independence of the
messages and their independence from queries and side information):
\begin{multline}
  H\!\left(\Yans^{[k+1]}_{1:N}\mid W_{1:k},Q^{[k+1]}_{1:N},S\right)\\
  \geq (H-\varepsilon)
  + H\!\left(\Yans^{[k+1]}_{1:N}\mid W_{1:k+1},Q^{[k+1]}_{1:N},S\right).
  \label{eq:app_peel}
\end{multline}

\emph{Unrolling.} Combining \eqref{eq:app_base}, \eqref{eq:app_shrink},
and~\eqref{eq:app_peel} for $k=1,\ldots,L-1$, and dropping the final
nonnegative term,
\begin{equation*}
  \sum_{i}H(\Yans^{[1]}_i\mid S)
  \geq (H-\varepsilon)\sum_{k=0}^{L-1}\left(\frac{P_c}{N}\right)^{\!k}
  =\frac{H-\varepsilon}{\CPIR(N,P_c,L)}.
\end{equation*}
Since $L$ is fixed and $\varepsilon=o(\log|\mathcal{W}|)$, the loss
$\varepsilon/\CPIR\leq L\varepsilon$ is $o(\log|\mathcal{W}|)$, which proves the
claim; hypothesis~(iii) gives $H=H(W_\theta\mid S)$, matching the statement of
Lemma~\ref{lem:kadhe}.
\end{proof}

\begin{lemma}[Answer locality under refined storage]
\label{lem:app_locality_refined}
Under hypotheses (i)--(iv) of Theorem~\ref{thm:refined_converse}---in
particular, $X_{1:L}\indep(\theta,\mathcal{R})$ given $S$, per-file coarsening
$W_\ell=\phi(X_\ell)$, queries $Q_i=f_i(\theta,\mathcal{R},S)$, and answers
$\Yans_i=g_i(Q_i,X_{1:L},S)$ deterministic in the refined storage---for every
$\mathcal{T}$ with $|\mathcal{T}|\leq P_c$, every $J\subseteq[L]$, and every
demanded index~$\theta$,
\begin{equation}
  \Yans^{[\theta]}_{\mathcal{T}}
  \;\indep\;
  Q^{[\theta]}_{\mathcal{T}^c}
  \quad\text{given}\quad
  \bigl(Q^{[\theta]}_{\mathcal{T}},\,W_J,\,S\bigr).
  \label{eq:locality_refined}
\end{equation}
The proof uses only the product structure of hypothesis~(i), the
\emph{per-file} form of the coarsening, and the fact that $g_{\mathcal{T}}$
takes $q_{\mathcal{T}}$ as its only query argument; no property of the residual
$X_\ell$ given $W_\ell$ is used.
\end{lemma}
\begin{proof}
Fix $s$ and a conditioning event
$E:=\{Q^{[\theta]}_{\mathcal{T}}=q_{\mathcal{T}},\,W_J=w_J,\,S=s\}$ of positive
probability (all alphabets are finite, so conditional laws are defined exactly
on such events, and no normalization subtlety arises). By hypothesis~(i) the
joint law factorizes as
$P(x_{1:L},\theta,r\mid s)=P(x_{1:L}\mid s)\,P(\theta,r\mid s)$. The two
constraints defining $E$ live in \emph{disjoint coordinates} of this product:
the coarsening event constrains $x_{1:L}$ alone, and per-file,
\begin{equation*}
  \mathbbm{1}\{W_J=w_J\}
  =\prod_{\ell\in J}\mathbbm{1}\{\phi(x_\ell)=w_\ell\},
\end{equation*}
while the query event constrains $(\theta,r)$ alone,
$\mathbbm{1}\{f_{\mathcal{T}}(\theta,r,s)=q_{\mathcal{T}}\}$. Hence
\begin{multline*}
  P(x_{1:L},\theta,r\mid E)
  =\frac{1}{P(E\mid s)}\,
  P(x_{1:L}\mid s)\prod_{\ell\in J}\mathbbm{1}\{\phi(x_\ell)=w_\ell\}\\
  \times\,
  P(\theta,r\mid s)\,\mathbbm{1}\{f_{\mathcal{T}}(\theta,r,s)=q_{\mathcal{T}}\},
\end{multline*}
which is again a product of a function of $x_{1:L}$ and a function of
$(\theta,r)$; therefore
\begin{equation}
  X_{1:L}\;\indep\;(\theta,\mathcal{R})
  \quad\text{given}\quad (Q^{[\theta]}_{\mathcal{T}},W_J,S).
  \label{eq:locality_product}
\end{equation}
Conditioned on $E$, the colluding answers are
\begin{equation}
  \Yans^{[\theta]}_{\mathcal{T}}
  = g_{\mathcal{T}}\bigl(q_{\mathcal{T}},\,X_{1:L},\,s\bigr),
  \label{eq:locality_crux}
\end{equation}
in which $g_{\mathcal{T}}$ takes $q_{\mathcal{T}}$ as its \emph{only} query
argument; so although the answers in~\eqref{eq:locality_crux} may read the
\emph{entire} refined library
$X_{1:L}$, including files outside~$J$, they are, on~$E$, a fixed
deterministic function of $X_{1:L}$ alone. Meanwhile
$Q^{[\theta]}_{\mathcal{T}^c}=f_{\mathcal{T}^c}(\theta,\mathcal{R},s)$ is a
function of $(\theta,\mathcal{R})$ alone. Functions of conditionally
independent variables are conditionally independent,
so~\eqref{eq:locality_product} gives~\eqref{eq:locality_refined}. Although
$\Yans^{[\theta]}_{\mathcal{T}}$ may depend on all refined files, including
those outside~$J$, \eqref{eq:locality_product} shows that the entire refined
library remains conditionally independent of $Q^{[\theta]}_{\mathcal{T}^c}$.
Reading those files therefore does not reintroduce a dependence on
$Q^{[\theta]}_{\mathcal{T}^c}$.
\end{proof}

\begin{proof}[Proof of Theorem~\ref{thm:refined_converse}]
We keep the refined-storage notation $X_\ell$ throughout; the hypotheses of
Proposition~\ref{prop:app_converse} are recovered in the special case
$X_\ell=W_\ell$ (exactly stored messages). Answers are
$\Yans^{[\theta]}_i=g_i(Q^{[\theta]}_i,X_{1:L},S)$; they need not be measurable
in $(Q,W_{1:L},S)$, so every step below is justified without message-level
answer determinism. Throughout, by hypothesis~(i) and the per-file
coarsening~(ii), $W_{1:L}\indep(\theta,\mathcal{R})$ given $S$, hence
$W_J\indep Q^{[\theta]}_{1:N}$ given $S$ for every $J$ and $\theta$; and by
hypothesis~(ii), $H(W_{k+1}\mid W_{1:k},S)=H$ for every $k$.

\emph{Base step.} By subadditivity and the chain identity (no determinism
used),
\begin{align}
  D_S=\sum_{i}H(\Yans^{[1]}_i\mid S)
  &\geq H\!\left(\Yans^{[1]}_{1:N}\mid Q^{[1]}_{1:N},S\right)\notag\\
  &= I\!\left(W_1;\Yans^{[1]}_{1:N}\mid Q^{[1]}_{1:N},S\right)\notag\\
  &\quad+H\!\left(\Yans^{[1]}_{1:N}\mid W_1,Q^{[1]}_{1:N},S\right)\notag\\
  &\geq (H-\varepsilon)
  + H\!\left(\Yans^{[1]}_{1:N}\mid W_1,Q^{[1]}_{1:N},S\right),
  \label{eq:ref_base}
\end{align}
since $I(W_1;\Yans^{[1]}_{1:N}\mid Q^{[1]}_{1:N},S)
=H(W_1\mid Q^{[1]}_{1:N},S)-H(W_1\mid\Yans^{[1]}_{1:N},Q^{[1]}_{1:N},S)
\geq H-\varepsilon$: the first term equals $H$ because
$W_1\indep Q^{[1]}_{1:N}$ given $S$, and the second is at most $\varepsilon$ by
correctness~(v) at $(t,J)=(1,\emptyset)$.

\emph{Induction step.} Fix $k\in[L-1]$ and let
$\mathcal{T}_n:=\{n,\ldots,n+P_c-1\}\ (\mathrm{mod}\ N)$ be the cyclic windows
of size $P_c$. For each $n$,
\begin{align}
  H\!\left(\Yans^{[k]}_{1:N}\mid W_{1:k},Q^{[k]}_{1:N},S\right)
  &\geq H\!\left(\Yans^{[k]}_{\mathcal{T}_n}\mid W_{1:k},Q^{[k]}_{1:N},S\right)
  \label{eq:ref_subset}\\
  &= H\!\left(\Yans^{[k]}_{\mathcal{T}_n}\mid W_{1:k},Q^{[k]}_{\mathcal{T}_n},S\right)
  \label{eq:ref_locality}\\
  &= H\!\left(\Yans^{[k+1]}_{\mathcal{T}_n}\mid W_{1:k},Q^{[k+1]}_{\mathcal{T}_n},S\right)
  \label{eq:ref_privacy}\\
  &\geq H\!\left(\Yans^{[k+1]}_{\mathcal{T}_n}\mid W_{1:k},Q^{[k+1]}_{1:N},S\right).
  \label{eq:ref_recondition}
\end{align}
Here \eqref{eq:ref_subset} drops answers; \eqref{eq:ref_locality} is
Lemma~\ref{lem:app_locality_refined} with $J=[k]$, which supplies exactly the
conditional independence
$\Yans^{[k]}_{\mathcal{T}_n}\indep Q^{[k]}_{\mathcal{T}_n^c}
\mid(Q^{[k]}_{\mathcal{T}_n},W_{1:k},S)$ that message-level determinism
supplied in the exact-storage proof; and \eqref{eq:ref_recondition} adds
conditioning.

\emph{The swap~\eqref{eq:ref_privacy}, with explicit marginalization.} By
privacy~(vi) and the full-support demand of hypothesis~(i), for every $s$ and
every $u\in[L]$,
$P(q_{\mathcal{T}_n},y_{\mathcal{T}_n}\mid w_{1:L},s,\theta{=}u)
=P(q_{\mathcal{T}_n},y_{\mathcal{T}_n}\mid w_{1:L},s)$---a demand of zero
conditional probability would be unconstrained by the mutual-information
form of~(vi), and this is the only step using full support. By
$W_{1:L}\indep\theta$ given $S$,
$P(w_{1:L}\mid s,\theta{=}u)=P(w_{1:L}\mid s)$. Multiplying, the joint
conditional law of
$(Q^{[u]}_{\mathcal{T}_n},\Yans^{[u]}_{\mathcal{T}_n},W_{1:L})$ given $S=s$ is
the same for every demand $u$; summing out $w_{k+1:L}$, so is the law of
$(Q^{[u]}_{\mathcal{T}_n},\Yans^{[u]}_{\mathcal{T}_n},W_{1:k})$ given $S=s$.
The conditional entropy in~\eqref{eq:ref_privacy} is a functional of this
common law (averaged over $S$), so its values at $u=k$ and $u=k+1$ coincide.

\emph{Shearer with common conditioning.} Averaging
\eqref{eq:ref_subset}--\eqref{eq:ref_recondition} over $n\in[N]$ and applying
Lemma~\ref{lem:app_shearer} with $t=P_c$, $|\mathcal{F}|=N$, and the common
conditioning $C=(W_{1:k},Q^{[k+1]}_{1:N},S)$---common across windows precisely
because \eqref{eq:ref_recondition} restored the full query vector first---
\begin{multline}
  H\!\left(\Yans^{[k]}_{1:N}\mid W_{1:k},Q^{[k]}_{1:N},S\right)\\
  \geq\frac{P_c}{N}\,
  H\!\left(\Yans^{[k+1]}_{1:N}\mid W_{1:k},Q^{[k+1]}_{1:N},S\right).
  \label{eq:ref_shrink}
\end{multline}

\emph{Peeling.} By the chain identity (again without determinism),
\begin{multline}
  H\!\left(\Yans^{[k+1]}_{1:N}\mid W_{1:k},Q^{[k+1]}_{1:N},S\right)\\
  \geq (H-\varepsilon)
  + H\!\left(\Yans^{[k+1]}_{1:N}\mid W_{1:k+1},Q^{[k+1]}_{1:N},S\right),
  \label{eq:ref_peel}
\end{multline}
since
$I(W_{k+1};\Yans^{[k+1]}_{1:N}\mid W_{1:k},Q^{[k+1]}_{1:N},S)\geq H-\varepsilon$:
$H(W_{k+1}\mid W_{1:k},Q^{[k+1]}_{1:N},S)=H$ by conditional mutual
independence~(ii) and $W_{1:L}\indep Q^{[k+1]}_{1:N}$ given $S$, and
correctness~(v) at $(t,J)=(k{+}1,[k])$ bounds the subtracted entropy
by~$\varepsilon$.

\emph{Unrolling and the final residual.} Iterating
\eqref{eq:ref_base}, \eqref{eq:ref_shrink}, \eqref{eq:ref_peel} for
$k=1,\ldots,L-1$,
\begin{multline*}
  D_S \geq (H-\varepsilon)\sum_{k=0}^{L-1}\left(\frac{P_c}{N}\right)^{\!k}\\
  +\left(\frac{P_c}{N}\right)^{\!L-1}
  H\!\left(\Yans^{[L]}_{1:N}\mid W_{1:L},Q^{[L]}_{1:N},S\right).
\end{multline*}
Under exact message storage the last entropy is zero (answers determined by
$(Q,W_{1:L},S)$); under refined storage it is merely nonnegative---residual
content of $X_{1:L}$ can survive conditioning on all functional messages---and
it sits additively on the lower-bound side, so dropping it only weakens the
bound. This yields $D_S\geq(H-\varepsilon)/\CPIR(N,P_c,L)$,
i.e.~\eqref{eq:refined_converse}.

The argument does not require the refinement variables $X_1,\ldots,X_L$ to be
independent: it uses only the conditional independence of the functional
messages $W_\ell$ given $S$, in the base and peeling steps, and the
storage--demand independence $X_{1:L}\indep(\theta,\mathcal{R})$ given $S$ of
hypothesis~(i), in the locality and swap steps. Refinements whose residuals are
correlated across files, but whose $W_\ell$ are mutually independent, are
covered verbatim; in particular every server may store the full refined library
of every file, and the bound is unchanged.
\end{proof}

\begin{remark}[Assumptions used in the converse]
\label{rem:app_nonuniform}
The proof uses mutual independence of the functional messages, equality of
their conditional entropies, conditional independence of the messages from
$(\theta,\mathcal{R})$ given $S$ (the message law may depend on $S$),
deterministic answers at the refined-storage level, and full conditional
support of the demand given $S$. It does not require uniform messages or
memoryless message coordinates. The full-support condition is what keeps the
mutual-information privacy hypothesis from leaving zero-probability demands
unconstrained in the privacy-swap steps~\eqref{eq:app_privacy_used}
and~\eqref{eq:ref_privacy}; in every
application (M4) makes $\theta$ uniform on $[L]$ and independent of the
conditioning, so it holds automatically. The message law enters only through
its entropy: no step invokes uniformity, typicality, or the cardinality of the
message alphabet beyond the slack bookkeeping. The bound therefore applies
without modification to the innovation library, whose messages are independent
and identically distributed with common entropy $H(\Ntheta)$ by (M2)--(M3).
The innovation messages may be nonuniform; whenever
$H(\Ntheta)<m^2\log q$, that entropy reduction is the source of the correlation
gain. In the per-realization application of
Theorems~\ref{thm:equivalence} and~\ref{thm:converse}, the side information is
the constant $S=\mathbf{a}$, the conditional-independence condition is trivial,
and privacy is supplied by the conditional privacy of the induced instance
(Theorem~\ref{thm:equivalence}).
\end{remark}

\bibliographystyle{IEEEtran}

\end{document}